\documentclass[12pt]{article}
\usepackage{epsfig}
\usepackage{amsfonts}
\usepackage{amscd}
\usepackage{latexsym}
\usepackage{amsmath,amssymb,amsthm}
\usepackage{verbatim}
\usepackage{setspace}
\usepackage{color}
\usepackage{cite}
\usepackage{graphicx}
\usepackage{mathtools}
\usepackage[all]{xy}
\usepackage{tikz}
\usetikzlibrary{patterns}

\usepackage[textheight=9in, textwidth=6.5in, letterpaper]{geometry}

\usepackage{color}   
\usepackage{hyperref}
\hypersetup{
    colorlinks=true,  
    linktoc=all,     
    linkcolor=black,  
    citecolor=black,
    filecolor=black,
    urlcolor=black,
}

\numberwithin{equation}{section}

\def\p{\partial}
\def\cl{{\cal L}}

\def\<{\langle}
\def\>{\rangle}

\def\cO{\mathcal{O}}

\def\be{\begin{equation}}
\def\ee{\end{equation}}
\def\beq{\be\begin{array}{c}}
\def\eeq{\end{array}\ee}
\def\bes{\be\begin{split}}
\def\ees{\end{split} \ee}
\def\bs{\begin{split}}
\def\es{\end{split} }
\def\nn{\nonumber}

\def\b{{\beta}}
\def\a{{\alpha}}
\def\g{{ \gamma}}
\def\d{{\delta}}
\def\e{{\epsilon}}

\def\v{{\varphi}}
\def\G{\Gamma}

   \makeatletter
  \let\over=\@@over \let\overwithdelims=\@@overwithdelims
  \let\atop=\@@atop \let\atopwithdelims=\@@atopwithdelims
  \let\above=\@@above \let\abovewithdelims=\@@abovewithdelims
\renewcommand\section{\@startsection {section}{1}{\z@}%
                                   {-3.5ex \@plus -1ex \@minus -.2ex}
                                   {2.3ex \@plus.2ex}%
                                   {\normalfont\large\bfseries}}

\renewcommand\subsection{\@startsection{subsection}{2}{\z@}%
                                     {-3.25ex\@plus -1ex \@minus -.2ex}%
                                     {1.5ex \@plus .2ex}%
                                     {\normalfont\bfseries}}

\begin{document}
\begin{titlepage}
\unitlength = 1mm

\vskip 1cm
\begin{center}

{ \Large {\textsc{Tropical Mirror Symmetry: Correlation functions}}} \\

\vspace{1cm}

Andrey Losev\\
\vspace{0.3cm}
{\it Wu Wen-Tsun Key Lab of Mathematics, Chinese Academy of Sciences, USTC,
No.96, JinZhai Road Baohe District, Hefei, Anhui, 230026, P.R.China \\
\vspace{0.3cm}
National Research University Higher School of Economics, Laboratory of Mirror Symmetry, NRU HSE, 6 Usacheva str., Moscow, Russia, 119048\\
}

\vspace{1cm}
 Vyacheslav Lysov\\
 \vspace{0.3cm}
{\it Okinawa Institute of Science and Technology,\\
 1919-1 Tancha, Onna-son, Okinawa 904-0495, Japan}
\vspace{1cm}

\begin{abstract}
We formulate the mirror symmetry for correlation functions of tropical observables. We prove the tropical  mirror  correspondence for correlation functions of evaluation observables on toric space. The key point of the proof  is the  localization of   correlation functions for mirror states  in type-B higher topological quantum mechanics on trees. The correlation functions localize  to the correlation functions of holomorphic functions, defined recursively in Landau-Ginzburg-Saito theory with exponential  mirror superpotential and  tropical  good section.
\end{abstract}

\vspace{1.0cm}

\end{center}

\end{titlepage}

\pagestyle{empty}
\pagestyle{plain}

\pagenumbering{arabic}

\tableofcontents

\newpage

\section{Introduction}

 The real  tropical numbers is a  set of real numbers, extended by   $\{-\infty\}$ with arithmetic operations:  tropical addition   $x+_Ty = \max(x,y)$  and tropical multiplication $x\ast_Ty = x+y $.
The set of tropical numbers is a semigroup with respect to tropical addition. The  tropical numbers appeared at different times at several branches of mathematics: Maslov introduced notion of tropical integration,  while computer scientist Imre Simon
introduced  and adjective {\it tropical}.

Mikhalkin  \cite{Mikh}, \cite{Mikhalkin2004},  \cite{mikhalkin2009tropical} used tropical numbers to study geometric problems. The collection of problems and methods become the  Tropical Geometry research area. 
In many cases the enumerative problems in algebraic geometry over complex numbers  have tropical counterparts. In particular, we can define the tropical Gromov-Witten invariants    by  counting  graphs  passing through some cycles, see Mikhalkin  \cite{Mikhalkin2004,Mikh}. 

Mikhalkin \cite{Mikhalkin2004} observed that the number of tropical curves in $\mathbb{P}^2$ of degree 3, passing through 8 points in general position is 12.  This number,   matches with counting of degree-3 algebraic curves  passing through 8  points. 
Gathmann and  Markwig \cite{Hannah}   showed that  equality generalizes for all Gromov-Witten invariants of $\mathbb{P}^2$.   The matching of tropical and complex invariants was formalized into:
\newline\newline
{\bf Theorem (tropical correspondence)}:  Gromov–Witten invariant  coincides with its tropical counterpart.
\newline\newline
We propose to analyze the tropical correspondence theorem in context of mirror symmetry.
\newline\newline
{\bf Theorem (mirror symmetry)}: The Gromov–Witten invariants for   $X$  equal  to the correlation functions in  Landau-Ginzburg-Saito theory on mirror space $X^{\vee}$.
\newline\newline
The mirror symmetry for toric space $X$ relates \cite{givental1995quantum} the  GW invariants to  Landau-Ginzburg-Saito theory with exponential superpotentials, see   \cite{vafa2003mirror} for review.   
\newline\newline
{\bf Theorem (tropical mirror symmetry)}: The tropical  Gromov–Witten invariants for toric space   $X$ of complex dimension $N$  equal  to the correlation functions in  Landau-Ginzburg-Saito theory on mirror space $X^{\vee} = {\mathbb{C}^\ast}^N$ with certain exponential superpotential, canonical holomorphic top form and mirror  K. Saito's   good section.
\newline\newline
Our main result is  the proof of the tropical mirror symmetry theorem. There are two possible application of our result: We can assume the  tropical correspondence theorem   and use our proof as a new proof of mirror symmetry for the toric spaces. Alternatively, we can assume the mirror symmetry theorem and use our proof as a proof for the tropical correspondence for toric spaces.

In our proof we use several key ideas: tropical numbers as a scaling of the exponential map,  quantum mechanics representation of the tropical GW invariants,  mirror relation for quantum mechanics as  summation over divisor states and localization for HTQM correlation functions. 

Real tropical numbers can be constructed from the field of real numbers with usual addition and multiplication by performing the exponential map $X = e^{\frac{x}{\e}}$ followed by the limit $\e \to 0$. The exponential map turns multiplication into addition  for any value of $\e$, while the tropical addition requires the limit
\be\nn
   \lim_{\e \to 0  } \;\; \e \ln (e^{\frac{x}{\e}}+e^{\frac{y}{\e}}) = \max (x,y) = x+_Ty.
\ee
The scaling of exponential map generalizes to the complex  geometry. The scaling construction also provides an heuristic proof of the  Tropical Correspondence theorem.

The Gromov-Witten invariants  over complex numbers can be described using $A$-type twisted  topological string theory \cite{Witten:1988xj}. In our work  \cite{Losev:2022tzr} we proved that the  tropical GW invariants  can be described using the  higher topological quantum mechanics (HTQM) with circle action on graphs, introduced in \cite{losev2019tqft}. Certainly, the HTQM  can be constructed as the $\e \to 0$ limit of maximally degenerate complex structure (very long strings) in topological string theory, but the fact that it captures all tree-level tropical  GW invariants is a novel result.  Moreover in \cite{Losev_2007} (see also  \cite{losev2019tqft})  it was shown that the quantum mechanics,  similar to the one we described in previous paper,  provides a   solution to the WDVV equations.  

Authors of  \cite{Frenkel:2005ku} used the 2d CFT to  relate the sum  over  holomortex insertions   in A-type topological string  with  superpotential 
deformation of the 2d CFT. In our paper   \cite{Losev:2022tzr} we showed that the  sum over  divisor states  in A-type HTQM amplitudes   equals to the amplitudes in B-type HTQM. The  B-type HTQM  is the deformation of the A-type HTQM by the boundary  divisor states. In particular we showed that for the case of toric spaces  the B-type HTQM has exponential superpotential.  

The topological  nature of the B-type topological string allows for the drastic simplification for correlation functions of certain observables.  In particular,   the correlation functions  of the  mirrored evaluation observables can be written using the  recursive construction, discussed in section \ref{sec_corr_funct} of present work. The base of the recursion,  the  3-points functions are evaluated  in terms of the residue formula, constructed from superpotential and holomorphic top form.  Such simplification is   a reflection of a  localization-like phenomenon.     Typically,  localization  in field theory and quantum mechanics    requires  path integral formulation, while we showed that the  localization construction can be realized   using the operator formalism in quantum mechanics. 

 The recursive construction for implies that the superpotential alone is not enough to evaluate the correlation functions for more than 3 observables! The additional data is given in the form of good section, introduced and studied by K. Saito  \cite{saito1983period}. The correlation functions in Landau-Ginzburg-Saito theory have non-trivial dependence on the choice of good section.  As a part of the proof for the main theorem we derive the good section for mirror superpotentials from the tropical mirror correspondence.

 The structure of our paper is as follows:   In section \ref{sec_mirr_corr} we briefly review the mirror symmetry and formulate it in the form of the equality for correlation functions. In section  \ref{sec_LG}  we review the  recursive construction for correlation functions in  Landau-Ginzburg-Saito theory. In section  \ref{sec_HTQM_tree} we define  the Higher Topological Quantum Mechanics on trees, describe the amplitudes and  describe a  deformation of   HTQM by  special state.  In section 5 we introduce the notion of correlation functions for HTQM and discuss their properties. In section 6 we briefly review the HTQM representation for tropical GW invariants and mirror relation between A- and B- types HTQMs. In section 7 we will introduce a notion of localization for mirror states, while in the last section we will use the localized states to evaluate the correlation functions in mirror HTQM.

\section{Mirror Correspondence}\label{sec_mirr_corr}

The mirror symmetry describes a  relation between the A- and B-models. The A-model of our interest is the theory of  the Gromov-Witten invariants for a Kahler manifold $X$ of dimension $N$.
The B-model  is the theory of complex structure deformations on the dual  complex manifold $X^\vee$ of same dimension $N$.    In well-known examples both $X$ and    $X^\vee$ are  compact Calabi-Yau 3-folds.

 We  can generalize the correspondence away from  three-dimensional Calabi-Yau spaces  by relaxing the compactness condition on  $X^{\vee}$.   The    complex structure deformations in that case are also parametrized    by a holomorphic a function  $W$  and the corresponding B-model  in commonly referred to as the Landau-Ginzburg (LG) theory with holomorphic superpotential $W$.  

In our paper we will discuss  the A-model on  a  toric space $X$ of dimension $N$. Furthermore, we will perform a certain scaling procedure to the geometry, which transforms  the usual GW invariants into  the tropical ones.     The corresponding mirror B-model becomes the LG theory on $X^{\vee} =\mathbb{C}^{\ast N} $ with exponential superpotential.

In the rest of this section we will briefly review the definition of toric manifolds, tropical GW invariants and give a detailed formulation of the mirror relation.

\subsection{Toric varieties}

Toric   manifold $X$ is a compactification of ${\mathbb{C^\ast}}^N$. We can represent $\mathbb{C}^{\ast N} = \mathbb{R}^N \times \mathbb{T}^N$  with the radial part $\mathbb{R}^N$, equipped  with standard coordinates $r^i, \; i=1,..,N$ and angular part,  $N$-dimensional torus  $\mathbb{T}^N = (S^1)^N$, with standard angular  coordinates $\phi^i$. Equivalently, we can say that the  $\mathbb{C}^{\ast N}$ is a trivial $N$-dimensional  toric fibration over  $\mathbb{R}^N$.  We describe the compactification of ${\mathbb{C^\ast}}^N$  using the fibraton data. 

\begin{itemize}
\item The radial part  is compactified by  the hyperplanes at infinity. Each hyperplane is given in terms of the inside-pointing $N$-dimensional   normal vector $\vec{b}_a$ with components $b_a^i$.   Each normal vector has integer components i.e. $b^i_a \in \mathbb{Z}$.  The two normal vectors with proportional components describe the same hypersurface. We can fix this ambiguity  by choosing the {\it primitive normal vector}, such that gcd$(b_a^1, b_a^2,...,b_a^N)=1$.
For toric space $X$ we will denote the set of such normal vectors by $B_X = \{ \vec{b}_a\}$.

\item In order to get a compactification of a complex manifold, we require that one of the circles $S^1 \subset \mathbb{T}^N$  inside the  toric fibration  shrinks to zero when we approach each of the compactifying hypersurfaces. The choice of a circle is given by a class in $\pi_1( \mathbb{T}^N) = H^1( \mathbb{T}^N, \mathbb{Z})$. For the hyperplane with normal vector $\vec{b}_a$ the corresponding class is the class of $ \sum b_a^i d\phi^i $. 
\end{itemize}

We will refer to the compacting hyperplanes as compactifying divisors,  and to $B_X$ as the set of all compactifying divisors for toric space $X$.

\subsection{Gromov-Witten theory}

The  Gromov-Witten invariant   $N^X_{\beta} (C_1,...,C_n)$ counts the number of  algebraic curves  of degree-$\beta$, genus-$0$ in complex  space $X$, passing   through the  cycles $ C_1,...,C_n $.  The integral representation
\be\label{top_compon_GW}
N^X_{\beta} (C_1,...,C_n) = \int_{\mathcal{M}_{0,n}   (X,\beta)} \bigwedge_{\alpha=1}^n \;\;  ev_\a^\ast \g_\a,
\ee
uses   $\mathcal{M}_{0,n}   (X,\beta) $, the  moduli space of degree $\beta$ curves in $X$ with $n$ marked points, compactified by quasi-maps, 
equipped with  the  evaluation map
\be
ev_\a: \mathcal{M}_{0,n}   (X,\beta)  \to X: (\phi: \mathbb{P}^1 \to X, z_1,...,z_n ) \mapsto \phi (z_\a).
\ee
The   $\g_\a$  are special representatives (smoothed out delta functions on cycles) of the  Poincare dual  to the cycles $C_\a$. In mathphysics literature the commonly used notation for the same GW invariant is   $\<\g_1,..,\g_n\>^X_{\beta,0}$ and we will use it in our paper.

For a given cycles $\g_k \in H^\ast (X)$ we can organize the genus-0 GW invariants of different degrees into a single expression 
\be
 \<\g_1,..,\g_n\>^X =  \sum_{\beta \in H_{1,1}(X)} q^\beta  \<\g_1,..,\g_n\>^X_{\beta,0}\;\;,
\ee
where $q$ describes the Kahler moduli of $X$.  For  generic $X$  the GW invariants are  formal series in $q$, but for toric $X$ they simplify to the    polynomials in $q$.
\newline\newline
{\bf Remark}: In present paper we will restrict our attention to  the GW invariants for 3 or more observables i.e $n\geq 3$.  There are interesting geometric invariants of $X$, with natural description in the form of   GW   invariants for two   observables, but most of them can be reformulated as invariants with 3 or more observables.

\subsection{Tropical  GW  invariants}

On toric manifold $X$ we can perform a tropical limit: a coordinate transformation in the form of scaling $(r^k,\phi^k) \to ( r^k/\e, \phi^k)$ followed by the  $\e \to 0$ limit.  For more details see our previous paper \cite{Losev:2022tzr} .  The limit of  a smooth algebraic curve of genus zero  in  toric space $X$  is  a circle bundle over   a tree. The tree is  embedded by a piece-wise linear map into radial part  of $X$.  The limiting curve is known as a tropical curve and was extensively studied by Mikhalkin  and collaborators \cite{Mikhalkin2004} in context of Tropical Geometry.    The moduli  of a tropical curve are  position of a root vertex, lengths  and twists of internal edges of a tree. We discussed the moduli space in our work \cite{Losev:2022tzr}, while for more detailed review see Mikhalkin's book \cite{mikhalkin2009tropical}. 

We can take the tropical limit for the  differential forms  $\g_\a$ to define the tropical A-model observables.  Note that the  tropical limit  turns smooth forms on toric space  $X$ into singular forms on radial part  of $X$.

We can define the tropical GW invariants as the integral over tropical moduli space of tropical observables. 
In our work  \cite{Losev:2022tzr}  we showed that the tropical GW invariants can be written as a sum of amplitudes in Higher Topological Quantum Mechanics (HTQM) on trees. We will provide the detailed definition of HTQM and describe the amplitudes in section \ref{sec_HTQM_tree}.  

In many examples it was observed that the tropical GW invariants match with the conventional GW invariants. This observation was formalized into:
\newline\newline
{\bf Theorem (tropical correspondence)}:  For the toric space $X$ and smooth differential forms $\g_k \in H_{dR}^\ast(X)$ the  Gromov-Witten invariant  $\<\g_1,..,\g_n\>^X_\beta$ matches with tropical Gromov-Witten invariant $\<\g^{trop}_1,..,\g^{trop}_n\>^X_\beta$ for the 
tropical limit  $\g_k^{trop}$ of the forms $\g_k$.
\newline\newline
{\bf Evidence}: There are several different types  of evidence for the theorem:

\begin{itemize}
\item Mikhalkin \cite{Mikhalkin2004}  counted the number  degree-3  tropical curves in $\mathbb{P}^2$ passing through the $8$ points in general position,  by presenting the  corresponding graphs, counted with proper multiplicities.  The total number he obtained was $12$, what  matched with the well known $N_3=12$  answer for the same problem for algebraic curves.

\item Gathmann and Markwig \cite{Markwig_N_d}  derived the   recursion formula  for number of tropical curves of degree-$d$, passing through the $3d-1$ points on $\mathbb{P}^2$.  Their result matches with the
 Kontsevich-Manin recursion formula \cite{kontsevich1994gromov} for algebraic curves.   Such matching  essentially gives a  prof of tropical correspondence  for  $ \mathbb{P}^2$.

\item There are several results on match for the descendant GW invariants: The case of $ \mathbb{P}^1$  was discussed by  Böhm, Goldner  and  Markwig   \cite{B_hm_2022}.   Markiw and Rau \cite{Markwig_2009}  proved the  equality for descendant invariants for $\mathbb{P}^2$. 
\item Our construction of tropical numbers and geometric objects as a scaling  limit $\e\to0$ in cylindrical coordinates serves as an  heuristic proof of the theorem. The GW invariants do not depend on the choice of coordinates, hence remain the same for any non-zero value of $\e$.  
\item  In  \cite{Losev:2022tzr} we derived  the tropical mirror superpotential for toric $X$ and it matches with   exponential   mirror superpotential  derived in \cite{givental1995quantum}, \cite{Hori:2000kt}   for the same toric space $X$. Given that the mirror symmetry relation holds for toric spaces we can use our result as an evidence in favor of the tropical correspondence theorem.
\end{itemize}

\subsection{Mirror relation}\label{sec_mirror_rel_corr}

The mirror of the $N$-dimensional  toric manifold $X$ is a non-compact $N$-dimensional Calabi-Yau  $X^{\vee} = \mathbb{C}^{\ast N} $ with holomorphic  superpotential.  We will use the  toric  representation $\mathbb{C}^{\ast N} = \mathbb{R}^N \times \mathbb{T}^N$
with radial coordinates $r^j$  and angular (holomorphic) coordinates $Y_j$.  The holomorphic top form in these coordinates is $\Omega = dY^1\wedge..\wedge dY^N$. Let us formulate several relations, implied by the mirror correspondence theorem. 
\newline\newline
{\bf Definition}: The Jacobi ring  for superpotential $W$ is
\be\label{def_Jacobi_ring}
J_W = R_{\mathbb{C}^{\ast N}} / I_W,
\ee
where $R_{\mathbb{C}^{\ast N}}$ is the ring  of holomorphic functions on $\mathbb{C}^{\ast N}$. In our coordinates $R_{\mathbb{C}^{\ast N}}$ is the ring of periodic functions of $Y$. The    $I_W$ is the  ideal generated by  the  partial derivatives  of  $W$
\be
I_W = \left\{  \frac{\p W}{\p Y_j} \right\}.
\ee
{\bf Remark}: If $W$ has isolated critical points then  $J_W$ is  finite-dimensional.
\newline\newline
{\bf Proposition (mirror for observables)}:  The  de Rahm cohomology of toric space $X$ isomorphic (as a vector spaces !) to the Jacobi ring 
\be
J: H_{dR}^\ast(X)  \to J_{W_X}: \gamma \mapsto J_\g
\ee
for mirror superpotential 
\be
W_X = \sum_{\vec{b}\in B_X}  q_{\vec{b}}\;\; e^{i \<\vec{b},\vec{Y}\>},
\ee
where the sum is taken over the  compactifying divisors $B_X$ of $X$.  
\newline\newline
We can refine the mirror relation for observables to the isomorphism of rings.
\newline\newline
{\bf Proposition (mirror for Frobenius rings)}:  The  quantum cohomology of toric space $X$ isomorphic (as graded-commutative Frobenius rings) to the Jacobi ring  for mirror superpotential.
\newline\newline
The Frobenius  ring structure $(C^A, g^A)$  on quantum cohomology ring of $X$ is determined by the 3-point GW invariants of $X$
\be
C^A_{\a\b\d} = \< \g_\a, \g_\b, \g_\d\>^X,\;\;\;  g^A_{\a\b} = \< \g_\a, \g_\b, 1\>^X = \int_X\g_\a\wedge \g_\b,
\ee
 while the Frobenius ring structure $(C^B, g^B)$ for Jacobi ring can be formulated using the residue formula.  Hence the mirror relation for rings can be formulated as equality 
\be\label{frob_jac_ring}
C^B_{\a\b\d}  =\sum_{dW_X=0} \frac{\Phi_{\g_\a}\;\Phi_{\g_\b}\;\Phi_{\g_\d} }{\det \p_k \p_l W_X},\;\; g^B_{\a\b}  =\sum_{dW_X=0} \frac{\Phi_{\g_\a}\;\Phi_{\g_\b}}{\det \p_k \p_l W_X},
\ee
where $\Phi_\g$ is a representative of a class $J_\g$. One can show that (\ref{frob_jac_ring}) is independent on the choice of representative. There  is a further generalization of the  mirror relation for rings, which includes all GW invariants of $X$.   

  Let $\g_1,..,\g_n$  be a basis in $H_{dR}^\ast (X)$ and $T^k$  $-$   linear coordinates  on this space in this basis. We can    organize all   genus-0  GW  invariants  for  $X$ 
 into the generating  function 
\be\label{a_model_gen_fun}
\mathcal{F}_{A} (T,q)  =  \sum_{j_1,..,j_n = 0}^\infty  \< \underbrace{\g_1,..,\g_1}_{j_1},...,\underbrace{\g_n,..,\g_n}_{j_n}\>^X\; \prod_{k=1}^{n}\frac{(T^k)^{j_k}}{j_k!} . 
\ee
Parameters $q$ represent the Kahler moduli dependence for  $X$ in A-model. 
\newline\newline
The B-model generating function was defined in  works of  K.Saito \cite{saito1983period},  Blok-Varchenko  \cite{blok1992topological}  and Dijkgraaf-Verlinde-Verlinde \cite{Dijkgraaf:1990dj}. Namely 
\be\label{def_b_model_gen_fun}
 \frac{ \p^3\mathcal{F}_{B} (T,q) }{ \p T^\a\; \p T^\b\; \p T^\d} =\sum_{dW=0} \frac{\Phi_\a(T)\cdot \Phi_\b(T) \cdot\Phi_\d(T)}{ \det \p_k \p_l W(T)},
\ee
where  $W(T)$ is the deformation of mirror superpotential   in special coordinates, introduced by K. Saito, the holomorphic functions $\Phi_\a(T)$ are partial derivatives of superpotential, i.e. 
\be
 \Phi_\a(T) = \frac{\p W}{\p T^\a}.
\ee
The partial derivatives $\p_k \p_l W$ of $W$ are taken in  coordinates $Y$, where the holomorphic top form is  $dY_1\wedge..\wedge dY_N$.
\newline\newline
{\bf Theorem (mirror correspondence)}: For toric space $X$ the generating function of the GW invariants equals to the B-model  generating function  (\ref{def_b_model_gen_fun}) for deformations of  mirror superpotential $W_X$  i.e.
\be
\mathcal{F}_{A} (T,q) =  \mathcal{F}_{B} (T,q).
\ee
{\bf Remark}:  The equality above is an equality for the   formal series in $q$ and $T$.  For the  toric    GW invariant  each coefficient in $T$-expansion is a  polynomial in $q$, rather than the   formal series.
Therefore  for the case of  toric A-model the corresponding mirror B-model expression should also be a polynomial. Hence we will focus on proving the equality of two  polynomials for the coefficients of formal series in $T$. 
In the next section we will give a  description for the B-model coefficients in the generating function expansion.

\section{Mirror for correlation functions}

For arbitrary three  holomorphic functions  we define  the 3-point correlation function 
\be
 \left\< \Phi_\a ,\Phi_\b, \Phi_\g \right\>_{W} = \sum_{dW=0} \frac{\Phi_\a\cdot \Phi_\b\cdot \Phi_\g}{ \det \p_k \p_l W}.
\ee
Then the generating function of B-model  (\ref{def_b_model_gen_fun}) can be written in the form 
\be\label{def_LG_gener_fun}
 \frac{ \p^3\mathcal{F}_{B} (T,q) }{ \p T^\a\; \p T^\b\; \p T^\g} = \left\< \Phi_\a(T) ,\Phi_\b(T), \Phi_\g(T) \right\>_{W(T)}. 
\ee
In particular, the relation above implies that the cubical term in $T$-expansion is given by 
\be
\mathcal{F}_{B} (T,q)  = \frac{1}{3!}T^\a T^\b T^\g  \left\< \Phi_\a ,\Phi_\b, \Phi_\g \right\>_{W}+\cO(T^4),
\ee
where   $\Phi_\a = \Phi_\a(0)$ are representatives for the classes   in Jacobi ring $J_W$ for $W = W(0)$.  The quartic term in the expansion  for $\mathcal{F}_B$ can be expressed in the form 
\be\label{gen_fun_quartic}
\begin{split}
& \frac{\p^4\mathcal{F}_B }{ \p T^\a\; \p T^\b\; \p T^\g  \;\p T^\d}  \Big|_{T=0} = \frac{\p}{\p T^\delta} \Big|_{T^\delta=0}  \left\< \Phi_{\a}, \Phi_{\b}, \Phi_{\g} \right\>_{W+T^\delta \Phi_\d}  + \left\< \frac{\p }{\p T^\delta} \Big|_{T^\delta=0} \Phi_{\a}(T), \Phi_{\b}, \Phi_{\g}\right\>_{W} \\
 &\qquad\qquad+ \left\<  \Phi_{\a},\frac{\p }{\p T^\delta} \Big|_{T^\delta=0}\Phi_\b (T), \Phi_{\g} \right\>_{W} + \left\<  \Phi_{\a}, \Phi_{\b}, \frac{\p }{\p T^\delta} \Big|_{T^\delta=0}\Phi_\g(T)   \right\>_{W}. 
  \end{split}
\ee
The expression above has the following interpretation: The first term is the change in 3-point correlation function under the change of $W$ in the direction of $\Phi_\delta$. The last three terms describe  the change of the  functions $\Phi_\a(T)$
under the transport in direction of $\Phi_\delta$.  We can introduce connection $C_W$, so the change of a function $\Phi_\a(T)$ in  direction $\Phi_\delta$   is given by
\be
 \frac{\p}{\p T^\delta} \Big|_{T^\delta=0} \Phi_{\a}(T)  = C_W(\Phi_\delta, \Phi_\a). 
\ee
 The combination of four  terms  in the formula  (\ref{gen_fun_quartic}) is known as the recursion formula for the 4-point  function  in 2-dimensional   topological theory. The commonly used form for the recursion relation is (see also  \cite{Losev:1998dv})
\be\label{def_4_pt_rec_rel}
\begin{split}
\<  \Phi_{\a},  \Phi_\b , \Phi_{\g}, \Phi_\d \>_{W} & = \frac{d}{d \e} \Big|_{\e=0}  \< \Phi_{\a}, \Phi_{\b}, \Phi_{\g}\>_{W+\e \Phi_\d}  + \< C_W(\Phi_\delta, \Phi_\a) , \Phi_{\b}, \Phi_{\g}\>_{W} \\
 &\qquad\qquad+ \<  \Phi_{\a}, C_W(\Phi_\delta, \Phi_\b)  , \Phi_{\g} \>_{W} + \<  \Phi_{\a}, \Phi_{\b}, C_W(\Phi_\delta, \Phi_\g)   \>_{W}. 
   \end{split}
\ee
Note that the 4-point function depends on a choice of holomorphic function representatives for the classes in Jacobi ring.

The connection coefficients $C_W(\Phi_\delta, \Phi_\a) $ are  referred to in \cite{Losev:1998dv}  as the contact terms.  The  fourth derivative of the generating function  can be written in terms of the  4-point function, defined by (\ref{def_4_pt_rec_rel})
\be
 \frac{\p^4\mathcal{F}_B }{ \p T^\a\; \p T^\b\; \p T^\g  \;\p T^\d}  \Big|_{T=0}  =\<  \Phi_{\a},  \Phi_\b , \Phi_{\g}, \Phi_\d \>_{W}. 
\ee
The recursion  relation (\ref{def_4_pt_rec_rel}) can be further generalized to the case of  $n$-point functions. In physics literature the recursion relations of that type are commonly discussed  in context of the Landau-Ginzburg theory \cite{Dijkgraaf:1990dj}, so we will adopt the same name for our review of it. We will give a detailed version of the recursion formula and contact terms later in this section.  The  expansion coefficients  of the generating function become the n-point functions, i.e.
\be\label{b_model_gen_fun}
\frac{\p^n \mathcal{F}_{B} (T,q)}{\p T^{\a_1}...\p T^{\a_n}}  \Big|_{T=0} = \< \Phi_{\a_1},...,\Phi_{\a_n}\>_{W}. 
\ee
The mirror relation for generating functions   implies the mirror relation for the correlation functions:
\newline\newline
{\bf Proposition (mirror for  correlation functions)}:  For toric space $X$ the GW invariants fo cycles $\g_\a$ equal to the correlation function of  special representatives $SJ_{\g_\a}$ of the corresponding the Jacobi ring  classes $J_{\g_\a}$ for the mirror superpotential.
\be
\< \g_1,...,\g_n\>^X  =  \< SJ_{\g_1},...,SJ_{\g_n}\>_{W^X}.
\ee

\subsection{Landau-Ginzburg-Saito theory}\label{sec_LG}

 The study of Landau-Ginzburg theory was motivated by the theory of critical phenomena, which later grown into 2D CFT and eventially become part of the topological string theory. For review see \cite{Hori:2000kt}, namely the $(2,2)$-supersymmetric 
 sigma models on non-compact spaces in B-type twisting.  After the B-type twisting, the theory is not superconformal and further requires setting the anti-holomorphic superpotential to zero, while keeping the holomorphic superpotential  $W$ fixed. 
 B-twisting is anomalous that results in appearence of the holomorphic top form.
\newline\newline
{\bf Definition}: The Landau-Ginzburg-Saito  theory  on complex  space $X$ with superpotential $W$ is a  collection of the following data:  
\begin{enumerate}
\item non-compact complex manifold $X $  of dimension $N$;
\item non-degenerate holomorphic top form
\be
\Omega \in \Omega^{N,0} (X);
\ee
Note that the pair $(X, \Omega)$   can be constructed from  non-compact Calabi-Yau manifold.
\item holomorphic function $W: X \to \mathbb{C}$, called superpotential;
\item Good section $S: J_W \to R_X$.  
\end{enumerate}
The readers, familiar with mathphysics literature on the Landau-Ginzburg models, might not be familiar with the notion of good section and its significance for the LGS theory.  Indeed all of the LGS correlation functions can be obtained from the $T$-expansion of the generating function   $\mathcal{F}_{LG} (T)$ defined via  the 3-point function  
 \be\label{def_LG_gener_fun_superpot}
 \frac{ \p^3\mathcal{F}_{LG} (T) }{ \p T^\a\; \p T^\b\; \p T^\g} = \left\< \frac{\p W}{\p T^\a} ,\frac{\p W}{\p T^\b}, \frac{\p W}{\p T^\g} \right\>_{W(T)}. 
\ee
However, in order to use the formula (\ref{def_LG_gener_fun_superpot}) we need  to define the good times $T$, which  typically have complicated functional relation to the deformations of $W$, linear on Jacobi ring.  The good section data is equivalent to the choice of good times $T$, but is has more straightforward meaning  for the recursive definition of  correlation functions. In some simple cases (polynomial superpotentials)  the good section can be  constructed from the superpotential   if we impose an  extra requirement of homogeneity.  For more information about the good times and good section relation see \cite{saito1983period} and \cite{Losev:1998dv}.

 The most studied LGS theories have complex manifold  $ \mathbb{C}^N$, with coordinates $x_j$, canonical holomorphic top form $\Omega = dx_1\wedge..\wedge dx_N$   and polynomial superpotential $W(x)$.  The simplest LGS model of this class has single variable ($N=1$) and polynomial superpotential of degree $k$.   The image of good section in this case is well known and  consists of monomials of degree up to $k-2$
 \be
 \hbox{Im}\; S\;  = \mathbb{C}\<1,x,..,x^{k-2}\>.
 \ee
Note that in  case of polynomial superpotential with  two variables, the good section is known only for limited classes of superpotentials.  

Our main interest is the mirror of the  A-model for toric manifold $X$ of complex dimension $N$, given in terms of compactifying divisors $B_X$. The corresponding LGS theory   has complex manifold   ${\mathbb{C}^\ast}^N$ 
equipped with  cylindrical coordinates: the radial coordinates   $r^j \in \mathbb{R}$ and  holomorphic angular coordinates $Y_j \in S^1$. The holomorphic top form in  this coordinates is
\be
\Omega =  dY_1\wedge...\wedge dY_N,
\ee
while the  superpotential is the exponential function, written using the primitive normal vectors 
\be
W_X = \sum_{\vec{b}\in B_X} q_{\vec{b}}\; e^{i \; b^kY_k}.
\ee
 The   good sections has not been constructed for all  LGS theories with exponential superpotentials. The simplest exponential superpotential  is   the  mirror superpotential for $X = \mathbb{P}^1$
\be
W_{\mathbb{P}^1} = e^{iY} + q e^{-iY}.
\ee
 The image of good section  is  
\be
\hbox{ Im}\; S_{\mathbb{P}^1} = \mathbb{C}\<1, e^{iY}\>.
\ee

\subsection{Correlation functions in Landau-Ginzburg-Saito theory}

{\bf Definition}: For holomorphic functions  $\Phi_\a,\; \a=1,..,n>2$  in LGS theory on $\mathbb{C}^{\ast N}$  with superpotential $W$  the {\it $n$-point  correlation function} $\<\Phi_1,...,\Phi_n\>_{W}$  
is defined  recursively:
\begin{itemize}
\item  the 3-point function is given by the residue formula 
\be\label{def_LG_3pt}
\begin{split}
 \< \Phi_1, &  \Phi_2 ,   \Phi_3\>_{W}= \sum_{d W=0} \frac{\Phi_1\Phi_2\Phi_3 }{ \det \p_j\p_k W}
  \end{split}
\ee

\item The $(n+1)$-point function    is defined  recursively  via $n$-point functions and their derivatives according to formula below
\be\label{n_pt_rec}
\begin{split}
\<&\Phi_1,\Phi_2,.., \Phi_n,\Phi_{n+1}\>_{W} = \frac{d}{d\e} \Big|_{\e=0}  \< \Phi_1,\Phi_2,..,\Phi_{n}\>_{  W+\e \Phi_{n+1}} +\< C_{W}(  \Phi_1, \Phi_{n+1}),\Phi_2,..,\Phi_{n}\>_{W}\\
&\qquad +\< \Phi_1,C_{W}( \Phi_2, \Phi_{n+1} ),..,\Phi_n\>_{W}+..+\< \Phi_1,\Phi_2,..,C_{W}(\Phi_{n}, \Phi_{n+1})\>_{W}
\end{split}
\ee

\end{itemize}
Earlier we saw that the  $n$-point correlation functions  represent the coefficient in the generating function (\ref{b_model_gen_fun}) hence they    are symmetric under permutation of  all arguments $\Phi_1, \Phi_2,..., \Phi_{n}$.  The 3-point function in our definition is manifestly  symmetric. The symmetry of higher point functions is rather obscure from the recursive definition and require  certain properties of the contact terms $C_W$. Following the literature \cite{saito1983period} and \cite{Losev:1998dv}, we will formulate this properties in terms of the K. Saito's connection on Brieskorn cohomology.
\newline\newline
{\bf Proposition}: The correlation functions in (\ref{n_pt_rec}) are symmetric if 
\begin{itemize}
\item connection is symmetric;
\item connection is flat;
\item connection preserves the metric.
\end{itemize}

In \cite{Losev:1998dv} was  proposed a construction of   $C_W$  in terms of the K. Saito's good section $S$. Such connection is  manifestly symmetric, while the flatness and metric preservation are derived from the 
properties of a good section.

\subsection{Cohomology and pairing}

In order to give a definition of good section  and contact terms we will  introduce a  cohomology theory, motivated by topological string theory of type B.

Let us consider a graded  vector space 
\be\label{def_lsg_vect_space}
V_{LGS} = R_{\mathbb{C}^{\ast N}} \otimes \mathbb{C}[\psi_\Phi^i]
\ee
for parity-odd variables $\psi_\Phi^i$. On $V_{LGS}$ there is a  pair of graded-commuting differentials 
\be\label{def_LGS_differntial}
 {\bf Q}_W = \frac{\p W}{\p Y_j} \frac{\p}{\p \psi_\Phi^j} ,\;\;\; {\bf G}_-  =\frac{\p}{\p Y_j} \frac{\p}{\p \psi_\Phi^j}.
\ee
{\bf Remark}: $V_{LGS}$ is isomorphic to the space of polyvector fields on $\mathbb{C}^{\ast N}$, hence it  is equipped with parity-odd symplectic structure and holomorphic top form. The  ${\bf G}_-$ is a Batalin–Vilkovisky (BV)  operator on $V_{LGS}$,
which generalizes  the divergence on vector fields to  polyvector fields.
\newline\newline
{\bf Remark}: The local holomorphic observables of dimension-0 in topological string theory of type B can be identified with   polyvector fields. The ${\bf G}_-$  is the action of the superpartner to certain $U(1)$-rotation, which preserves insertion positions for  these observables.
\newline\newline
{\bf Definition}: On $V_{LGS}$ there is   $\mathbb{C}[[z]]$-valued   {\it Saito's higher residue   pairing}  
\be\label{Pairing_hz_cohom}
 {\bf K} (v_1, v_2) = \int_{{S^1}^N} d^N Y \int_{\mathbb{R}^N} d^Nr \int d^N \psi_\Phi d^N\psi_R\;  \; v_1  \wedge  e^{-i\Lambda \{{\bf Q}_W+z{\bf G}_-+d_R, L\} } v_2,
\ee
where $\Lambda$ is a real parameter, 
\be
L= \sum_{k=1}^N r^k \psi_\Phi^k
\ee
is the localization function and 
\be
d_R = \psi_R^j \frac{\p}{\p r^j}
\ee
is the   radial de Rham operator.
\newline\newline  
{\bf Remark}: The integral  form of the K. Saito's  pairing (\ref{Pairing_hz_cohom})  was proposed in \cite{Losev:1998dv}.
\newline\newline
{\bf Definition}: The $\mathbb{C}$-valued   higher pairings ${\bf K}^{(k)}$ are  defined as expansion coefficients in $z$-expansion of ${\bf K}$ i.e.
\be\label{def_lg_high_pairings_k}
{\bf K} (v_1, v_2) = \sum_{k=0}^\infty  z^k\; {\bf K}^{(k)}  (v_1, v_2).
\ee
In section \ref{sec_b_model_high_pairing} we will discuss a similar pairing in details, so for now  let us list some properties of the pairing without the proof:
\begin{itemize} 
\item  The operators ${\bf Q}_W-z{\bf G}_-$  and ${\bf Q}_W+z{\bf G}_-$ are conjugated with respect to the pairing (\ref{Pairing_hz_cohom}) i.e.
\be
{\bf  K}(({\bf Q}_W-z{\bf G}_-)v_1, v_2) =-(-1)^{|v_1|} {\bf K} (v_1, ({\bf Q}_W+z{\bf G}_-)  v_2). 
\ee
Hence we can descend the pairing to the pairing on   cohomology $H^\ast ({\bf Q}_W-z {\bf G}_-)\otimes H^\ast ({\bf Q}_W+z{\bf G}_-)$. One can show that all cohomology for ${\bf Q}_W\pm z{\bf G}_-$ are holomorphic functions.
\item  The pairing on $H^\ast ({\bf Q}_W-z{\bf G}_-)\otimes H^\ast ({\bf Q}_W+z {\bf G}_-)$ is independent of $\Lambda$. Hence we can choose $\Lambda \to \infty$, what localizes  the pairing  on a sum over critical points of $W$.
In particular, the first two pairings on holomorphic functions 
\be\label{LGS_0_pairing}
 {\bf K}^{(0)}( \Phi_1,   \Phi_2) =(2\pi i)^N   \sum_{dW=0} \frac{\Phi_1\Phi_2}{  \det  \p_i \p_j W} 
\ee
and 
\be\label{LGS_1_pairing}
 {\bf K}^{(1)}( \Phi_1,   \Phi_2) = (2\pi i)^N \frac12 \sum_{dW=0}  (\p_{k} \p_{{l}} W)^{-1}   \frac{ \left(  \Phi_1\; \p_{k}  \p_{l}\Phi_2  - \Phi_2\; \p_{k} \p_{l}  \Phi_1 \right)}{\det \p_{m} \p_{n} W}. 
\ee

\item We can use the pairing $ {\bf K}^{(0)}$ to establish an isomorphism between the Jacobi ring $J_W$  and $H^\ast ({\bf Q}_W)$.

\end{itemize}
  {\bf Remark}: The cohomology of ${\bf Q}_W+z{\bf G}_-$  were introduced by K. Saito under the name of Brieskorn lattice.

\subsection{Contact terms from good section}\label{good_sect}

The construction of Jacobi ring comes with canonical projection $\pi_W: R_{\mathbb{C}^{\ast N}} \to J_W$. Given a pair of homolorphic functions $\Phi_1$  and $\Phi_2 $ we can project their  product $\Phi_1 \Phi_2$ to the class $\pi_W(\Phi_1\Phi_2)$ in Jacobi ring $J_W$.  The section (which inverts $\pi_W$) $S_W: J_W \to R_{\mathbb{C}^{\ast N}}$ turns this class into holomorphic function $S_W\; \pi_W(\Phi_1\Phi_2)$. The difference 
\be
\Phi_1\Phi_2 - S_W\; \pi_W(\Phi_1\Phi_2) 
\ee
 is trivial in Jacobi ring.  An isomorphism between the $J_W$  and $H^\ast ({\bf Q}_W)$ means that   there exists a map ${\bf \Sigma}_W: R_{\mathbb{C}^{\ast N}} \to V_{LGS}$ such that 
\be
\Phi_1 \Phi_2 - S_W \pi_W (\Phi_1 \Phi_2) = {\bf Q}_W {\bf \Sigma}_W(\Phi_1\Phi_2),
\ee
and 
\be
{\bf \Sigma}_W S_W  =  0.
\ee
The choice of such ${\bf \Sigma}_W$ is known as the choice of homotopy for ${\bf Q}_W$.
\newline\newline
{\bf Definition}: We define a contact term  fo  $\Phi_1$ and $\Phi_2$  in LGS theory with  section $S_W$
\be\label{def_conn_good_section}
C^{S}_W (\Phi_1, \Phi_2) = \pm {\bf G}_- {\bf \Sigma}_W(\Phi_1\Phi_2).  
\ee
In other terms the product of two functions $\Phi_1\Phi_2$ can be decomposed into  the sum of the image of $S_W$ and  a linear combination of $\p^1 W,..,\p^N W$, i.e.
\be\label{decom_s_w_sigma}
\Phi_1\Phi_2 = S_W \pi_W( \Phi_1\Phi_2)+\sigma_k \p^k W 
\ee
The ${\bf \Sigma}_W(\Phi_1\Phi_2)$ has the form $\sigma_k(Y) \psi_{\Phi}^k$, so  ${\bf G}_-$-action on it is 
\be
{\bf G}_-{\bf \Sigma}_W(\Phi_1\Phi_2) = \frac{\p\sigma_k(Y) }{\p Y_k},
\ee
i.e. just a divergence of the vector field $\sigma_k(Y) \p_{Y_k}$. Note that for a given $S_W$ the decomposition  in  (\ref{decom_s_w_sigma}) does not uniquely fixes the $\sigma_k(Y)$. The freedom of choice $\sigma$ is fixed by the choice of
 homotopy ${\bf \Sigma}_W$.

Note that the dependence of contact term $C_W$ on the choice of homotopy ${\bf \Sigma}_W$ is $({\bf Q}_W + z {\bf G}_-)$-exact. It was shown that the  correlation functions are well-defined in $H^\ast ({\bf Q}_W + z {\bf G}_-)$, so the choice of homotopy does not affect the recursion formula.

 We can define a  natural projection $\pi: V_{LGS}\otimes \mathbb{C}[[z]] \to V_{LGS}$, given by an evaluation  at $z=0$.  The projection $\pi$ is a chain map, hence it  induces projection on cohomology
\be
\pi:  H^\ast ({\bf Q}_W + z {\bf G}_-) \to H^\ast ({\bf Q}_W ).
\ee
The section $S_W$  induces   a section  $S_{Saito}: J_W = H^\ast ({\bf Q}_W ) \to H^\ast ({\bf Q}_W + z {\bf G}_-)  $.  Indeed, every holomorphic function is both  ${\bf Q}_W$- and $ {\bf G}_-$-closed, hence it describes a class in $H^\ast ({\bf Q}_W + z {\bf G}_-)$, which we take as an image of the $S_{Saito}$-map. 
\newline\newline
{\bf Definition}: The {\it good section}  $S_{Saito}: H^\ast ({\bf Q}_W ) \to H^\ast ({\bf Q}_W + z {\bf G}_-)  $  is 
\begin{itemize}
 \item a section  for $\pi$ i.e. 
\be
\pi\circ S_{Saito} = Id_{H^\ast ({\bf Q}_W )};
\ee
\item the higher pairings (\ref{def_lg_high_pairings_k})  vanish  for all pairs $\Phi_1, \Phi_2 \in \hbox{Im}\; S_{Saito}$ i.e
\be\label{def_good_section}
 {\bf K}^{(k)} (  \Phi_1, \; \Phi_2)  = 0,\;\; \forall\; k>0;
\ee
\item  For a given section $S_{Saito}$  we can construct the corresponding contact term and connection. The  good section $S_{Saito}$  is preserved  under the parallel transport  respect to this connection.  
\end{itemize}

\section{HTQM on trees}\label{sec_HTQM_tree}
In previous work  \cite{Losev:2022tzr} we showed that the tropical Gromow-Witten invariants can be described using the higher topological quantum mechanics (HTQM) on tree graphs. In this section we will briefly review the definition on the HTQM on trees, describe the 
amplitudes and construct a family of HTQMs as a deformation of HTQM by a certain type of states.

\subsection{Higher topological quantum mechanics}\label{sec_htqm_def}

{\bf Definition}: The  {\it higher topological quantum mechanics}, HTQM (with the circle action)    $(V, Q, G_\pm)$  is a collection of the following data: 
 \begin{enumerate}
 \item $\mathbb{Z}_2$-bi-complex $(V,Q, G_-)$, namely:
 \begin{itemize}
 \item $\mathbb{Z}_2$-graded vector space $V$, can be infinite-dimensional. There is a  decomposition of  $V = V_0 \oplus V_1$ into even $V_0$ and   odd $V_1$  under the grading. We will use the  notation $|v| \in \mathbb{Z}_2$  to describe the grading of a vector $v \in V$;
 \item pair of differentials $Q, G_-: V \to V$, such that 
 \begin{itemize}
  \item grading-odd operators:  $|Qv| = |G_- v|= |v|+1$,
  \item square to zero:   $Q^2 =G_-^2 = 0$;
\end{itemize}
  \item two differentials graded-commute, i.e.  $\{Q, G_-\} = 0$.
\end{itemize}
\item Unnormalized  homotopy
 $G_{+}: V \to V$, such that 
 \begin{itemize}
  \item grading-odd operator:  $|G_+ v| = |v|+1$,
  \item  squares to zero  $G_+^2=0$,
  \item     $\{G_+, G_-\}  = 0$.  
\end{itemize}
\end{enumerate}
 In case $V$ is infinite-dimensional we  impose certain consistency conditions on HTQM data  $(V,Q,G_\pm)$.  We   define the Hamiltonian  operator  $H = \{ Q, G_+\} : V \to V$. The consistency conditions are formulated in terms of Hamiltonian:  
 \begin{itemize}
\item   The hamiltonian  $H$ is  such that the evolution operator $e^{-tH}$ is well defined for $t\geq0$  in the following sense:
 \begin{itemize}
  \item it  is a  solution to the ODE
  \be\label{np_exp}
  (\p_t +H) e^{-tH} =0,\;\;\; e^{-0\cdot  H} =1,\;\;\;t \in \mathbb{R}^+\cup\{0\};
  \ee
  \item  forms a 1-parameter semi-group with multiplication 
  \be\label{semi-grp}
e^{-t_1H} e^{-t_2 H} = e^{-(t_1+t_2)H},\;\; \forall \;t_1, t_2 \in \mathbb{R}^+\cup\{0\}.
\ee
\end{itemize}
\item  We require  that  the $t \to \infty$ limit  of the  evolution  operator  exists   and is the  projector on $ \ker H$, i.e.
\be
 \lim_{t\to+\infty} e^{-tH} = \Pi_0.
\ee

\item The projector $\Pi_0$ obeys 
\be\label{def_Proj_G_+}
\Pi_0 G_\pm =G_\pm\Pi_0 =  0.
\ee
\end{itemize}

 \subsection{HTQM on  trees}\label{sec_htqm_tree_def}

{\bf Definition}: The HTQM $(V, Q, G_\pm, \mu_2, g)$  on a connected  tree  $\Gamma$ with distinct root vertex   is the collection of the following data 
 \begin{enumerate}
 \item 1-valent vertices  are assigned  the HTQM states  i.e. $v_a \in V,\;\; a=1,..,n_1=|V_1(\Gamma)|$.
 \item 2-valent  vertices are assigned  observables $\cO_\a \in V\otimes V^\ast$  $\a = 1,..,n_2 = |V_2(\Gamma)|$.
 \item 3-valent vertices are assigned the multiplication  $\mu_2:  V\otimes V \to   V $ such that the triple $(Q, G_-,  \mu_2)$ obeys
 \begin{itemize}
 \item $\mu_2$ is  graded commutative  
 \be
 \mu_2 (v,w) = (-1)^{|v||w|} \mu_2 (w,v);
 \ee
 \item  $\mu_2$ is  associative 
 \be
 \mu_2 (\mu_2 (v,w),u) = \mu_2 (v, \mu_2(w,u));
 \ee
 \item  Leibniz   rule for $(\mu_2, Q)$
 \be\label{def_Leibniz}
 Q \mu_2(v,w) = \mu_2 (Qv, w)+ (-1)^{|v|} \mu_2 (v, Qw);
 \ee
 \item  the pair $(G_-, \mu_2)$ obeys the 7-term relation  for all $v,u,w\in V$
 \be\label{def_7_term}
 \begin{split}
 G_-& \mu_2(\mu_2(v,w),u)  =  \mu_2(G_-\mu_2(v,w),u) + ( -1)^{|w|(|v|-1)} \mu_2 (w, G_- \mu_2(v,u)) \\
 &\qquad+ ( -1)^{|v|} \mu_2 (v, G_- \mu_2(w,u)) - \mu_2(G_- v, \mu_2(w,u))  - (-1)^{|v|} \mu_2(v, \mu_2 (G_-w, u))\\
 & \qquad- (-1)^{|u|+|v|} \mu_2(v, \mu_2(w, G_-u)).
  \end{split}
 \ee
 \end{itemize}
 \item the root 3-valent vertex assigned the multiplication 
 \be
 \mu_3^0  = g\circ \mu_2: V^{\otimes 3}  \to \mathbb{R}
 \ee 
 constructed from  Frobenius structure $(g, \mu_2, Q)$, where the scalar product, commonly referred to as the {\it pairing}, $g: V\otimes V \to \mathbb{R}$ obeys the following properties:
 \begin{itemize}
 \item  non-degeneracy on $V$;
 \item the graded-symmetry
 \be
 g(v,w) = (-1)^{|v||w|} g(w,v);
 \ee
 \item $Q$-invariance 
 \be\label{Q_preserve}
 g (Qv, w)  +  (-1)^{|v|} g (v,Qw) = 0;
 \ee
   \item   $G_\pm$-invariance
 \be\label{def_G_inv_pairing}
 g (G_\pm v, w) = (-1)^{|v|} g (v, G_\pm w);
 \ee
 \item evolution  invariance
 \be\label{def_evol_inv_pairing}
 g(e^{-tH}v,w) =  g(v,e^{-tH}w).
 \ee
 
  \end{itemize}
  \end{enumerate}

 \subsection{Amplitudes  on trees}
 
{\bf Definition}:  The  {\it evolution operator} in HTQM $(V,Q, G_{\pm})$  is 
 \be
 U(t, dt, d\v) = e^{-tH + G_+ dt +G_-d\v} \in \Omega^\ast (\mathbb{R}^+ \times S^1) \otimes End (V). 
 \ee
 {\bf  Definition}: For   each  tree $\Gamma$  we  define {\it  pre-amplitude}
 \be
\mathcal{PA}_\Gamma : V^{\otimes n_1(\G)} \otimes (V \otimes V^\ast)^{\otimes n_2(\G)} \to \Omega^\ast(\mathcal{M}(\Gamma)),
\ee
where $\mathcal{M}(\Gamma)$ is the moduli space of trees $\G$, defined in (\ref{graph_moduli_internal_only}).
Each connected    tree  $\Gamma$   defines a   contraction  in tensor algebra
\be
\<\;\;\>_\Gamma : (V\otimes V^\ast )^{\otimes E} \otimes V^{\otimes n_1} \otimes (V^\ast \otimes  V)^{\otimes n_2}   \otimes (V^{\ast \otimes 2} \otimes   V )^{\otimes (n_3-1)} \otimes V^{\ast \otimes 3}   \to \mathbb{R},
\ee
where $n_3$ is the number of 3-valent vertices in $\G$. The pre-amplitude on a tree $\G$   with  states   $v_a$  and   operators $\cO_\a$ is 
\be
\mathcal{PA}_\Gamma (v_a; \cO_\a) =  \left\<U^{\otimes I(\G)}\otimes 1^{\otimes n_1} \bigotimes\limits_{a=1}^{n_1}  v_a\bigotimes\limits_{\a=1}^{n_2} \cO_\a \otimes \mu_2^{\otimes (n_3-1)}\otimes \mu_3^0  \right\>_\Gamma\;.
\ee
Note that the   pre-amplitude on tree  $\Gamma$ has  no  evolution operators  on  external edges (edges attached to leaves of a tree $\G$).
\newline\newline
{\bf Definition}: The  amplitude on connected  tree $\G$ is an  integral 
\be\label{def_ampl}
\mathcal{A}_\Gamma = \int_{\mathcal{M}^{0}(\Gamma)} \mathcal{PA}_\Gamma,
\ee
over moduli space of connected tree 
 \be\label{graph_moduli_internal_only}
 \mathcal{M}^0(\Gamma) = \left(\mathbb{R}^+ \times S^1\right)^{I(\Gamma)}.
 \ee

\subsection{Amplitudes in homotopy notation}
{\bf Definition}:  The  {\it propagator} $K: V \to V$  for HTQM $(V, Q, G_{\pm})$ is
\be\label{def_HTQM_propagator}
K    = \lim_{T\to \infty}  \int^T_0dt\; e^{-t H}\; G_+ =  \int^\infty_0dt\; e^{-t H}\; G_+.
\ee
Note, that the integral has potential divergence, when the exponent vanishes for states from $\ker H$. The $G_+$ in the expression (\ref{def_HTQM_propagator})  and the HTQM property  (\ref{def_Proj_G_+}) evaluates   $G_+ v =0$ on all  $v \in \ker H$, hence $Kv=0$ for such states.

The propagator $K$ is  a  homotopy i.e.
\be\label{normalized_homotopy}
\{ Q,K\}   =    \int^\infty_0dt\;  e^{-t H}\; \{Q,G_+\} =- \int_{0}^\infty d \left( \;e^{-tH} \right)  = e^{-tH}\Big|_{0}-e^{-tH}\Big|_{\infty}  =   1-\Pi_0. 
\ee
We can perform the moduli space integral in the amplitude definition (\ref{def_ampl}) and  express the amplitude using propagators
\be
\mathcal{A}_\Gamma (v_a; \cO_\a; K)= \left\<(2\pi KG_-)^{\otimes I(\G)}\bigotimes\limits_{a=1}^{n_1(\G)}  v_a\bigotimes\limits_{\a=1}^{n_2(\G)} \cO_\a \otimes \mu_2^{\otimes (n_3(\G)-1)}\otimes \mu_3^0  \right\>_\Gamma\;.
\ee
Note that the factors $2\pi G_-$ originate from the angular parts of the moduli space integral.

It is very convenient to introduce a graphical representation for amplitudes:  we use solid lines   for  edges, equipped with propagator $2\pi KG_-$ and dashed lines for  edges without the propagator. We label  1-valent vertices by the corresponding states  $v_a$,  2-valent vertices by the corresponding operators  $\cO_\a$. For each tree there is a single special vertex, responsible for the pairing in HTQM, which can be either  the 3-valent special vertex  equipped with  $\mu_3^0$-multiplication, or the
2-valent  vertex, equipped with the pairing $g$.   The graphical representation is not unique. Below we present three graphical representations for the same amplitude.
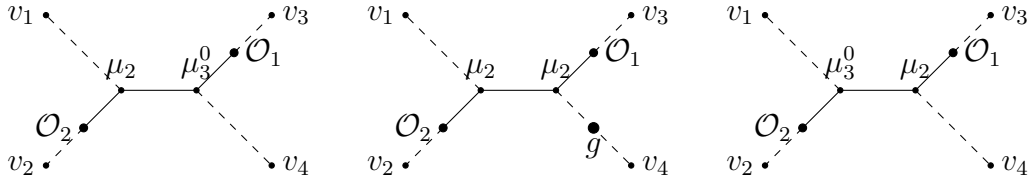
\newline\newline
  \begin{tikzpicture}[scale=1]
\draw  [black]  (0, 0) -- (1, 0);
\draw  [black, dashed]  (0, 0) -- (-1, 1);
\draw  [black]  (0, 0) -- (-0.5, -0.5);
\draw  [black, dashed]  (-.5, -0.5) -- (-1, -1);
\draw  [black, dashed]  (1, 0) -- (2, -1);
\draw  [black]  (1, 0) -- (1.5, 0.5);
\draw  [black, dashed]  (1.5, 0.5) -- (2, 1);
\filldraw[black] (0,0) circle (1pt) node[anchor=south]{$\mu_2$};
\filldraw[black] (1,0) circle (1pt) node[anchor=south]{$\mu^0_3$};
\filldraw[black] (-1,1) circle (1pt) node[anchor=east]{$v_1$};
\filldraw[black] (-1,-1) circle (1pt) node[anchor=east]{$v_2$};
\filldraw[black] (2,1) circle (1pt) node[anchor=west]{$v_3$};
\filldraw[black] (2,-1) circle (1pt) node[anchor=west]{$v_4$};
\filldraw[black] (1.5,0.5) circle (1.5pt) node[anchor=west]{$\cO_1$};
\filldraw[black] (-0.5,-0.5) circle (1.5pt) node[anchor=east]{$\cO_2$};
\end{tikzpicture}\;\;\;
  \begin{tikzpicture}[scale=1]
\draw  [black]  (0, 0) -- (1, 0);
\draw  [black, dashed]  (0, 0) -- (-1, 1);
\draw  [black]  (0, 0) -- (-0.5, -0.5);
\draw  [black, dashed]  (-.5, -0.5) -- (-1, -1);
\draw  [black, dashed]  (1, 0) -- (2, -1);
\draw  [black]  (1, 0) -- (1.5, 0.5);
\draw  [black, dashed]  (1.5, 0.5) -- (2, 1);
\filldraw[black] (0,0) circle (1pt) node[anchor=south]{$\mu_2$};
\filldraw[black] (1,0) circle (1pt) node[anchor=south]{$\mu_2$};
\filldraw[black] (-1,1) circle (1pt) node[anchor=east]{$v_1$};
\filldraw[black] (-1,-1) circle (1pt) node[anchor=east]{$v_2$};
\filldraw[black] (2,1) circle (1pt) node[anchor=west]{$v_3$};
\filldraw[black] (2,-1) circle (1pt) node[anchor=west]{$v_4$};
\filldraw[black] (1.5,0.5) circle (1.5pt) node[anchor=west]{$\cO_1$};
\filldraw[black] (-0.5,-0.5) circle (1.5pt) node[anchor=east]{$\cO_2$};
\filldraw[black] (1.5,-0.5) circle (2pt) node[anchor=north]{$g$};
\end{tikzpicture}\;\;\;
  \begin{tikzpicture}[scale=1]
\draw  [black]  (0, 0) -- (1, 0);
\draw  [black, dashed]  (0, 0) -- (-1, 1);
\draw  [black]  (0, 0) -- (-0.5, -0.5);
\draw  [black, dashed]  (-.5, -0.5) -- (-1, -1);
\draw  [black, dashed]  (1, 0) -- (2, -1);
\draw  [black]  (1, 0) -- (1.5, 0.5);
\draw  [black, dashed]  (1.5, 0.5) -- (2, 1);
\filldraw[black] (0,0) circle (1pt) node[anchor=south]{$\mu_3^0$};
\filldraw[black] (1,0) circle (1pt) node[anchor=south]{$\mu_2$};
\filldraw[black] (-1,1) circle (1pt) node[anchor=east]{$v_1$};
\filldraw[black] (-1,-1) circle (1pt) node[anchor=east]{$v_2$};
\filldraw[black] (2,1) circle (1pt) node[anchor=west]{$v_3$};
\filldraw[black] (2,-1) circle (1pt) node[anchor=west]{$v_4$};
\filldraw[black] (1.5,0.5) circle (1.5pt) node[anchor=west]{$\cO_1$};
\filldraw[black] (-0.5,-0.5) circle (1.5pt) node[anchor=east]{$\cO_2$};
\end{tikzpicture}
\newline
The amplitude, evaluated from   the left representation  is 
 \be\label{ex_ampl_1}
 \mathcal{A}_\Gamma  = (2\pi)^3\; \mu_3^0 (v_4,  KG_- \cO_1 v_3,   KG_- \mu_2 (v_1,  KG_- \cO_2 v_2)). 
 \ee
We can use the Frobenius structure for $(g, \mu_2)$ to evaluate the same amplitude, while  moving  the 3-valent special vertex with $\mu_3^0$ to the edge with $v_4$-state and turning it into 2-valent vertex with pairing $g$, i.e.
  \be
 \mathcal{A}_\Gamma  = (2\pi)^3\; g (  v_4,  \mu_2(  KG_- \cO_1 v_3,  KG_- \mu_2 (v_1,  KG_- \cO_2 v_2))). 
 \ee
We can revaluate the same amplitude by moving the $\mu_3^0$ to the different 3-vertex
  \be\label{ex_ampl_3}
 \mathcal{A}_\Gamma  =(2\pi)^3 \;\mu_3^0 (  v_1,  KG_- \cO_2 v_2, KG_- \mu_2(v_4,  KG_- \cO_1 v_3)). 
 \ee
The two representation (\ref{ex_ampl_1}) and (\ref{ex_ampl_3})  are related by the  $KG_-$-flip 
\be\label{KG_flip}
g (KG_-v, w) = g (v, KG_- w).
\ee
We can derive the flip formula using the definition (\ref{def_HTQM_propagator}) of propagator 
\be
\begin{split}
g ( KG_- v,  w)&  = \lim_{T\to \infty } \int^T_0 dt\;  g \left( e^{-tH} G_+ G_- v, w\right) \\
& = \lim_{T\to \infty } \int^T_0 dt\;  g \left(G_+ G_- v, e^{-tH}  w \right)   = \lim_{T\to \infty } \int^T_0 dt\;  g \left(v,  G_+G_-e^{-tH}  w \right)  \\
&= g (v, KG_- w). 
\end{split}
\ee
In the equality we used the integral representation (\ref{def_HTQM_propagator}) for the propagator, the  $G_\pm$-invariance of the pairing (\ref{def_G_inv_pairing}) and the  evolution invariance  of the pairing (\ref{def_evol_inv_pairing}).

\subsection{Deformation of HTQM by a state}\label{sec_htqm_deform}

In our work  \cite{Losev:2022tzr} on tropical mirror  we argued that the HTQM on trees admits a ``state-operator correspondence"-type  relation.  We can turn a HTQM state $\Psi \in V$ into an operator
\be
\cO_\Psi = \mu_2(\Psi, \cdot): V \to V
\ee
acting as the $\mu_2$-multiplication  by $\Psi$.  Such relation allowed us to indirectly study the HTQM deformation by a state, by the means of turning state into an operator first. For the discussion in later parts of the paper we introduce the notion of the 
HTQM, deformed by a state below. 
\newline\newline
{\bf Proposition (HTQM deformation by a state)}: Given HTQM  $(V, Q, G_\pm, \mu_2,g)$ on trees  and an even state $\e\Psi$ such that 
\be
Q \Psi = G_- \Psi = 0
\ee
there is  an one-parameter family    $(V, Q^\e, G_\pm, \mu_2,g)$ of HTQMs on trees   with differential
 \be
 Q^\e  =  Q- [G_-,  \mu_2(\e\Psi, \cdot)] +\cO(\e^2),
 \ee
i.e. the  action on states is given by
 \be
Q^\e v =  Q v-  G_- \mu_2(\e\Psi, v) +  \mu_2 (\e\Psi, G_- v)+\cO(\e^2).
 \ee
 {\bf Proof}: For a proof we need to  check  that the  family $(V, Q^\e, G_{\pm}, \mu_2, g)$  satisfies the   definitions from sections \ref{sec_htqm_def} and \ref{sec_htqm_tree_def}. The gradings of  $G_-$ and $Q$ are odd, while the grading of $\e\Psi$  is even hence the  grading of  $Q_\e$ is odd i.e.
 \be
 |Q^\e| = |[G_-,  \mu_2(\e \Psi, \cdot)] | = 1+|\e\Psi| =1.
 \ee
 The   $Q^\e$ is a differential  to the  leading order in $\e$, what follows from the square-evaluation
 \be
 \begin{split}
 Q^\e Q^\e v &= Q^2 v-  Q G_- \mu_2(\e \Psi, v) +  Q \mu_2 (\e\Psi, G_- v) -  G_- \mu_2(\e\Psi, Q v) +  \mu_2 (\e\Psi, G_- Q v )  \\
 &\quad\qquad+   G_- \mu_2(\e\Psi,   G_- \mu_2(\e\Psi, v) )  -   G_- \mu_2(\e\Psi,  \mu_2 (\e\Psi, G_- v))   -  \mu_2 (\e\Psi,  G_- \mu_2 (\e\Psi, G_- v))  \\
 &=     \frac12 \mu_2(  G_-  \mu_2(\e\Psi, \e\Psi), G_-v )  -  \frac12  G_- \mu_2(    G_- \mu_2(\e\Psi,\e\Psi), v )  = \cO(\e^2). 
 \end{split}
 \ee
 In the equality we used Leibniz  rule (\ref{def_Leibniz}), associativity of the multiplication $\mu_2$  and the 7-term relation (\ref{def_7_term}). 
\newline\newline
 {\bf Remark}:  The $Q^\e Q^\e =0$ holds for all orders in $\e$ for the states  $\e\Psi$, such that 
 \be
 G_- \mu_2 (\e\Psi, \e\Psi) = 0.
 \ee
 The condition above does not follow from the $G_- \Psi = 0$, since the pair  $(G_-, \mu_2)$ does not obey the Leibniz rule. 
  \newline\newline
 The  $Q^\e$ and  $G_-$ form  a pair of graded-commuting differentials.  Indeed, we can simplify the graded commutator
 \be
   \begin{split}
 \{ Q^\e, G_-\} v  &=Q G_-v-  G_- \mu_2(\e\Psi, G_-v) +  \mu_2 (\e\Psi, G_- G_-v)+ G_-Q v\\
& \qquad-   G_- G_- \mu_2(\e\Psi, v) +   G_- \mu_2 (\e\Psi, G_- v)=0.
   \end{split}
   \ee
The pair $(Q^\e, \mu_2)$ obey DGA: By construction $\mu_2$ is an associative, graded commutative multiplication and $Q^\e$ is the differential, so we just need to check the Leibniz  rule for $(\mu_2, Q^\e)$. Indeed, we can evaluate  
   \be
 \begin{split}
 & \mu_2(Q^\e v,w ) + (-1)^{|v|} \mu_2 (v, Q^\e w)- Q^\e \mu_2(v,w) \\
& =  - \mu_2(  G_- \mu_2(\e\Psi, v), w)  + \mu_2( \mu_2 (\e\Psi, G_- v) ,w )-(-1)^{|v|} \mu_2 (v,   G_- \mu_2(\e\Psi, w)) \\
 & \qquad\qquad  +(-1)^{|v|} \mu_2(v,  \mu_2 (\e\Psi, G_- w))  -  \mu_2 (\e\Psi, G_- \mu_2(v,w)) +  \mu_2(G_-\mu_2(v,w),\e \Psi)  \\
  &\qquad\qquad+ ( -1)^{|w|(|v|-1)} \mu_2 (w, G_- \mu_2(v,\e \Psi)) + ( -1)^{|v|} \mu_2 (v, G_- \mu_2(w,\e \Psi)) \\
   &\qquad\qquad - \mu_2(G_- v, \mu_2(w,\e \Psi))  - (-1)^{|v|} \mu_2(v, \mu_2 (G_-w, \e \Psi)) =0
\end{split}
\ee
In the equality we used the Leibniz  rule  (\ref{def_Leibniz}) and 7-term relation (\ref{def_7_term}).

The $Q^\e$-invariance of the pairing follows from 
 \be\label{Q_e_inv_par}
  \begin{split}
 &g (Q^\e v, w) + (-1)^{|v|} g (v, Q^\e w)   = -g (  G_- \mu_2(\e\Psi, v), w) +g(  \mu_2 (\e\Psi, G_- v), w)  \\
 &\qquad\qquad+(-1)^{|v|}  g(v,  \mu_2 (\e\Psi, G_- w)) -(-1)^{|v|}   g (v,  G_- \mu_2(\e\Psi, w))=0.
  \end{split}
 \ee
 We used the $Q$-invariance of the pairing (\ref{Q_preserve}) and  $G_\pm$ invariance of the pairing (\ref{def_G_inv_pairing}) and Frobenius structure for $(g, \mu_2)$.
 
The  Hamiltonian in deformed HTQM  is given by
 \be\label{def_deform_ham}
 H^\e = \{Q^\e, G_+\} = H-  \{ G_+,[ G_-, \mu_2 (\e \Psi, \cdot)]\} =  H-\e\; V_\Psi.
 \ee
The last equality introduces $V_{\Psi}$,  the linear in $\e$ correction to the Hamiltonian in deformed theory.   The Cauchy problem  for the evolution operator 
   \be
  (\p_t +H -\e V_\Psi) e^{-tH^\e} =0,\;\;\; e^{-0\cdot  H^\e} =1,\;\;\;t \in \mathbb{R}^+
  \ee
  can be solved  in  power  series in $\e$.  The evolution operator for deformed theory is 
 \be\label{deform_bos_evol}
   \begin{split}
 e^{-t H^\e} & =  e^{-tH} +   \int^t_0ds\;  e^{(s-t)H}\; \e V_\Psi \; e^{-sH} + \cO(\e^2) \\
 &= e^{-tH} + \int^t_0 ds\; e^{(s-t)H} \{G_+, [G_-, \mu_2(\e \Psi, \cdot)] \} e^{-s H} +\cO(\e^2). 
   \end{split}
 \ee
 We can rewrite  the  integral  in more symmetric form 
\be\label{proj_symm_repr}
\int^t_0 ds\; e^{s H-t H} \;  \e V_\Psi \; e^{-s H} =  \int_{t_1>0,\; t_2>0,\; t_1+t_2=t}  dt_1dt_2\; e^{-t_2 H}  \;  \e V_\Psi  \; e^{-t_1 H}.
  \ee
The symmetric formula is a common formula for perturbation theory  in quantum mechanics and has the following interpretation: The $t_1$  and $t_2$ describe the decomposition of the interval $[0,t]$ into two sub-intervals:  $[0, t_1]$  of length $t_1$  and $[t_1, t]$ of length $t_2$. Each interval is equipped  with the evolution operator $e^{-t_1 H}$ and  $e^{-t_2 H}$, while the splitting point carries the insertion of the deformation $ \e V_\Psi $  of  the  Hamiltonian.  

 The evolution operators in deformed theory  form a semi-group, what follows from the analysis of  the composition of  two   perturbative solutions  (\ref{deform_bos_evol}) with times $t$  and $s$
 \be\nn
   \begin{split}
 e^{-s H^\e} e^{-t H^\e} &= e^{-(t+s)H} +  \int^t_0 dt_1\; e^{-(t+s-t_1) H} \; \e V_\Psi \; e^{-t_1 H}\\
 &\qquad\qquad+ \int^{s+t}_t dt_1\; e^{-(s+t-t_1) H}\; \e  V_\Psi  \;e^{-t_1 H}  +\cO(\e^2)  \\
  &= e^{-(t+s)H^\e} +\cO(\e^2).
   \end{split}
 \ee
The evolution with respect to $H^\e$  preserves   the pairing. Indeed, we can evaluate   
 \be
  \begin{split}
 g(e^{-tH^\e}v,w) &=  g(v,e^{-tH}w) +\int_{t_1+t_2=t} dt_1dt_2\;  g \left( e^{- t_2H}\; \e V_\Psi \; e^{-t_1H}v, w\right) \\
    &=  g(v,e^{-tH}w) +   \int_{t_1+t_2=t} dt_1dt_2\; g \left( v,    e^{-t_1H}\; \e V_\Psi \;  e^{- t_2H}  w \right) \\ 
  & = g(v,e^{-tH^\e}w).
  \end{split}
 \ee
 We used the evolution invariance of the pairing (\ref{def_evol_inv_pairing}) and the  invariance of the pairing with respect to the $\e V_\Psi$-action, what follows from the $Q^\e$-invariance of the pairing (\ref{Q_e_inv_par})  and  $G_{\pm}$-invariance of the pairing (\ref{def_G_inv_pairing})  and the following relation
   \be
   \begin{split}
 g(\e V_\Psi \; v, w) & = g( \{G_+, Q^\e -Q \} v, w)  =  g( G_+( Q^\e -Q) v, w) +g( (Q^\e -Q ) G_+v, w) \\
& =(-1)^{2|v|+2}g(  v, ( Q^\e -Q) G_+w) +(-1)^{2|v|+2} g( v, G_+(Q^\e -Q )  w) \\
&=  g( v,\e V_\Psi \; w).
   \end{split}
 \ee

We use the semi-continuity of the kernel: The kernel of $H$ can only decrease under the small deformation, hence $\dim \ker H^\e \leq \dim \ker H$. The possible decrease of the kernel is due to the  obstruction for deformation of  $v^0\in \ker H$
into $v^{0\e}\in \ker H^\e $. Given a  state $v_0 \in \ker H$ we can  deform   it by  a $\cO(\e)$-term to construct a state 
\be
v^{0\e} = v^0 +\e v^1+ \cO(\e^2) \in \ker H^\e,
\ee 
where the deformation $\e v^1$  is a solution to 
\be
 H^\e v^{0\e}  = -\e V_\Psi   v^0 + \e H  v^1 +\cO(\e^2)  = \cO(\e^2). 
\ee
The solution for $\e v^1$ can be written in the form 
\be\label{def_vac_state}
\begin{split}
\e v^1 & =  \int^\infty_0 dt\; e^{-t H} G_+ G_- \mu_2 (\e \Psi, v^0)  = KG_- \mu_2 (\e \Psi, v^0).
\end{split}
\ee
Indeed, we can check 
\be
\begin{split}
H\e v^1  &= \int^\infty_0 dt\;  H\; e^{-t H}G_+G_- \mu_2 (\e\Psi, v^0)= - \int^\infty_0  d \left( e^{-tH} G_+G_-\mu_2 (\e\Psi, v^0)\right) \\
& = - e^{-tH}G_+G_-\mu_2 (\e\Psi, v^0) \Big|^{t=\infty}_{t=0}  = G_+G_-\mu_2 (\e\Psi, v^0) - \Pi_0 G_+G_-\mu_2 (\e\Psi, v^0) \\
& =G_+G_-\mu_2 (\e\Psi, v^0)  =  \e \; V_\Psi  v^0,
\end{split}
\ee
where we replaced the $t\to \infty$-limit by a projector $\Pi_0$  and used (\ref{def_Proj_G_+}) to eliminate the term.  The  deformation (\ref{def_vac_state}) exists for all $v^0 \in \ker H$, hence $\dim \ker H^\e = \dim \ker H$.
\newline\newline
{\bf Remark}:  The integral in  (\ref{def_vac_state})  acquires most of its value near $t=0$, since for large $t$, the exponential operator   $e^{-tH}$ is close to the projector $\Pi_0$.
Hence we can write an   approximation for the integral in the form of the finite region integral
\be
\e v_1 = \int^\infty_0 dt\; e^{-t H} G_+G_- \mu_2 (\e \Psi, v_0)\approx  \int^T_0 dt\; e^{-t H} G_+G_- \mu_2 (\e \Psi, v_0).
\ee
Let us choose a basis $v^0_k$ in $\ker H$. Since $\ker H$  and $\ker H^\e$ are of the same dimension   then  the corresponding deformed states $v^{0\e}_k$ form a basis in $\ker H^\e$.   We  use the non-degeneracy  of the pairing $g$ to identify the vector space  $V$ and its dual  $V^\ast$  to express the projector 
\be
\Pi_0 = \sum v^0_k {v_k^0}^\ast.
\ee
The projector $\Pi_0^\e$  on $\ker H^\e$  is written using  the deformed states  $v^{0\e}_k$
\be\label{proj_def_state}
\begin{split}
\Pi^\e_0 &= \sum v^{0\e}_k \; {v_k^{0\e}}^\ast  = \Pi_0 +  \sum \e v^{1}_k\; {v_k^{0}}^\ast  +\e  \sum v^{0}_k\;  {\e v_k^{1}}^\ast  +   \cO(\e^2).  
\end{split}
\ee
The $t\to \infty$ limit of the deformed evolution operator (\ref{deform_bos_evol}) is
 \be
  \begin{split}
  \lim_{t\to \infty} &e^{-t H_\e}  = \Pi_0+ \lim_{t\to \infty} \int^t_0 ds\; e^{s H-t H}\;   \e V_\Psi \;e^{-s H} +\cO(\e^2). 
\end{split}
 \ee
  We  decompose the  integration interval $[0,t]$  into three regions  and evaluate the limit for the integration over each region:
\begin{itemize}
\item {\bf Right side of the interval}: The $t_2$ is small, i.e  $t_2 \in [0,T]$ for some finite $T$, while  $t_1\approx t$ is very large and we can replace the corresponding exponential factor by the projector $\Pi_0$. The integral (\ref{proj_symm_repr}) in this region  evaluates   into
 \be
   \begin{split}
  \lim_{t\to \infty}& \int^T_0 dt_2\; e^{-t_2 H} \;   \e V_\Psi \; e^{ -tH}e^{t_2 H} = \int^T_0 dt_2\; e^{-t_2 H} \;   \e V_\Psi   \cdot  \Pi_0 \\
  & =  \int^T_0 dt_2\; e^{-t_2 H}    \e V_\Psi    \sum v^0_k {v_k^0}^\ast  = \sum  \int^T_0 dt_2\; e^{-t_2 H}   G_+  G_- \mu_2  (\e \Psi, v^0_k) {v_k^0}^\ast  \\
  & \approx   \sum  \int^\infty_0 dt_2\; e^{-t_2 H}   G_+  G_- \mu_2  (\e \Psi, v^0_k) {v_k^0}^\ast   = \sum    K  G_- \mu_2  (\e \Psi, v^0_k) {v_k^0}^\ast \\
  & =  \sum \e v^1_k \;{v_k^0}^\ast. 
    \end{split}
\ee
\item{\bf Left side of the interval}: The $t_1$ is such that  $t_1<T$, while  $t_2 \approx t$ is very large and we can replace the corresponding exponential factor by the projector. The integral (\ref{proj_symm_repr}) in this region  evaluates   into
 \be
   \begin{split}
  \lim_{t\to \infty}& \int^T_0 dt_1\; e^{-t H} e^{t_1H}  \;  \e V_\Psi \; e^{ -t_1H}  =    \int^T_0 dt_1\; \Pi_0  \;  \e V_\Psi  \; e^{ -t_1H} \\
  & =  \int^T_0 dt_1\;   \sum v^0_k {v_k^0}^\ast \; \e V_\Psi   e^{ -t_1H} = \sum v^0_k  \left[ \int^T_0 dt_1\;  e^{ -t_1H} G_+  G_- \mu_2  (\e \Psi, v^0_k)\right]^\ast  \\
  &\approx  \sum v^0_k  \left[ \int^\infty_0 dt_1\;  e^{ -t_1H} G_+  G_- \mu_2  (\e \Psi, v^0_k)\right]^\ast =  \sum  v^0_k \left[  K  G_- \mu_2  (\e \Psi, v^0_k) \right]^\ast \\
  &=\sum v^0_k\; {\e v_k^1}^\ast.
   \end{split}
\ee
\item{\bf Middle  of the interval}: For the middle  region $t_1 \in [t/2-T, t/2+T]$, i.e. both $t_1$ and $ t_2$ are large. The integral (\ref{proj_symm_repr}) in this region  evaluates   into
   \be
   \begin{split}
  \lim_{t\to \infty}& \int^{t/2+T}_{t/2-T} dt_2\; e^{-t_2 H}   \; \e V_\Psi \;  e^{ -t_1H} = \int^T_{-T} dt_1\; \Pi_0  \;  \e V_\Psi  \;  \Pi_0 \\
  &= \int^T_{-T} dt_1\;  0   =\cO(T)\cdot 0 =  0.
    \end{split}
\ee
\end{itemize}
The sum over three contributions 
 \be
   \begin{split}
   \lim_{t\to \infty} e^{-t H_\e} =\Pi_0+  \sum \e v^1_k\; {v_k^0}^\ast+\sum v^0_k\; {\e v_k^1}^\ast = \Pi_0^\e
      \end{split}
\ee
matches with our expression  (\ref{proj_def_state}) for projector on $\ker H^\e$.
 
 We can use the (\ref{proj_def_state})-representation  to check the properties of projector in deformed theory 
 \be
   \begin{split}
 \Pi_0^\e G_\pm v& =\Pi_0 G_\pm v +  \sum  \left(  \e v^1_k\; {v_k^0}^\ast+  v^0_k \; {\e v_k^1}^\ast \right) G_\pm v \\
 & =    \sum   \e v^1_k \; g (v_k^0, G_\pm v) +  \sum   v^0_k \; g(\e v_k^1,  G_\pm v) = 0
   \end{split}
 \ee
 and 
  \be
   \begin{split}
G_\pm \Pi_0^\e v & = G_\pm \Pi_0  v +G_\pm  \sum  \left(  \e v^1_k\; {v_k^0}^\ast+  v^0_k\; {\e v_k^1}^\ast \right)v\\
 & = \sum (G_\pm \;\e v^1_k) \;  {v_k^0}^\ast+ \sum  (G_ \pm v^0_k) \; {\e v_k^1}^\ast  = 0.
   \end{split}
 \ee
In equalities we  used  the $G_\pm $-invariance of the pairing,  $G_\pm v^0_k=0$ and  expression (\ref{def_vac_state}) for $v^1_k$  to evaluate 
\be
G_\pm \; \e v^1_k  = G_\pm   \int^\infty_0 dt\; e^{-t H} G_+ G_- \mu_2 (\e \Psi, v_k^0)  = 0.
\ee
$\hfill\blacksquare$
\newline\newline
{\bf Definition}: For any  state $v$ in HTQM $(V, Q, G_{\pm})$   its  {\it leading order  deformation  by $Q$-, $G_-$-closed state $\Psi$ }is
\be\label{state_deform_htqm}
v^{\e} = v +KG_- \mu_2 (\e\Psi, v)+\cO(\e^2).
\ee
\newline
{\bf Proposition (preservation of closeness)}: If $v$ is $Q$- and $G_-$-closed, then  the  deformed state $v^\e$  is  $Q^\e$- and $G_-$- closed  i.e.
\be\label{prop_def_state_close}
Q^\e v^{\e} =G _- v^{\e} = \cO(\e^2).
\ee
{\bf Proof}:  The $G_-$-closeness if fairly straightforward 
\be
G_-v^{\e} = G_- v +G_- KG_- \mu_2 (\e\Psi, v)= - G_-^2 K  \mu_2 (\e\Psi, v)  = 0.
\ee
The $Q^\e$-action on the deformed state 
\be
\begin{split}
Q^\e v^\e &= Q^\e v +   Q KG_- \mu_2 (\e\Psi, v) + \cO(\e^2)  \\
&= Q v  -  G_- \mu_2 (\e\Psi,   v) +   Q KG_-  \mu_2 (\e\Psi, v)  + \cO(\e^2) \\
&=  -  G_- \mu_2 (\e\Psi,  v) +   (1-\Pi_0)G_-  \mu_2 (\e\Psi, v)  + \cO(\e^2) = \cO(\e^2).
\end{split}
\ee
In the equality we used the homotopy formula (\ref{normalized_homotopy}),  Leibniz rule  (\ref{def_Leibniz}) for even state $\e \Psi$ and the projector $\Pi_0$ property (\ref{def_Proj_G_+}) from the HTQM definition.
$\hfill\blacksquare$
\newline\newline
{\bf Remark}: The formula (\ref {state_deform_htqm}) for deformation of a state $v$, describes a connection on $Q+zG_-$-cohomology, fibered over the space of HTQM deformations.

\subsection{Diagrammatic representation of  deformed theory}
We can express the  propagator for deformed theory as expansion  in $\e$  in terms of propagators in original theory 
 \be\nn
 \begin{split}
 K^\e  v &= \int^\infty_0dt\; e^{-tH^\e}G_+v =  \int^\infty_0dt\;  e^{-tH} G_+v \\
 &\qquad\qquad +  \int^\infty_0dt\; \int_{t_1+t_2 =t,\; t_1,\; t_2>0}dt_1dt_2\;  e^{-t_1H} \{G_+, [G_-, \mu_2(\e\Psi, \cdot)] \} e^{-t_2H}G_+v + \cO(\e^2) \\
  & = Kv + \int_{(\mathbb{R}^+)^2} dt_1dt_2\;   e^{-t_1H} G_+ [G_-, \mu_2(\e\Psi, \cdot)] e^{-t_2H}G_+v + \cO(\e^2) \\
    & = Kv +  K  [G_-, \mu_2(\e\Psi, \cdot)] K v + \cO(\e^2).  
 \end{split}
 \ee
 The propagator for deformed theory  further simplifies  if we use it in $KG_-$-combination i.e.
 \be\label{prop_deform_htqm}
 K^\e G_-v   = KG_-v +  K G_- \mu_2 (\e\Psi,  K G_- v )  +\cO(\e^2). 
 \ee
We can give a graphical representation for a  propagator in deformed theory (denoted as thick solid line) in terms of the  diagrams in the original theory
\newline\newline
\begin{tikzpicture}[scale=0.9]
\draw[ultra thick]  (0,0)-- (2, 0) ;
\filldraw[black] (0,0) circle (2pt) ;
\filldraw[black] (2,0) circle (2pt) ;
\node  at (3,0) {$=$};
\draw[thick]  (4,0)-- (6, 0) ;
\filldraw[black] (4,0) circle (2pt) ;
\filldraw[black] (6,0) circle (2pt) ;
\node  at (7,0) {$+$};
\draw[thick]  (8,0)-- (10, 0) ;
\draw[thick, dashed]  (9,0)-- (9, 0.5) ;
\filldraw[black] (8,0) circle (2pt) ;
\filldraw[black] (10,0) circle (2pt) ;
\filldraw[black] (9,0) circle (1pt) ;
\filldraw[black] (9,0.5) circle (2pt) node[anchor=south]{$\e\Psi$};
\end{tikzpicture}
\newline\newline
In case $G_- \mu_2 (\e\Psi, \e\Psi)=0$ the $KG_-$ in deformed theory  can be  written to all orders in $\e$ in the form
 \be
 K^\e G_- v = KG_- + KG_- \mu_2 (\e \Psi, KG_-v) +K G_- \mu_2 (\e \Psi,  KG_- \mu_2(\e \Psi,   KG_- v))  + \dots  
 \ee
with   graphical representation 
 \newline\newline
\begin{tikzpicture}[scale=0.9]
\draw[ultra thick]  (0,0)-- (2, 0) ;
\filldraw[black] (0,0) circle (2pt) ;
\filldraw[black] (2,0) circle (2pt) ;
\node  at (3,0) {$=$};
\draw[thick]  (4,0)-- (6, 0) ;
\filldraw[black] (4,0) circle (2pt) ;
\filldraw[black] (6,0) circle (2pt) ;
\node  at (7,0) {$+$};
\draw[thick]  (8,0)-- (10, 0) ;
\draw[thick, dashed]  (9,0)-- (9, 0.5) ;
\filldraw[black] (8,0) circle (2pt) ;
\filldraw[black] (10,0) circle (2pt) ;
\filldraw[black] (9,0) circle (1pt) ;
\filldraw[black] (9,0.5) circle (2pt) node[anchor=south]{$\e\Psi$};
\node  at (11,0) {$+$};
\draw[thick]  (12,0)-- (15, 0) ;
\draw[thick, dashed]  (13,0)-- (13, 0.5) ;
\draw[thick, dashed]  (14,0)-- (14, 0.5) ;
\filldraw[black] (12,0) circle (2pt) ;
\filldraw[black] (15,0) circle (2pt) ;
\filldraw[black] (13,0) circle (1pt) ;
\filldraw[black] (14,0) circle (1pt) ;
\filldraw[black] (13,0.5) circle (2pt) node[anchor=south]{$\e\Psi$};
\filldraw[black] (14,0.5) circle (2pt) node[anchor=south]{$\e\Psi$};
\node  at (16,0) {$+\dots$};
\end{tikzpicture}
\newline\newline
The diagrammatic expression for the  deformed  state (\ref{state_deform_htqm}), denoted as the thick dashed line,  is the sum of two terms
\newline\newline
\begin{tikzpicture}[scale=0.9]
\draw[ultra thick, dashed]  (0,0)-- (2, 0) ;
\filldraw[black] (2,0) circle (2pt) ;
\node  at (2,0)[anchor=north] {$v^\e$};
\node  at (3,0) {$=$};
\draw[thick, dashed]  (4,0)-- (6, 0) ;
\filldraw[black] (6,0) circle (2pt) ;
\node  at (6,0) [anchor=north] {$v$};
\node at (7,0) {$+$};
\draw[thick]  (8,0)-- (9, 0) ;
\draw[thick, dashed]  (9,0)-- (10, 0) ;
\draw[thick, dashed]  (9,0)-- (9, 0.5) ;
\filldraw[black] (10,0) circle (2pt) ;
\filldraw[black] (9,0.5) circle (2pt) node[anchor=south]{$\e\Psi$};
\node  at (10,0) [anchor=north] {$v$};
\end{tikzpicture}
\newline
In case $G_-\mu_2(\e \Psi, \e\Psi) =0$ the higher order terms of the state deformation take the form
\newline\newline
\begin{tikzpicture}[scale=0.9]
\draw[ultra thick, dashed]  (0,0)-- (2, 0) ;
\filldraw[black] (0,0) circle (2pt) ;
\filldraw[black] (2,0) circle (2pt) ;
\node  at (2,0)[anchor=north] {$v^\e$};
\node  at (3,0) {$=$};
\draw[thick, dashed]  (4,0)-- (6, 0) ;
\filldraw[black] (4,0) circle (2pt) ;
\filldraw[black] (6,0) circle (2pt) ;
\node  at (6,0)[anchor=north] {$v$};
\node  at (7,0) {$+$};
\draw[thick]  (8,0)-- (9, 0) ;
\draw[thick, dashed]  (9,0)-- (10, 0) ;
\draw[thick, dashed]  (9,0)-- (9, 0.5) ;
\filldraw[black] (8,0) circle (2pt) ;
\filldraw[black] (10,0) circle (2pt) ;
\filldraw[black] (9,0.5) circle (2pt) node[anchor=south]{$\e\Psi$};
\node  at (10,0)[anchor=north] {$v$};
\node  at (11,0) {$+$};
\draw[thick]  (12,0)-- (14, 0) ;
\draw[thick, dashed]  (14,0)-- (15, 0) ;
\draw[thick, dashed]  (13,0)-- (13, 0.5) ;
\draw[thick, dashed]  (14,0)-- (14, 0.5) ;
\filldraw[black] (12,0) circle (2pt) ;
\filldraw[black] (15,0) circle (2pt) ;
\filldraw[black] (13,0.5) circle (2pt) node[anchor=south]{$\e\Psi$};
\filldraw[black] (14,0.5) circle (2pt) node[anchor=south]{$\e\Psi$}; 
\node  at (15,0)[anchor=north] {$v$};
\node (+) at (16,0) {$+\dots$};
\end{tikzpicture}
\newline
The diagrammatic expression above  describes the sum 
\be\label{obs_deform}
\begin{split}
v^\e &=v +  K G_-\mu_2( \e \Psi,  v )+KG_- \mu_2 (\e \Psi , K G_-\mu_2( \e \Psi,  v )) +...
\end{split}
\ee

\section{Correlation functions in HTQM}
In this section we introduce the  correlation functions for the states  in   HTQM. The correlation functions obey certain nice properties such as symmetry in all arguments,  $Q$-invariance and recursion relation.
\newline\newline
{\bf Definition}: The {\it  total  amplitude on spacial} with 3 valent vertices $\Psi_1, ..., \Psi_n$ in  HTQM $(V,Q, G_{\pm}, g, \mu_2)$   is 
\be\label{def_qm_corr_func}
\< \Psi_{1},...,\Psi_{n}\>_Q  = \sum_{\G, \sigma\in S_n}  \frac{  \mathcal{A}_\G (\Psi_{\sigma(1)},...,\Psi_{\sigma(n)}; K) }{|Aut(\G)|},
\ee
where $|Aut(\G)|$ is the symmetry factor for the tree $\G$ and $K$ is a homotopy (\ref{def_HTQM_propagator}). The summation is taken over all distinct  3-valent  connected trees  $\Gamma$ with $n$ leaves and over possible assignment of    states   $\Psi_1,...,\Psi_n$ on  leaves of $\G$. 
\newline\newline
{\bf Remark}:  The sum over permutations  in (\ref{def_qm_corr_func}) makes the   correlation function  $\< \Psi_{1},...,\Psi_{n}\>_Q $ manifestly  symmetric in all arguments.
\newline\newline
{\bf Remark}: The sum over  trees, weighted with the symmetry factors, in correlation functions  (\ref{def_qm_corr_func})  is a signature of their   relation   to amplitudes in certain Quantum Field Theory. We conjecture that the QFT is a   BCOV-like  theory  \cite{Bershadsky:1993cx}, see also   \cite{Losev_2007}. 
We are working on further investigation of this conjecture.
\newline\newline
{\bf Remark}: The HTQM  states $\Psi_a$,  relevant for the mirror symmetry,  are even, i.e.  $|\Psi_a|=0$.  Hence, for simplicity,  we  will assume that $|\Psi_a|=0$ for the rest of this section.   Such assumption will drastically reduce the    complexity of the sign factors  in various expressions.

\subsection{3-point function}
There is a single tree with 3-valent vertices and three leaves. Hence, the  3-point correlation function  is just a sum over   3!  possible  permutations of states $\Psi_1, \Psi_2, \Psi_3$ on the leaves of a tree below
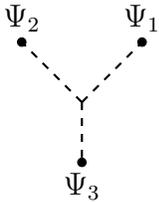
\newline\newline
\begin{tikzpicture}[scale=0.8]
\draw   [black,   thick, dashed]   (0,0)-- (0, -1) ;
\draw   [black,   thick, dashed]   (0,0)--(1, 1) ;
\draw   [black,   thick, dashed]   (0,0)--(-1, 1) ;
\filldraw[black] (1,1) circle (2pt) node[anchor=south]{$\Psi_1$};
\filldraw[black] (-1,1) circle (2pt) node[anchor=south]{$\Psi_2$};
\filldraw[black] (0,-1) circle (2pt) node[anchor=north]{$\Psi_3$};
\end{tikzpicture}
\newline
The amplitudes for each permutation are identical, hence we have a sum of   $3! =6$ identical  terms. The symmetry factor of the graph above is $|Aut\;\G_3| = 3! =6$,  so the 3-point correlation function simplifies to
\be\label{3pt_simplif}
\< \Psi_{1},\Psi_2,\Psi_{3}\>_Q  = \frac{1}{3!} \cdot 3! \; \mu_3^0 (\Psi_{1},\Psi_2,\Psi_{3}) = \mu_3^0 (\Psi_{1},\Psi_2,\Psi_{3}).
\ee
The 3-point function (\ref{3pt_simplif}) is invariant under the shift $\Psi_3 \to \Psi_3 + Q\chi$, given that $Q\Psi_1 = Q\Psi_2 = 0$.  Indeed, we can  evaluate  the difference
\be
\begin{split}
\<\Psi_1,\Psi_2, Q\chi  \>_Q &= g (Q\chi, \mu_2( \Psi_1, \Psi_2)) = g (\chi, Q\mu_2(\Psi_1, \Psi_2)) \\
&=   g (\chi, \mu_2( Q\Psi_1, \Psi_2)) + g (\chi, \mu_2(\Psi_1, Q\Psi_2))= 0. 
\end{split}
\ee
We used the $Q$-invariance of the pairing (\ref{Q_preserve}) and  Leibniz rule (\ref{def_Leibniz}) to simplify the expression.

\subsection{4-point function}\label{sec_4pt_inv}

There is a single tree with 3-valent vertices and four leaves. Hence the  4-point correlation function  is just a sum over   4!  possible  distributions of states $\Psi_1, \Psi_2, \Psi_3, \Psi_4$ on the leaves of a tree.  The 24 terms in a sum of three groups of $8$, with equal amplitudes in each group.       The 3  (independent) amplitudes  are depicted below and are commonly  referred to as the $s-,t-,u-$diagrams
 \newline\newline
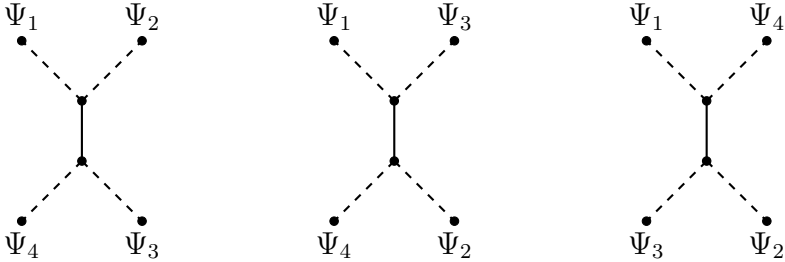
 \begin{tikzpicture}[scale=0.8]
\draw  [color=black, dashed,  thick]  (0, -1) -- (-1, -2);
\draw   [color=black, dashed,  thick]    (0, -1) -- (1, -2);
\draw  [color=black,   thick]   (0, -1) -- (0, 0);
\draw  [color=black, dashed, thick]   (0, 0) -- (1, 1);
\draw   [color=black, dashed,  thick]  (0, 0) -- (-1, 1);
\filldraw[black] (0,0) circle (2pt);
\filldraw[black] (0,-1) circle (2pt);
\filldraw[black] (-1,-2) circle (2pt) node[anchor=north]{$\Psi_4$};
\filldraw[black] (-1,1) circle (2pt) node[anchor=south]{$\Psi_1$};
\filldraw[black] (1,-2) circle (2pt) node[anchor=north]{$\Psi_3$};
\filldraw[black] (1,1) circle (2pt) node[anchor=south]{$\Psi_2$};
\end{tikzpicture}
\qquad\qquad
 \begin{tikzpicture}[scale=0.8]
\draw  [color=black, dashed,  thick]  (0, -1) -- (-1, -2);
\draw   [color=black, dashed,  thick]    (0, -1) -- (1, -2);
\draw  [color=black,  thick]   (0, -1) -- (0, 0);
\draw  [color=black, dashed,  thick]   (0, 0) -- (1, 1);
\draw   [color=black, dashed,  thick]  (0, 0) -- (-1, 1);
\filldraw[black] (0,0) circle (2pt);
\filldraw[black] (0,-1) circle (2pt);
\filldraw[black] (-1,-2) circle (2pt) node[anchor=north]{$\Psi_4$};
\filldraw[black] (-1,1) circle (2pt) node[anchor=south]{$\Psi_1$};
\filldraw[black] (1,-2) circle (2pt) node[anchor=north]{$\Psi_2$};
\filldraw[black] (1,1) circle (2pt) node[anchor=south]{$\Psi_3$};
\end{tikzpicture}
\qquad\qquad
 \begin{tikzpicture}[scale=0.8]
\draw  [color=black, dashed,  thick]  (0, -1) -- (-1, -2);
\draw   [color=black, dashed, thick]    (0, -1) -- (1, -2);
\draw  [color=black,   thick]   (0, -1) -- (0, 0);
\draw  [color=black, dashed,  thick]   (0, 0) -- (1, 1);
\draw   [color=black, dashed, thick]  (0, 0) -- (-1, 1);
\filldraw[black] (0,0) circle (2pt);
\filldraw[black] (0,-1) circle (2pt);
\filldraw[black] (-1,-2) circle (2pt) node[anchor=north]{$\Psi_3$};
\filldraw[black] (-1,1) circle (2pt) node[anchor=south]{$\Psi_1$};
\filldraw[black] (1,-2) circle (2pt) node[anchor=north]{$\Psi_2$};
\filldraw[black] (1,1) circle (2pt) node[anchor=south]{$\Psi_4$};
\end{tikzpicture}
\newline\newline  
The 4-point correlation function is  the sum of three contributions 
 \be\label{4pt_function_gen}
\< \Psi_{1},\Psi_2,\Psi_{3}, \Psi_4\>_Q  =   \mathcal{A}_{\G_4} (\Psi_{1},\Psi_2,\Psi_{3}, \Psi_4)+\mathcal{A}_{\G_4} (\Psi_{1},\Psi_3,\Psi_{2}, \Psi_4)+\mathcal{A}_{\G_4} (\Psi_{1},\Psi_4,\Psi_{2}, \Psi_3). 
\ee
 The number $8$ equals to   the symmetry factor $| Aut \;\G_4|=8$ of a tree and is constructed as $2\cdot 2\cdot 2$.   The    amplitudes on graphs, related by symmetry, are the same. Indeed, the amplitude for the first tree 
\be
\begin{split}
  \mathcal{A}_{\G_4} &(\Psi_{1},\Psi_2,\Psi_{3}, \Psi_4) = g (\mu_2(\Psi_3, \Psi_4), 2\pi KG_-\mu_2(\Psi_1,\Psi_2))\\
  &=g (\mu_2(\Psi_3, \Psi_4), 2 \pi KG_-\mu_2(\Psi_2,\Psi_1)) =g (\mu_2(\Psi_4, \Psi_3), 2\pi KG_-\mu_2(\Psi_1,\Psi_2)).
\end{split}
\ee
is invariant under the exchange of two pairs of states $\Psi_1\leftrightarrow\Psi_2$ and $\Psi_3\leftrightarrow\Psi_4$. Indeed, the  $\Psi_a$ are even states and $\mu_2$ is graded-symmetric. The last factor of 2 is related to the 
reflection of the tree, i.e. exchange of  two pairs $\Psi_1, \Psi_2$  and $\Psi_3, \Psi_4$ i.e. 
\be
g (\mu_2(\Psi_4, \Psi_3), KG_-\mu_2(\Psi_1,\Psi_2))=g (KG_-\mu_2(\Psi_4, \Psi_3), \mu_2(\Psi_1,\Psi_2)).
\ee
The invariance of the amplitude follows from the $KG_-$-flip relation (\ref{KG_flip}).

The 4-point amplitude is invariant under the shift $\Psi_4 \to \Psi_4 +Q\chi$, given that 
\be
Q \Psi_a = G_-\Psi_a = G_-\chi =0,\;\;\; a=1,2,3.
\ee
Indeed, we can evaluate 
\be\label{4pt_inv_details}
\begin{split}
\frac{2}{2\pi}&\<\Psi_1, \Psi_2, \Psi_3, Q \chi\>_Q = \sum_{\sigma\in S_3}  g ( Q \chi , \mu_2(\Psi_{\sigma(3)}, KG_-\mu_2(\Psi_{\sigma(1)}, \Psi_{\sigma(2)})) ) \\
 &  = \sum_{\sigma\in S_3}  g (  \chi , Q \mu_2(\Psi_{\sigma(3)}, KG_-\mu_2(\Psi_{\sigma(1)}, \Psi_{\sigma(2)})) ) \\
 & =   \sum_{\sigma\in S_3}  g (  \chi ,  \mu_2(\Psi_{\sigma(3)}, \{Q, K\} G_-\mu_2(\Psi_{\sigma(1)}, \Psi_{\sigma(2)})) ) =   \sum_{\sigma\in S_3}  g (  \chi ,  \mu_2(\Psi_{\sigma(3)},G_-\mu_2(\Psi_{\sigma(1)}, \Psi_{\sigma(2)})) )\\
    & = 2 g(\chi, G_-\mu_2(\mu_2(\Psi_1,\Psi_2),\Psi_3) ) =  - 2 g(G_-\chi, \mu_2(\mu_2(\Psi_1,\Psi_2),\Psi_3) )=0.
  \end{split}
\ee
In rewriting the equality we  used the  $Q$-invariance of the pairing (\ref{Q_preserve}), Leibniz rule (\ref{def_Leibniz}), the homotopy formula (\ref{normalized_homotopy}), the projector property (\ref{def_Proj_G_+}),  the  7-term relation (\ref{def_7_term}) for even $G_-$-closed states 
in the form 
  \be
 \begin{split}
 G_- \mu_2(\mu_2(\Psi_1,\Psi_2),\Psi_3)  &=  \mu_2(G_-\mu_2(\Psi_1,\Psi_2),\Psi_3) + \mu_2 (\Psi_2, G_- \mu_2(\Psi_1,\Psi_3)) \\
 &\qquad+ \mu_2 (\Psi_1, G_- \mu_2(\Psi_2,\Psi_3)). 
  \end{split}
 \ee
The last equality in (\ref{4pt_inv_details})  uses the  $G_\pm$-invariance of pairing (\ref{def_G_inv_pairing}) and completes the proof of invariance.

\subsection{Generating function}
We   introduce  formal  parameters $t_k$,  such that  $t_k^2=0$ and combine $\Psi_k$ into  even state
\be
\Psi =\Psi(t)=  \sum t_k \Psi_k.
\ee
The  $n$-point correlation  function of $\Psi$  has $t$-expansion with coefficients being the $n$-point correlation functions i.e.
\be
\begin{split}
\<\underbrace{\Psi, ...,\Psi}_n\>_Q  =n!\cdot  \<t_1\Psi_1,t_2\Psi_2, .., t_n \Psi_n  \>_{Q}.  
\end{split}
\ee
{\bf Definition}: The  n-point  correlation function of $\Psi$   can be organized into generating function 
\be\label{gen_funct_ampl}
\mathcal{F}_0(\Psi, K)  = \sum_{k=3}^\infty \frac{1}{k!} \<\underbrace{\Psi, ...,\Psi}_k\>_Q. 
\ee
The generating function  equals to the  the sum over connected 3-valent  trees
\be
\mathcal{F}_0(\Psi, K) =  \sum_{\G}  \frac{ \mathcal{A}_\G (\Psi,...,\Psi; K) }{|Aut(\G)|}.  
\ee
The diagrammatic expression for generating function is 
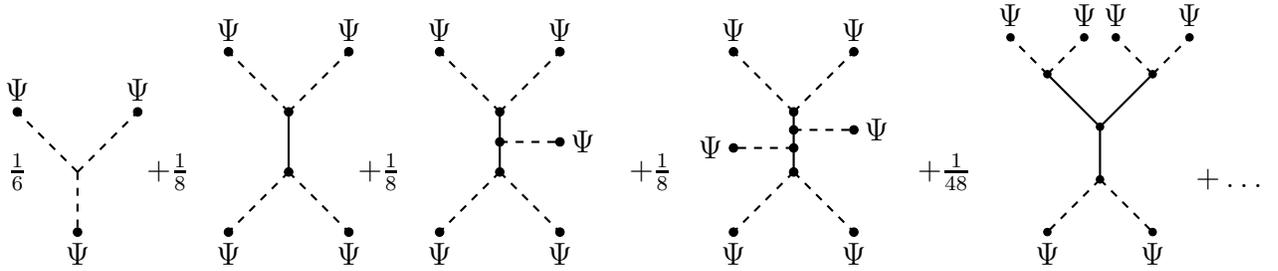
\newline\newline
\begin{tikzpicture}[scale=0.8]
\draw   [black,   thick, dashed]   (0,0)-- (0, -1) ;
\draw   [black,   thick, dashed]   (0,0)--(1, 1) ;
\draw   [black,   thick, dashed]   (0,0)--(-1, 1) ;
\filldraw[black] (1,1) circle (2pt) node[anchor=south]{$\Psi$};
\filldraw[black] (-1,1) circle (2pt) node[anchor=south]{$\Psi$};
\filldraw[black] (0,-1) circle (2pt) node[anchor=north]{$\Psi$};
\node  at (1.5, 0) {$+\frac18$};
\node  at (-1, 0) {$\frac16$};
\end{tikzpicture}
 \begin{tikzpicture}[scale=0.8]
\draw  [color=black, dashed,  thick]  (0, -1) -- (-1, -2);
\draw   [color=black, dashed,  thick]    (0, -1) -- (1, -2);
\draw  [color=black,   thick]   (0, -1) -- (0, 0);
\draw  [color=black, dashed, thick]   (0, 0) -- (1, 1);
\draw   [color=black, dashed,  thick]  (0, 0) -- (-1, 1);
\filldraw[black] (0,0) circle (2pt);
\filldraw[black] (0,-1) circle (2pt);
\filldraw[black] (-1,-2) circle (2pt) node[anchor=north]{$\Psi$};
\filldraw[black] (-1,1) circle (2pt) node[anchor=south]{$\Psi$};
\filldraw[black] (1,-2) circle (2pt) node[anchor=north]{$\Psi$};
\filldraw[black] (1,1) circle (2pt) node[anchor=south]{$\Psi$};
\node  at (1.5, -1) {$+\frac18$};
\end{tikzpicture}
 \begin{tikzpicture}[scale=0.8]
\draw  [color=black, dashed,  thick]  (0, -1) -- (-1, -2);
\draw   [color=black, dashed,  thick]    (0, -1) -- (1, -2);
\draw  [color=black,   thick]   (0, -1) -- (0, 0);
\draw  [color=black, dashed,  thick]   (0, -0.5) -- (1, -0.5);
\draw  [color=black, dashed, thick]   (0, 0) -- (1, 1);
\draw   [color=black, dashed,  thick]  (0, 0) -- (-1, 1);
\filldraw[black] (0,0) circle (2pt);
\filldraw[black] (0,-1) circle (2pt);
\filldraw[black] (0,-0.5) circle (2pt);
\filldraw[black] (-1,-2) circle (2pt) node[anchor=north]{$\Psi$};
\filldraw[black] (-1,1) circle (2pt) node[anchor=south]{$\Psi$};
\filldraw[black] (1,-2) circle (2pt) node[anchor=north]{$\Psi$};
\filldraw[black] (1,1) circle (2pt) node[anchor=south]{$\Psi$};
\filldraw[black] (1,-0.5) circle (2pt) node[anchor=west]{$\Psi$};
\node  at (2.5, -1) {$+\frac18$};
\end{tikzpicture}
 \begin{tikzpicture}[scale=0.8]
\draw  [color=black, dashed,  thick]  (0, -1) -- (-1, -2);
\draw   [color=black, dashed,  thick]    (0, -1) -- (1, -2);
\draw  [color=black,   thick]   (0, -1) -- (0, 0);
\draw  [color=black, dashed,  thick]   (0, -0.3) -- (1, -0.3);
\draw  [color=black, dashed,  thick]   (0, -0.6) -- (-1, -0.6);
\draw  [color=black, dashed, thick]   (0, 0) -- (1, 1);
\draw   [color=black, dashed,  thick]  (0, 0) -- (-1, 1);
\filldraw[black] (0,0) circle (2pt);
\filldraw[black] (0,-1) circle (2pt);
\filldraw[black] (0,-0.3) circle (2pt);
\filldraw[black] (0,-0.6) circle (2pt);
\filldraw[black] (-1,-2) circle (2pt) node[anchor=north]{$\Psi$};
\filldraw[black] (-1,1) circle (2pt) node[anchor=south]{$\Psi$};
\filldraw[black] (1,-2) circle (2pt) node[anchor=north]{$\Psi$};
\filldraw[black] (1,1) circle (2pt) node[anchor=south]{$\Psi$};
\filldraw[black] (1,-0.3) circle (2pt) node[anchor=west]{$\Psi$};
\filldraw[black] (-1,-0.6) circle (2pt) node[anchor=east]{$\Psi$};
\node  at (2.5, -1) {$+\frac{1}{48}$};
\end{tikzpicture}
\begin{tikzpicture}[scale=0.7]
\draw   [black,  thick, dashed]   (0,-1)-- (-1, -2) ;
\draw   [black,  thick, dashed]   (0,-1)-- (1, -2) ;
\draw   [black,  thick]   (0,0)-- (0, -1) ;
\draw   [black,  thick]   (0,0)--(-1, 1) ;
\draw   [black,  thick]   (0,0)--(1, 1) ;
\draw   [black,  thick, dashed]   (1,1)--(0.3, 1.7) ;
\draw   [black,  thick, dashed]   (1,1)--(1.7, 1.7) ;
\draw   [black,  thick, dashed]   (-1,1)--(-0.3, 1.7) ;
\draw   [black,  thick, dashed]   (-1,1)--(-1.7, 1.7) ;
\filldraw[black] (0,0) circle (2pt);
\filldraw[black] (0,-1) circle (2pt);
\filldraw[black] (1,1) circle (2pt);
\filldraw[black] (-1,1) circle (2pt);
\filldraw[black] (0.3,1.7) circle (2pt) node[anchor=south]{$\Psi$};
\filldraw[black] (1.7,1.7) circle (2pt) node[anchor=south]{$\Psi$};
\filldraw[black] (-0.3,1.7) circle (2pt) node[anchor=south]{$\Psi$};
\filldraw[black] (-1.7,1.7) circle (2pt) node[anchor=south]{$\Psi$};
\filldraw[black] (1,-2) circle (2pt) node[anchor=north]{$\Psi$};
\filldraw[black] (-1,-2) circle (2pt) node[anchor=north]{$\Psi$};
\node  at (2.5, -1) {$+\dots$};
\end{tikzpicture}
\newline\newline

\subsection{Invariance theorem}

In our analysis for the 3- and 4-point functions we observed the invariance under the shift  by a $Q$-exact term, given that the states were $Q$- and $G$-closed. The  invariance generalizes to the n-point function.
\newline\newline
 {\bf Theorem (invariance)}\label{corr_inv}: Given the  states  $\Psi_1, \Psi_2,...,\Psi_n, \chi$ of HTQM on trees $(V, Q, G_{\pm}, \mu_2, g)$ such that 
 \be
 Q\Psi_\a = G_-\Psi_\a = G_-\chi=0,\;\; \a=1..n
 \ee
the $(n+1)$-point correlation function vanishes i.e.
\be
\<\Psi_1,...,\Psi_n, Q\chi  \>_{Q} = 0.
\ee
{\bf Proof}:   We use the generating function (\ref{gen_funct_ampl})  to prove the theorem. The change in generating function  can be expressed 
\be
\delta \mathcal{F}_0(\Psi)  =  \mathcal{F}_0(\Psi +\delta \Psi) -  \mathcal{F}_0(\Psi ) = g( \delta \Psi, \tilde{\g})  +\cO(\delta \Psi)^2
\ee
via the state $\tilde{\g}$, defined as a  sum over 3-valent rooted trees, weighted with symmetry factors. The first several  trees  of the sum $\tilde{\g}$ are   presented below
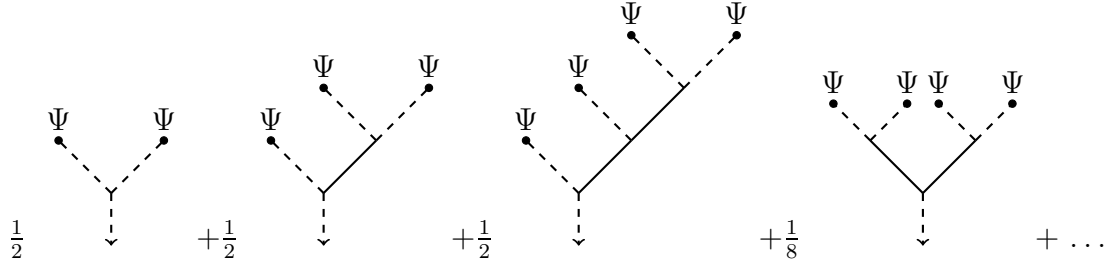
\newline\newline
$\frac12$
\begin{tikzpicture}[scale=0.7]
\draw   [black,  thick, dashed, ->]   (0,0)-- (0, -1) ;
\draw   [black,  thick, dashed]   (0,0)--(1, 1) ;
\draw   [black,  thick, dashed]   (0,0)--(-1, 1) ;
\filldraw[black] (1,1) circle (2pt) node[anchor=south]{$\Psi$};
\filldraw[black] (-1,1) circle (2pt) node[anchor=south]{$\Psi$};
\end{tikzpicture}
$+\frac12$ \begin{tikzpicture}[scale=0.7]
\draw   [black,  thick, dashed, ->]   (0,0)-- (0, -1) ;
\draw   [black,  thick, dashed]   (0,0)--(-1, 1) ;
\draw   [black,  thick]   (0,0)--(1, 1) ;
\draw   [black,  thick, dashed]   (1,1)--(0, 2) ;
\draw   [black,  thick, dashed]   (1,1)--(2, 2) ;
\filldraw[black] (2,2) circle (2pt) node[anchor=south]{$\Psi$};
\filldraw[black] (0,2) circle (2pt) node[anchor=south]{$\Psi$};
\filldraw[black] (-1,1) circle (2pt) node[anchor=south]{$\Psi$};
\end{tikzpicture}$+\frac12$
\begin{tikzpicture}[scale=0.7]
\draw   [black,  thick, ->, dashed]   (0,0)-- (0, -1) ;
\draw   [black,  thick, dashed]   (0,0)--(-1, 1) ;
\draw   [black,  thick]   (0,0)--(1, 1) ;
\draw   [black,  thick, dashed]   (1,1)--(0, 2) ;
\draw   [black,  thick]   (1,1)--(2, 2) ;
\draw   [black,  thick, dashed]   (2,2)--(1, 3) ;
\draw   [black,  thick, dashed]   (2,2)--(3, 3) ;
\filldraw[black] (3,3) circle (2pt) node[anchor=south]{$\Psi$};
\filldraw[black] (1,3) circle (2pt) node[anchor=south]{$\Psi$};
\filldraw[black] (0,2) circle (2pt) node[anchor=south]{$\Psi$};
\filldraw[black] (-1,1) circle (2pt) node[anchor=south]{$\Psi$};
\end{tikzpicture}$+\frac18$
\begin{tikzpicture}[scale=0.7]
\draw   [black,  thick, ->, dashed]   (0,0)-- (0, -1) ;
\draw   [black,  thick]   (0,0)--(-1, 1) ;
\draw   [black,  thick]   (0,0)--(1, 1) ;
\draw   [black,  thick, dashed]   (1,1)--(0.3, 1.7) ;
\draw   [black,  thick, dashed]   (1,1)--(1.7, 1.7) ;
\draw   [black,  thick, dashed]   (-1,1)--(-0.3, 1.7) ;
\draw   [black,  thick, dashed]   (-1,1)--(-1.7, 1.7) ;
\filldraw[black] (0.3,1.7) circle (2pt) node[anchor=south]{$\Psi$};
\filldraw[black] (1.7,1.7) circle (2pt) node[anchor=south]{$\Psi$};
\filldraw[black] (-0.3,1.7) circle (2pt) node[anchor=south]{$\Psi$};
\filldraw[black] (-1.7,1.7) circle (2pt) node[anchor=south]{$\Psi$};
\end{tikzpicture}+ $\dots$
\newline\newline
The diagrammatic sum  for $\tilde{\g}$ takes the form 
\be
\begin{split}\label{tree_state}
\tilde{\g}&=\frac12  \mu_2(\Psi, \Psi) + \frac12  \mu_2 (\Psi, 2\pi KG_- \mu_2(\Psi, \Psi)) + \frac12 \mu_2(\Psi, 2\pi KG_- \mu_2 (\Psi, 2\pi KG_- \mu_2(\Psi, \Psi))) \\
&+ \frac18  \mu_2 (2\pi KG_- \mu_2(\Psi, \Psi), 2\pi KG_- \mu_2(\Psi, \Psi))+\dots
\end{split}
\ee
Note that $\tilde{\g}$ is an even state since $\Psi$ is even and $KG_-$ is even operator.  We also introduce a related   even state 
\be\label{tree_sum_relations}
\gamma = \Psi + 2\pi KG_- \tilde{\g},
\ee
used in \cite{Losev_2007}, which is $G_-$-closed
\be
G_- \g = G_-\Psi -2\pi  KG^2_- \tilde{\g} = 0.
\ee
The $\tilde{\g}$ and $\g$ obey the ``root-cutting" relation
\be\label{g_g_tilda}
\begin{split}
\tilde{\g} = \frac12  \mu_2(\g, \g).
\end{split}
\ee

The  representation  (\ref{g_g_tilda}) for $\tilde{\g}$  is   very convenient for  deriving the recursive formula
\be
\begin{split}
Q \tilde{\g}&=  \frac12   Q\mu_2(\g, \g) = \mu_2 (\g, Q\g) =\mu_2 (\g, Q \Psi + 2\pi Q KG_- \tilde{\g})   \\
& = \mu_2 (\g,  2\pi G_- \tilde{\g})  +\mu_2 (\g,   2\pi KG_- Q\tilde{\g})    =\pi  \mu_2 (\g,  G_-   \mu_2(\g, \g) )  +\mu_2 (\g,   2\pi KG_- Q\tilde{\g})  \\
& =  \frac \pi 3 G_- \mu_2 (\g,   \mu_2(\g, \g) )  +\mu_2 (\g,   2\pi KG_- Q\tilde{\g}).  
\end{split}
\ee
In our derivation we used Leibniz rule  (\ref{def_Leibniz}),  homotopy formula (\ref{normalized_homotopy}), the projector property (\ref{def_Proj_G_+}) and the 7-term relation  (\ref{def_7_term}) for even, $G_-$-closed  state $\g$  in the form 
 \be
G_-  \mu_2 (\g, \mu_2(\g, \g) )  =   3 \mu_2 (\g, G_- \mu_2 (\g, \g)).
 \ee
We can replace the $Q \tilde{\g}$ in the last expression and use $G_-^2 =0 $ to get 
\be
\begin{split}
Q \tilde{\g} =  \frac\pi 3 G_- \mu_2 (\g,   \mu_2(\g, \g) )  + \mu_2 (\g, 2\pi  KG_-\mu_2 (\g,   2\pi KG_- Q\tilde{\g})  ).  
\end{split}
\ee
We can iterate the process for of the  $Q\tilde{\g}$-replacement to arrive into
\be
Q \tilde{\g} = \frac\pi 3 G_- \mu_2 (\g,   \mu_2(\g, \g) ). 
\ee
The invariance of the correlation functions follows from
\be
\begin{split}
\mathcal{F}_0(\Psi +Q \chi)   - \mathcal{F}_0(\Psi) & = g (Q\chi, \tilde{\g}) =  g (\chi, Q\tilde{\g}) =  \frac\pi 3 g (\chi, G_- \mu_2 (\g,   \mu_2(\g, \g) ) )  \\
&=   - \frac\pi 3 g ( G_- \chi , \mu_2 (\g,   \mu_2(\g, \g) ) ) = 0.
\end{split}
\ee
We used the $Q$-invariance of the  pairing  (\ref{Q_preserve}), the $G_\pm$-invariance of the pairing (\ref{def_G_inv_pairing})  and the assumption of the theorem that   $G_-\chi =0$. 
 $\hfill\blacksquare$

\subsection{Recursion relation for  correlation functions}\label{corr_recurs}

Let us consider the  HTQM $(V, Q, G_{\pm}, \mu_2, g)$ and four $Q$-, $G_-$-closed states $\Psi_a, a=1,..,4$. The  $4$-point correlation function   (\ref{4pt_function_gen})  of such states can be written in the following form 
\be\label{4p_3pt_recursion}
\begin{split}
\< \Psi_{1},\Psi_2,&\Psi_{3}, \Psi_4\>_Q   =    \mu_3^0\left( 2\pi K G_- \mu_2 (\Psi_4, \Psi_1),\Psi_{2},\Psi_3\right)+ \mu_3^0\left( \Psi_{1}, 2\pi K G_- \mu_2 (\Psi_4, \Psi_2),\Psi_3\right) \\
&+  \mu_3^0\left( \Psi_{1},\Psi_{2}, 2\pi K G_- \mu_2 (\Psi_4, \Psi_3)\right)  \\
&  = \frac{d}{d\e}\Big|_{\e=0}  \mu_3^0( \Psi^\e_{1},\Psi^\e_{2},\Psi^\e_3)  = \frac{d}{d\e}\Big|_{\e=0}   \< \Psi^\e_{1},\Psi^\e_{2},\Psi^\e_{{3}}\>_{Q_\e}, 
\end{split}
\ee
where we used   the leading order deformation (\ref{state_deform_htqm}) of states $\Psi_1, \Psi_2, \Psi_3$ by the state $\Psi_4$   i.e.
\be
\Psi_a^\e = \Psi_a+  2\pi  K G_- \mu_2 (\e\Psi_{4}, \Psi_a)  + \cO(\e^2),\;\; a=1,2,3.
\ee 
We can describe the relation (\ref{4p_3pt_recursion}) using  diagrammatic representation 
\newline\newline
\begin{tikzpicture}[scale=0.8]
\draw   [black,   ultra thick, dashed]   (0,0)-- (0, -1.5) ;
\draw   [black,  ultra thick, dashed]   (0,0)--(1, 1) ;
\draw   [black,  ultra thick, dashed]   (0,0)--(-1, 1) ;
\filldraw[black] (1,1) circle (2pt) node[anchor=south]{$\Psi_1^\e$};
\filldraw[black] (-1,1) circle (2pt) node[anchor=south]{$\Psi_2^\e$};
\filldraw[black] (0,-1.5) circle (2pt) node[anchor=north]{$\Psi_3^\e$};
\node  at (-2,-.5) {$\frac{d}{d\e}|_{\e=0}$};
\node  at (2,-.5) {$=$};
\end{tikzpicture}
\begin{tikzpicture}[scale=0.8]
\draw   [black,   thick, dashed]   (0,0)-- (0, -1.5) ;
\draw   [black,   thick ]   (0,0)--(0.5, 0.5) ;
\draw   [black,   thick, dashed]   (0.5,0.5)--(1, 1) ;
\draw   [black,   thick, dashed]   (0.5,0.5)--(1, 0) ;
\draw   [black,   thick, dashed]   (0,0)--(-1, 1) ;
\filldraw[black] (1,1) circle (2pt) node[anchor=south]{$\Psi_1$};
\filldraw[black] (-1,1) circle (2pt) node[anchor=south]{$\Psi_2$};
\filldraw[black] (0,-1.5) circle (2pt) node[anchor=north]{$\Psi_3$};
\filldraw[black] (1,0) circle (2pt) node[anchor=north]{$\Psi_4$};
\node  at (2,-.5) {$+$};
\end{tikzpicture}
\quad
\begin{tikzpicture}[scale=0.8]
\draw   [black,   thick, dashed]   (0,0)-- (0, -1.5) ;
\draw   [black,   thick, dashed]   (0,0)--(1, 1) ;
\draw   [black,   thick]   (0,0)--(-0.5, 0.5) ;
\draw   [black,   thick, dashed]   (-0.5,0.5)--(-1, 1) ;
\draw   [black,   thick, dashed]   (-0.5, 0.5)--(-1, 0) ;
\filldraw[black] (1,1) circle (2pt) node[anchor=south]{$\Psi_1$};
\filldraw[black] (-1,1) circle (2pt) node[anchor=south]{$\Psi_2$};
\filldraw[black] (0,-1.5) circle (2pt) node[anchor=north]{$\Psi_3$};
\filldraw[black] (-1,0) circle (2pt) node[anchor=north]{$\Psi_4$};
\node  at (2,-.5) {$+$};
\end{tikzpicture}
\quad
\begin{tikzpicture}[scale=0.8]
\draw   [black,   thick]   (0,0)-- (0, -0.5) ;
\draw   [black,   thick, dashed]   (0,-0.5)-- (0, -1.5) ;
\draw   [black,   thick, dashed]   (0,-0.5)-- (1, -0.5) ;
\draw   [black,   thick, dashed]   (0,0)--(1, 1) ;
\draw   [black,   thick, dashed]   (0,0)--(-1, 1) ;
\filldraw[black] (1,1) circle (2pt) node[anchor=south]{$\Psi_1$};
\filldraw[black] (-1,1) circle (2pt) node[anchor=south]{$\Psi_2$};
\filldraw[black] (0,-1.5) circle (2pt) node[anchor=north]{$\Psi_3$};
\filldraw[black] (1,-0.5) circle (2pt) node[anchor=west]{$\Psi_4$};
\end{tikzpicture}
\newline\newline
 {\bf Theorem (recursion relation)}:  The   $(n+1)$-point correlation function for the $Q$-, $G_-$-closed states $\Psi_0, \Psi_1,..,\Psi_n$ in HTQM on trees $(V, Q, G_{\pm}, \mu_2, g)$ can be expressed as a derivative of  $n$-point correlation function in 
  HTQM, deformed by the state $\Psi_{0}$,  i.e.
\be
\< \Psi_{1},...,\Psi_{n}, \Psi_0\>_Q = \frac{d}{d\e}\Big|_{\e=0}  \< \Psi^\e_{1},...,\Psi^\e_{{n}}\>_{Q^\e}. 
\ee
The deformed  HTQM $(V, Q^\e, G_\pm, \mu_2, g )$ has   differential 
\be
Q^\e = Q- 2\pi  [G_-, \mu_2 (\e\Psi_{0},\cdot) ]+\cO(\e^2),
\ee
while  deformed   states are
\be
\Psi_a^\e = \Psi_a +2\pi K G_- \mu_2 (\e\Psi_{0}, \Psi_a)  + \cO(\e^2).
\ee 
{\bf Proof}: The key idea in our proof is  to use the generating function for the amplitudes (\ref{gen_funct_ampl}).  The generating function  in deformed  theory is
\be
 \mathcal{F}_0 ( \Psi^\e, K^\e) =  \sum_{k=3}^\infty \frac{1}{k!} \<\underbrace{\Psi^\e, ...,\Psi^\e}_k\>_{Q^\e}. 
\ee
 Let us recall the formula  for the change of   generating function  under variation of the external state   $\Psi$  
\be
 \mathcal{F}_0 ( \Psi + \e \delta \Psi, K) = \mathcal{F}_0 ( \Psi, K ) +   g (\tilde{\gamma}, \e \delta \Psi)  +\cO(\e^2),
\ee
where  the state $\tilde{\g}$ is a sum over rooted trees (\ref{tree_state}). Similarly we can derive the change of generating function under the change  of  propagator $K$
\be
 \mathcal{F}_0 ( \Psi, K+\e \delta K) = \mathcal{F}_0 (\Psi, K ) +\pi  g (\tilde{\gamma}, \e \delta K \tilde{\gamma})  +\cO(\e^2).
\ee
The full change of the generating function is presented on the picture below
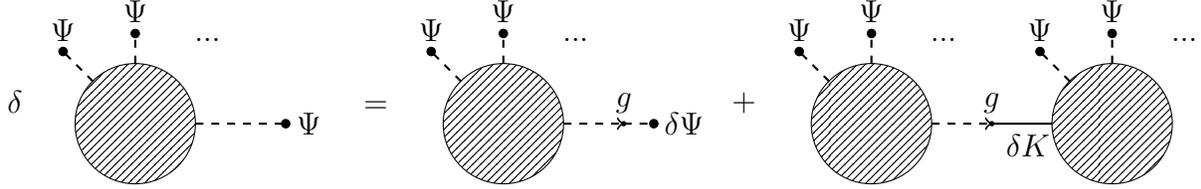
\newline\newline
\begin{tikzpicture}[scale=0.8]
  \draw[pattern=north east lines, pattern color=black] (0,0) circle (1);
  \draw   [black,   thick, dashed]   (1,0)-- (2.5, 0) ;
  \filldraw[black] (2.5,0) circle (2pt) node[anchor=west]{$\Psi$};
    \draw   [black,   thick, dashed]   (0,1)-- (0,1.5) ;
      \filldraw[black] (0,1.5) circle (2pt) node[anchor=south]{$\Psi$};
    \draw   [black,   thick, dashed]   (-0.7,0.7)-- (-1.2,1.2) ;
  \filldraw[black] (-1.2,1.2) circle (2pt) node[anchor=south]{$\Psi$};
    \filldraw[black] (1.2,1.2) node[anchor=south]{$...$};
    \filldraw[black] (-2,0)  node[anchor=south]{$\delta$};
        \filldraw[black] (4,0)  node[anchor=south]{$=$};
\end{tikzpicture}
\begin{tikzpicture}[scale=0.8]
  \draw[pattern=north east lines, pattern color=black] (0,0) circle (1);
  \draw   [black,   thick, dashed, ->]   (1,0)-- (2, 0) ;
    \draw   [black,   thick, dashed]   (2,0)-- (2.5, 0) ;
  \filldraw[black] (2.5,0) circle (2pt) node[anchor=west]{$\delta\Psi$};
    \filldraw[black] (2,0) circle (1pt) node[anchor=south]{$g$};
    \draw   [black,   thick, dashed]   (0,1)-- (0,1.5) ;
      \filldraw[black] (0,1.5) circle (2pt) node[anchor=south]{$\Psi$};
    \draw   [black,   thick, dashed]   (-0.7,0.7)-- (-1.2,1.2) ;
  \filldraw[black] (-1.2,1.2) circle (2pt) node[anchor=south]{$\Psi$};
    \filldraw[black] (1.2,1.2) node[anchor=south]{$...$};
        \filldraw[black] (4,0)  node[anchor=south]{$+$};
\end{tikzpicture}
\begin{tikzpicture}[scale=0.8]
  \draw[pattern=north east lines, pattern color=black] (0,0) circle (1);
    \draw[pattern=north east lines, pattern color=black] (4,0) circle (1);
  \draw   [black,   thick]   (2,0)-- (3, 0) ;
    \draw   [black,   thick, dashed, ->]   (1,0)-- (2, 0) ;
        \filldraw[black] (2,0) circle (1pt) node[anchor=south]{$g$};
                \filldraw[black] (2.6,0) node[anchor=north]{$\delta K$};
    \draw   [black,   thick, dashed]   (0,1)-- (0,1.5) ;
      \filldraw[black] (0,1.5) circle (2pt) node[anchor=south]{$\Psi$};
    \draw   [black,   thick, dashed]   (-0.7,0.7)-- (-1.2,1.2) ;
  \filldraw[black] (-1.2,1.2) circle (2pt) node[anchor=south]{$\Psi$};
     \filldraw[black] (1.2,1.2) node[anchor=south]{$...$};
     
      \draw   [black,   thick, dashed]   (4,1)-- (4,1.5) ;
      \filldraw[black] (4,1.5) circle (2pt) node[anchor=south]{$\Psi$};
    \draw   [black,   thick, dashed]   (3.3,0.7)-- (2.8,1.2) ;
  \filldraw[black] (2.8,1.2) circle (2pt) node[anchor=south]{$\Psi$};
    \filldraw[black] (5.2,1.2) node[anchor=south]{$...$};
\end{tikzpicture}
 \newline\newline
We use   the state deformation formula  (\ref{state_deform_htqm}) for  $\delta \Psi$ and  the propagator in deformed theory (\ref{prop_deform_htqm}) for $\delta K$  in terms of the state $\Psi_0$, i.e.
\be
\begin{split}
 \e\delta KG_-&=K^\e G_- -KG_-=  2\pi KG_- \mu_2 (\e\Psi_0,  KG_- \cdot ), \\
\e\delta \Psi &=\Psi^\e  -\Psi =   2\pi KG_- \mu_2 (\e\Psi_0,  \Psi).
\end{split}
\ee
Using the $KG_-$-flip (\ref{KG_flip}) and relations for the sum over rooted   trees  (\ref{tree_sum_relations}) we can  rewrite the  generating function  in deformed theory
\be
\begin{split}
 \mathcal{F}_0& ( \Psi^\e, K^\e)  - \mathcal{F}_0 ( \Psi, K)  =  g (\tilde{\gamma}, \e\delta \Psi)  +\pi  g (\tilde{\gamma}, \e\delta K \tilde{\gamma})+\cO(\e^2)\\
 & = g (\e\Psi_0,  \mu_2(2\pi KG_-\tilde{\gamma},  \Psi)) + \pi g ( \e\Psi_0, \mu_2( 2\pi KG_-\tilde{\gamma} ,  KG_- \tilde{\gamma} ) )+\cO(\e^2) \\
        & =   g (\e \Psi_0,\tilde{\g})- \frac12  g (\e\Psi_0,  \mu_2 (\Psi,  \Psi)) +\cO(\e^2) \\
        & =  \mathcal{F}_0 ( \Psi +\e  \Psi_0, K)  - \mathcal{F}_0 (\Psi, K )  - \frac12  g (\e \Psi_0,  \mu_2 (\Psi,  \Psi)) +\cO(\e^2). 
 \end{split}
\ee
The derivative of the relation becomes
\be
\begin{split}
 \sum_{k=3}^\infty & \frac{d}{d\e}\Big|_{\e=0}  \frac{1}{k!} \<\underbrace{\Psi^\e, ...,\Psi^\e}_k\>_{Q_\e}   =  \frac{d}{d\e}\Big|_{\e=0} \mathcal{F}_0 ( \Psi +\e  \Psi_0, K)- \frac12 \; g (\Psi_0,  \mu_2 (\Psi,  \Psi)) \\
    &= \frac{3}{3!} \<\Psi, \Psi,  \Psi_0\>_Q + \sum_{k=4}^\infty \frac{1}{k!} k \<\underbrace{\Psi,...,\Psi}_{k-1},  \Psi_0\>_Q - \frac12 \<\Psi, \Psi,  \Psi_0\>_Q \\
    &  = \sum_{k=3}^\infty \frac{1}{k!}  \<\underbrace{\Psi,...,\Psi}_k,  \Psi_0\>_Q.
 \end{split}
\ee
The equality holds at each order in $\Psi$ so 
\be
\frac{d}{d\e}\Big|_{\e=0}  \<\underbrace{\Psi^\e, ...,\Psi^\e}_k\>_{Q_\e}= \<\underbrace{\Psi,...,\Psi}_k,  \Psi_0\>_Q
\ee
what completes the proof of the theorem. $\hfill\blacksquare$

\section{Mirror for HTQM}

\subsection{A-model}

Tropical Gromov-Witten theory on toric manifold $X$ of complex dimension $N$  defines the  A-type HTQM  on trees, also denoted as the A-model in this section. In this section we will describe the   HTQM data $(V, Q, G_\pm,g, \mu_2)$ for A-model.

Let us consider pair  $|\omega, \vec{m}\>$,  of tropical form $\omega$  on $\mathbb{C}^{\ast N}$, and $N$-dimensional  integer-valued vector $\vec{m}$. 
  The  pairing on $|\omega, \vec{m}\>$ is the  integration of  corresponding form forms 
 \be
g( |\omega_1,\vec{m}_1\>, | \omega_2,\vec{m}_2\>)  = \delta_{\vec{m}_1+\vec{m}_2, \vec{0}}  \int_{\mathbb{C}^{\ast N}} \omega_1\wedge \omega_2.
 \ee
 The vector space $V_A$  is the space of tropical differential forms on $\mathbb{C}^{\ast N}$, equipped with the integer vector,  i.e.
 \be
 V_A =  \Omega^\ast_{trop}(\mathbb{C}^{\ast N}) \otimes  \mathbb{R}\< \vec{m}\;| \;\vec{m}\in \mathbb{Z}^N\>.
  \ee
The $\mathbb{Z}_2$-grading of the state $| \omega, \vec{m}\>$ is  the grading of the form $\omega$, a degree of the differential  form mod $2$. The differential $Q$  is the  de Rham operator on $\mathbb{C}^{\ast N}$
\be
Q|\omega, \vec{m}\>  = |d\omega, \vec{m}\> .
\ee 
The  $G_\pm$ on states is  a contraction with the  constant radial (angular) vector field $\vec{m}$
 \be
 \begin{split}
  G_+ |\omega, \vec{m}\> & = | \iota^R_{\vec{m}}\omega, \vec{m}\> =  | \iota_{m^i \p_{r^i}}\omega, \vec{m}\>,  \\
 G_-|\omega, \vec{m}\>&= |\iota^\Phi_{\vec{m}} \omega, \vec{m}\>.
 \end{split}
 \ee
The multiplication $\mu_2: V \otimes V \to V$  is  the wedge product on differential forms  supplemented with addition of corresponding vectors 
 \be
  \mu_2 ( |\omega_1,\vec{m}_1\> ,|\omega_2,\vec{m}_2\>) =|\omega_1\wedge\omega_2, \vec{m}_1+\vec{m}_2\>.
  \ee
The pair  $(Q, \mu_2)$ is essentially an external derivative and the wedge product,  hence obeys the  DGA  properties. The pair $(G_-, \mu_2)$  obeys the 7-term relation.

The Hamiltonian 
\be
H = \{ Q, G_+\}
\ee
is  the Lie derivative along the constant radial vector field $\vec{m}$
\be\label{hamiltonian_a_mod}
 H  |\omega, \vec{m}\> = \{ Q, G_+\} |\omega, \vec{m}\> =  | \{ d, \iota^R_{\vec{m}}\} \omega, \vec{m}\> =  | \cl^R_{\vec{m}}\omega, \vec{m}\> .  
\ee
 The evolution operator $ e^{-tH}$, defined as solution  to 
\be\label{diff_eq_exponent}
(\p_t +H) e^{-tH} = (\p_t + \cl^R_{\vec{m}}) e^{-tH} = 0
\ee
is  a 1-parameter family of diffeomorphisms $\Phi^t_{\vec{m}}: r^i  \mapsto r^i - m^i t$
 \be
 e^{-tH}  | \omega, \vec{m}\> =  |(\Phi^t_{\vec{m}})^\ast \omega, \vec{m}\>.
 \ee
The  composition property (\ref{semi-grp}) naturally holds for diffeomorphisms. Since the vector fields are constant vector fields the corresponding flows do not develop any singularities, hence composition is valid for all values of $t$.

\subsection{Correlation functions in A-model}

In our paper  \cite{Losev:2022tzr} we showed that the  tropical  GW  invariant of genus-0 and degree-$\beta$ on toric space $X$ is the sum of the A-type  HTQM  amplitudes 
\be
\<\g_1,...,\g_n \>^X_\beta =  \sum_\G  \mathcal{A}_\G (\Psi_{\g_1},..,\Psi_{\g_n},  \Psi_{b_1},..,\Psi_{b_B})
\ee
with two types of states: 
\begin{itemize}
\item Evaluation states $\Psi_{\g_k} = |\g_k, \vec{0}\>$, constructed  from the tropical forms $\g_k$ on X.  The space $\Omega^{trop}(X)$ of  tropical forms on $X$  is a  subspace of tropical forms on 
$\mathbb{C}^{\ast N}$ with good behaviour at compactifying divisors. 
\item Divisor states $\Psi_{\vec{b}_a} = |1, \vec{b}_a\>$,  where $\vec{b}_a$ are primitive normal vectors  for  compactifying divisors  of $X$.
\end{itemize}

Using the definition (\ref{def_qm_corr_func}) we can replace the sum over the amplitudes by the correlation function  and  formulate the HTQM representation for the tropical GW invariants in the form: 
\newline\newline
{\bf Theorem (HTQM representation of tropical GW)}: For toric space $X$, given in terms of boundary divisors $B_X$, and tropical cycles $\g_k$,  the tropical GW invariant matches with the HTQM correlation function i.e.
\be\label{trop_GW_htqm_rep}
\<\g_1,...,\g_n \>^X_\beta = \frac{1}{d_1!\cdot..\cdot d_{B}!} \< \Psi_{\g_1},..,\Psi_{\g_n},  \underbrace{\Psi_{\vec{b}_1},..,\Psi_{\vec{b}_1}}_{d_1},.., \underbrace{\Psi_{\vec{b}_B},..,\Psi_{\vec{b}_B}}_{d_B}\>_Q.
\ee
Here $\Psi_{\g_a}$ are the evaluation states for tropical cycles $\g_a$, $\Psi_{\vec{b}}$ are divisor states  for boundary divisors from $B_X$ of  cardinality $B = | B_X|$.  The degree of the map $\beta  \in H_{1,1}(X) $ determines the number $d_a$ of divisor states  $\Psi_{\vec{b}_a}$ of a given type via 
the corresponding  tropical intersection number of $\beta$ and boundary divisor with normal vector $\vec{b}_a$.
\newline\newline
The  GW theory is defined on classes of cycles $C_\a$. The change of a cycle within the same class  leads to the shift of an evaluation observable (Poincare-dual form) $\g$ by an exact form 
\be
\g  \to \g  + d \lambda.
\ee 
The shift of evaluation observable $\g$ changes the corresponding  A-model state  $\Psi_\g$ by a $Q$-exact term
\be
\Psi_\g \to \Psi_{\g + d\lambda} = \Psi_\g +  Q \Psi_\lambda.
\ee
Furthermore, since the states $\Psi_\g$ and $\Psi_\lambda$ carry trivial  integer vector, they are $G_-$-closed, i.e.  
\be
G_-\Psi_\lambda =G_-|\lambda, \vec{0}\> =  0,\;\; G_-\Psi_\g =G_-|\g, \vec{0}\> =  0
\ee
The forms $\g_a$ are closed forms so is the corresponding HTQM states $\Psi_{\g_1}$. Hence we can use the invariance theorem (\ref{corr_inv}) to verify that the tropical GW invariants in HTQM representation (\ref{trop_GW_htqm_rep}) are  defined on cohomology classes of $\g_a$.

\subsection{Dual variables}

It is  convenient to introduce  angular variables   $Y_j \in S^1 $,  dual to the  integer vector components  $m^i \in \mathbb{Z}$. 
We  introduce  a Fourier transform of a state 
\be
\sum_{\vec{m}\in \mathbb{Z}^N}  c_{\vec{m}}|\omega_{\vec{m}}, \vec{m}\> \mapsto \Psi= \sum_{\vec{m}\in \mathbb{Z}^N} e^{i \< \vec{m},  \vec{Y} \>} c_{\vec{m}}|\omega_{\vec{m}}, \vec{m}\> \in V_B =  \Omega^{trop}(\mathbb{C}^{\ast N}) \otimes C^{\infty} (\mathbb{T}^N).
\ee
For convenience we  describe the differential forms on $X$ using  Grassmann variables: 
\be
dr^j  = \psi_R^j,\;\;\; d\phi^j = \psi_\Phi^j.
\ee 
The differentials     on mirror states in new notations  become first and second order differential operators 
\be\label{diff_op_repres}
\begin{split}
Q\Psi &=\sum_{\vec{m}\in \mathbb{Z}^N} e^{i \< \vec{m},  \vec{Y} \>}c_{\vec{m}}| d\omega_{\vec{m}}, \vec{m}\> = d_R \Psi = \psi_R^k \frac{\p}{\p r^k} \Psi,\\
G_-\Psi &= \sum_{\vec{m}\in \mathbb{Z}^N} e^{i \< \vec{m},  \vec{Y} \>}c_{\vec{m}}| \iota^\Phi_{\vec{m}}\omega, \vec{m}\>   = -i \frac{\p}{\p Y_k} \frac{\p}{\p \psi_{\Phi}^k}\Psi,\\
 G_+\Psi & = \sum_{\vec{m}\in \mathbb{Z}^N} e^{i \< \vec{m},  \vec{Y} \>}c_{\vec{m}}| \iota^R_{\vec{m}}\omega, \vec{m}\>   = -i \frac{\p}{\p Y_k} \frac{\p}{\p \psi_{R}^k}\Psi.
 \end{split}
\ee
The  multiplication $\mu_2$,  on forms  becomes multiplication of functions on superspace with coordinates $r, Y, \psi_R, \psi_\Phi$
 \be
  \mu_2 (\Psi_1 ,\Psi_2) =\Psi_1\cdot \Psi_2,
  \ee
while the pairing is the integration over superspace
 \be\label{B_htqm_pairing}
g(\Psi_1, \Psi_2) =\int d\mu  \;\; \Psi_1\Psi_2,
  \ee
where Berezin integration measure  for $\dim_{\mathbb{C}} X = N$ is
\be\label{pairing_int_measure}
d\mu = d^N r\; d^N Y\; d^N\psi_{\Phi} d^N \psi_{R}.
\ee
The integrating region (for Grassmann-even variables) is the $N$-dimensional torus $(S^1)^N$ for $Y$-variables  and Euclidean space $\mathbb{R}^N$ for $r$-variables.
\newline\newline
{\bf Remark}: The differential operator representation (\ref{diff_op_repres}) of the HTQM  data $(V, Q, G_{\pm}, \mu_2, g)$ allows for an easy check of   HTQM definitions from sections \ref{sec_htqm_def} and \ref{sec_htqm_tree_def}.
In particular  the 7-term relation (\ref{def_7_term}) for $G_-$ is a property of the second order differential operator in representation (\ref{diff_op_repres}). 
\newline\newline
The  divisor states  $ |1, \vec{b}\>$    become exponential   functions  of $Y$
  \be
\Psi_{\vec{b}} (Y) =  e^{i\<\vec{b},\vec{Y}\>}=e^{i b^kY_k}.
  \ee
 For the tropical form 
 \be
 \g = \g_{i_1..i_kj_1..j_l}(r) \;d\phi^{i_1}\wedge..\wedge d\phi^{i_k}\wedge dr^{j_1}\wedge..\wedge dr^{j_l}
 \ee
 the corresponding evaluation state  is 
 \be
 \Psi_\g  =  \g_{i_1..i_kj_1..j_l}(r)\; \psi_{\Phi}^{i_1}\cdot..\cdot \psi_{\Phi}^{i_k} \cdot\psi_R^{j_1}\cdot..\cdot\psi_R^{j_l}.
 \ee
 {\bf Example}:   The evaluation state for $U(1)$-invariant  Poincare dual of the point on $\mathbb{P}^1$ is
  \be
\g = \frac{1}{2\pi}\delta (r-r_0) d\phi dr   \mapsto  \Psi_\g = | \g, 0\> =  \frac{1}{2\pi} \delta (r-r_0) \psi_\Phi \psi_R.
  \ee

\subsection{B-model}\label{sec_B_model}
In our work  \cite{Losev:2022tzr}  we showed that the deformation of A-model, the HTQM  $(V_A,Q, G_\pm, \mu_2, g)$, by divisor states $\Psi_{\vec{b}}$ for toric $X$  is also an  HTQM $(V_B,Q_X, G_\pm, \mu_2, g)$, which we will denote as the B-type HTQM or B-model for short. 
\newline\newline
{\bf Remark}: The two-dimensional version of the deformation by compactifying divisors of toric space was discussed in A-I-B mirror paper \cite{Frenkel:2005ku}.  However,   the deformation of observables was not  discussed there. 
\newline\newline
Divisor states obey 
\be
G_- \mu_2 (\Psi_{\vec{b}_1}, \Psi_{\vec{b}_2}) = 0,
\ee 
hence the  deformation of the differential $Q$, which we discussed in section \ref{sec_htqm_deform}  holds beyond the linearized level.   The deformation of the A-model  differential by all divisors $B_X$ of $X$ is
\be
Q_X=Q - 2\pi \sum_{\vec{b} \in B_X} [G_-, \mu_2 (q_{\vec{b}} \Psi_{\vec{b}}, \cdot)]= \psi_R^j \frac{\p}{\p r^j}+ 2\pi  i \sum_{\vec{b} \in B_X}  q_{\vec{b}}  \frac{\p \Psi_{\vec{b}}}{ \p Y_j }\frac{\p}{\p \psi_{\Phi}^j}  .
 \ee
The same differential  can be written as 
 \be
 Q_X  =   Q + 2\pi i  \frac{\p W_X(Y)}{ \p Y_j }\frac{\p}{\p \psi_{\Phi}^j} =Q_{W_X},
 \ee
using  the  {\it  mirror superpotential} 
 \be
 W_X (Y) = \sum_{\vec{b}\in B_X} q_{\vec{b}}\; \Psi_{\vec{b}} =  \sum_{\vec{b} \in B_X} q_{\vec{b}}\;  e^{i\<\vec{b} ,\vec{Y}\>}.
 \ee
 We can absorb some   $q_{\vec{b}}$ by the redefinition of $Y_j$ to obtain more familiar (at least for the $\mathbb{P}^N$ case)  form of the superpotential with fewer parameters $q_{\vec{b}}$.
\newline\newline
{\bf Remark}: The exponential mirror superpotentials for toric spaces were derived by Givental \cite{givental1995quantum} and by Hori and Vafa \cite{Hori:2000kt} using different methods.
\newline\newline 
{\bf Definition}: For  toric space $X$  the deformation (\ref{state_deform_htqm}) of an A-model  evaluation state   $\Psi_\g$ by the  divisor states $\Psi_{\vec{b}}$   is a  {\it mirror state} $\Psi_\g^X$. We can   write the mirror state  using the  A-model notations
in the form
\be\label{B_type_QM_obs_deform}
\begin{split}
\Psi_\g^X &=\Psi_\g  + 2\pi K G_-\mu_2 (W_X,   \Psi_\g)  +  (2\pi)^2 K G_- \mu_2 (W_X,    K G_- \mu_2 (W_X,   \Psi_\g) )+...
 \end{split}
\ee
 For an A-model state  $\Psi_\g$ corresponding to the tropical form  $\g \in \Omega^{k,l}(X)$ the sum (\ref{B_type_QM_obs_deform})  terminates after  min$(k,l)+1$ terms. Indeed, the action of $KG_-$ lowers the degree of the form by $(1,1)$, hence it can be applied to $(k,l)$-form $\Psi_\g$ at most  min$(k,l)$ times. 
 
 By preservation of closeness proposition (\ref{prop_def_state_close}),  the mirror state $\Psi_\g^X$ is $Q_X$-  and $G_-$-closed, i.e.
 \be
Q_X \Psi_\g^X  = G_- \Psi_\g^X =0.
 \ee
{\bf Example}: For  $X = \mathbb{P}^1$   the space of boundary normal vectors is   $B_{\mathbb{P}^1} = \{+1, -1\}$, hence the mirror superpotential   equals
 \be
 W_{\mathbb{P}^1} (Y) =  \sum_{b \in B_{\mathbb{P}^1} } q_{b}\;  e^{ ibY} =  q_+ e^{iY} + q_- e^{-iY}.
 \ee
 We can absorb  the   $q_+$ by constant shift of $Y$ to arrive into  more familiar form of the superpotential 
 \be
 W_{\mathbb{P}^1} =  e^{iY} + q e^{-iY}
 \ee
 with $q=q_+q_-$. The B-model  differential  is
 \be
 Q_{\mathbb{P}^1} = \psi_R \frac{\p}{\p r} +  2\pi i  \frac{\p W_{\mathbb{P}^1}}{ \p Y }\frac{\p}{\p \psi_{\Phi}} = \psi_R \frac{\p}{\p r}  -2\pi ( q_+ e^{iY} - q_- e^{-iY})   \frac{\p}{\p \psi_{\Phi}}.
 \ee
The mirror state  $\Psi^{\mathbb{P}^1}_\g$   for the A-model evaluation observable $\g \in H^\ast_{dR} (\mathbb{P}^1)$  contains three terms
\be
\begin{split}
\Psi^{\mathbb{P}^1}_\g&=\Psi_\g  + 2\pi KG_- \mu_2(W_{\mathbb{P}^1}, \Psi_\g )  \\
&=\Psi_\g  + 2\pi q_+\int^\infty_0 dt\; e^{-tH}\; G_+G_-  (e^{iY}\Psi_\g)  + 2\pi q_- \int^\infty_0dt\; e^{-tH}\; G_+G_- ( e^{-iY} \Psi_\g ).
\end{split}
\ee
Indeed, the top form on $\mathbb{P}^1$ has degree $(1,1)$, hence we can apply the deformation only once for each of two boundary divisors.
The mirror state for the  constant form  $\g= 1 \in H_{dR}^{0}(\mathbb{P}^1) $  has a trivial deformation 
\be
\Psi^{\mathbb{P}^1}_1 = \Psi_1 = 1.
\ee
Consider the Poincare dual of the point $\phi=\phi_0$ and $r=r_0$ in tropical coordinates on $\mathbb{P}^1$
\be
P_0 =  \delta (\phi-\phi_0)\delta(r-r_0)\; d\phi dr.
\ee
The   $U(1)$-averaging over $\phi_0$ of the form $P_0$ is 
\be
P  = \frac{1}{2\pi} \delta(r-r_0)\; d\phi dr  \in H_{dR}^{1,1}(\mathbb{P})
\ee 
The corresponding  A-model state  equals
\be
\Psi_P = \frac{1}{2\pi} \delta(r-r_0)  \psi_\Phi \psi_R.
\ee
The corresponding  mirror  state is  
\be\label{P_1_mirror_state}
\begin{split}
\Psi^{\mathbb{P}^1}_P&= \frac{1}{2\pi} \delta (r-r_0) \psi_\Phi \psi_R +q_+ e^{iY}\int\limits^{\infty}_{0}dt\; \delta(r-r_0-t) +  q_- e^{-iY} \int\limits^{\infty}_{0}dt\; \delta(r-r_0+t)   \\
&  = \frac{1}{2\pi}   \delta (r-r_0) \psi_\Phi \psi_R+ q_+ e^{iY} \Theta (r-r_0) +  q_- e^{-iY}  \Theta(r_0-r).
\end{split}
\ee
\newline\newline
In our work  \cite{Losev:2022tzr}  we proved the mirror symmetry for the A-type HTQM $(V_A, Q, G_{\pm}, \mu_2, g)$  and the B-type HTQM $(V_B, Q_X, G_{\pm}, \mu_2, g)$. The proof was essentially a summation over the divisor states $\Psi_{\vec{b}}$
for all divisors of the toric space $X$.  The HTQM mirror implies the following statement for the correlation functions.
\newline\newline
{\bf Theorem (tropical mirror for HTQMs)}: On toric space $X$, with boundary divisors $\vec{b}_1,...,\vec{b}_B$  the sum over divisor states in the    A-model's   correlation function of evaluation states  $\Psi_{\g_a}$ equals to the correlation function of the corresponding mirror states  $\Psi^X_{\g_a}$ in  B-model   with mirror superpotential $W_X$ i.e.
\be
\sum_{k=0}^\infty \frac{1}{k!} \;\< \Psi_{\g_1},...,\Psi_{\g_n},  \underbrace{W_X,..,W_X}_k\>_Q = \< \Psi^X_{\g_1},...,\Psi^X_{\g_n} \>_{Q_{W_X}}.
\ee
Note that the $q$-dependence in the A-model is due to the divisor states in $W_X$, while in B-model both evaluation states and differential $Q_X$ have nontrivial $q$-dependence.

\section{Localization of  mirror states}

In the  A-model formulation    evaluation states and  divisor states look very different and does not have any simple relation. In case  $X = \mathbb{P}^1$ the evaluation  state for the point observable at  $r=r_0$   and divisor states are 
\be
\Psi_P  =\frac{1}{2\pi}\delta (r-r_0)\psi_\Phi \psi_R,\;\;\; \Psi_{+}  =  e^{iY},\;\;   \Psi_- =   e^{-iY}.
\ee
In section \ref{sec_B_model} we constructed the  $\mathbb{P}^1$-mirror state for the  evaluation state of a point observable
\be\label{ex_point_morr_p1}
\Psi^{\mathbb{P}^1}_P  =\frac{1}{2\pi} \delta (r-r_0)\psi_\Phi \psi_R +  q_+e^{iY} \Theta (r-r_0) +  q_- e^{-iY}  \Theta(r_0-r).
\ee
The expression (\ref{ex_point_morr_p1}) for mirror state   turns into a  pure  divisor state $\Psi_{\pm}$ in  the limit  $r_0 \to \pm \infty$, i.e.
\be
\begin{split}
 \lim_{r_0 \to -\infty} &\Psi^{\mathbb{P}^1}_P  = q_+ e^{iY} =q_+ \Psi_+ \\
\lim_{r_0 \to +\infty} & \Psi^{\mathbb{P}^1}_P =  q_- e^{-iY} =q_- \Psi_-.
\end{split} 
\ee
There is a natural geometric interpretation of this relation: The point at finite position $r=r_0$ becomes a compactifying divisor (a point at infinity in case of $\mathbb{P}^1$) when we move $r_0$ to infinity. 

The relation  between  mirror states and divisor states plays a key role for the localization of the B-model correlation functions. 
 Note, that the limit of the state, often referred to as the point-wise limit,   in general,  may  not commute  with the amplitude/correlation function evaluation for the same state. We will discuss this potential problem  carefully in the next section.

\subsection{Mirror states vs divisor states}

 The pair of  A-model states, corresponding to the Poincare duals of    points $r_0$ and  $r_1$ are related by a $Q$-exact term
\be
\begin{split}
\Psi_1-\Psi_0 &=\frac{1}{2\pi} \delta (r-r_1)\psi_\Phi \psi_R-\frac{1}{2\pi}\delta (r-r_0)\psi_\Phi \psi_R  \\
&= Q \left(-\frac{1}{2\pi}\Theta (r-r_1)\psi_\Phi + \frac{1}{2\pi}\Theta (r-r_0) \psi_\Phi \right) = Q \chi_{01}.
\end{split}
\ee
The tropical  form 
\be\label{chi_01}
\begin{split}
\chi_{01} &  =\frac{1}{2\pi} \left[ \Theta (r-r_0)-\Theta (r-r_1) \right]  \psi_\Phi. 
\end{split}
\ee
is a tropical form on $\mathbb{P}^1$ since the difference  of two $\Theta$-functions has finite support.   We can take a limit $r_1 \to -\infty$  for  $\chi_{01}$  to define 
\be
\begin{split}
\chi_+ = \lim_{r_1 \to -\infty} \chi_{01} & = \frac{1}{2\pi} (\Theta (r-r_0)-1)\psi_\Phi  =- \frac{1}{2\pi} \Theta (r_0-r) \psi_{\Phi}. 
\end{split}
\ee
The   tropical form  $\chi_+$  can be used to turn the    mirror state into  the boundary divisor state,  i.e.
\be\label{cp_1_dress_+}
\begin{split}
q_+\Psi_+ = \Psi^{\mathbb{P}^1}_P +Q_{\mathbb{P}^1} \chi_+.  
\end{split}
\ee
The relation (\ref{cp_1_dress_+}) is equivalent to the  statement that the  mirror state $\Psi^{\mathbb{P}^1}_P$  and the holomorphic function $\Psi_+  =  e^{iY}$ represent the same cohomology as a forms in $V_B$, i.e.
\be
 \Psi^{\mathbb{P}^1}_P = q_+\Psi_+ \in H^\ast (Q_{\mathbb{P}^1}, V_B).
\ee
Note that  the tropical form $\chi_+$ is $G_-$-closed, i.e.
\be
G_-\chi_+ = G_- \left(- \frac{1}{2\pi}\Theta (r_0-r)   \psi_\Phi \right) = 0,
\ee
hence we can formulate a stronger equality in cohomology
\be
\Psi^{\mathbb{P}^1}_P  =q_+ \Psi_+   \in H^\ast (Q_{\mathbb{P}^1} +z  G_-, V_B\otimes\mathbb{R}[[z]]).
\ee
We can take $r_1 \to +\infty$ limit for (\ref{chi_01})  to define  a different tropical form   
\be
\chi_-=\frac{1}{2\pi} \Theta (r-r_0)\psi_\Phi  
\ee
 such that  
\be\label{cp_1_dress_-}
q_-\Psi_- =\Psi^{\mathbb{P}^1}_P +Q_{\mathbb{P}^1}  \chi_-  
\ee
Similarly to the previous case 
\be
G_-\chi_-  = G_-\left( \frac{1}{2\pi} \Theta (r-r_0)   \psi_\Phi \right)= 0,
\ee
hence 
\be
\Psi^{\mathbb{P}^1}_P  =q_- \Psi_-   \in H^\ast (Q_{\mathbb{P}^1}+z G_-, V_B\otimes\mathbb{R}[[z]]).
\ee
 We can choose  more general   $Q_{\mathbb{P}^1}$-exact term, so that the mirror state will become a  holomorphic function, which is not one of the compactifying divisor states  for $\mathbb{P}^1$
\be\label{cp_1_dress_descend}
\begin{split}
\hat{\Psi}_P &=\Psi^{\mathbb{P}^1}_P +Q_{\mathbb{P}^1} \hat{\chi}  =  \Psi^{\mathbb{P}^1}_P +Q_{\mathbb{P}^1} \left( \frac{1}{2\pi} \Theta (r-r_0)   \psi_\Phi +  \frac{1}{2\pi}  q e^{-2iY}  \psi_{\Phi} \right)  \\
& = q^2 e^{-3iY}. 
\end{split}
\ee
The key difference from the $\chi_\pm$  is that the tropical form $\hat{\chi}$ is not $G_-$-closed, i.e.
\be
G_- \hat{\chi} =  \left( \frac{1}{2\pi} \Theta (r-r_0)   \psi_\Phi +  \frac{q}{2\pi} \;  e^{-2iY}  \psi_{\Phi} \right)  =- \frac{q}{\pi } \; e^{-2iY}\neq 0.
\ee
hence  $\hat{\Psi}_P$  and $\Psi_P$ belong to different  classes in $(Q_{\mathbb{P}^1} +z G_-)$-cohomology.   Indeed, a simple computation shows that  
\be
q^2e^{-3iY} = e^{iY}    +  \frac{qz}{\pi}  \; e^{-2iY}  \in  H^\ast ( Q_{\mathbb{P}^1} +z  G_-, V_B\otimes\mathbb{R}[[z]]).
\ee

In the remaining part of  this section we will show that all mirror states admit  holomorphic function  representatives in   $H^\ast (Q_W+z G_-)$  and discuss some properties of these representatives.
 
\subsection{Spectral sequence for $Q_W$-cohomology }\label{sec_spec_seq}

A mirror state describes a class in $H^\ast (Q_W, V_B)$. The $Q_W$-differential 
\be
Q_W = \psi^j_R \frac{\p}{\p r^j} + 2\pi i \frac{\p W}{\p Y_j} \frac{\p}{\p \psi_\Phi^j} = d_R + 2\pi i \; {\bf Q}_W
\ee
 is a sum of two (graded-) commuting  differentials, the radial de Rham differential 
 \be
 d_R = \psi^j_R \frac{\p}{\p r^j},\;\;\; d_R^2=0
 \ee 
 and the LGS  differential  (\ref{def_LGS_differntial})
\be
{\bf Q}_W = \frac{\p W}{\p Y_k} \frac{\p}{\p \psi_\Phi^k},\;\;\; {\bf Q}_W^2=0. 
\ee
Hence we  can use a spectral sequence to evaluate the cohomology of the $Q_W$. The spectral sequence converges at the second step 
\be
H^{\ast}(Q_W, V_B) = H^\ast({\bf Q}_W,  H^\ast ({d_R}, ¬V_B)),
\ee
since the only cohomology of the radial de Rham operator on $\mathbb{R}^N$ are constant ($r$-independent) forms of degree 0 in $\psi_R$ and is isomorphic  to  the LSG vector space (\ref{def_lsg_vect_space}).
 The ${\bf Q}_W$  is differential in LGS theory,  hence we can express the cohomology via the Jacobi ring (\ref{def_Jacobi_ring}) for superpotential W
\be\label{isom_Q_W_jacobi}
H^{\ast}(Q_W, V_B) =  J_W.
\ee
An isomorphism (\ref{isom_Q_W_jacobi}) means  that for every $Q_W$-closed B-model state we can find a holomorphic function from the same $Q_W$-cohomology class on $V_B$.

\subsection{Pairing and localization of states}\label{sec_pairing}
{\bf Definition}: The  {\it holomorphic germ}  for  a  B-model state $\Psi$  is an  evaluation of the state $\Psi$  at  $r=\psi=0$, i.e.
\be
\Phi(Y) = \Psi(r, Y, \psi_\Phi, \psi_R)\Big|_{\psi=r=0}.
\ee
\newline\newline
{\bf Definition}: For states $\Psi_1, \Psi_2$ in  B-model with  superpotential $W$  we   introduce a   pairing  
\be\label{QM_W-pairing}
g^\Lambda_W(\Psi_1, \Psi_2) =g\left(\Psi_1, e^{\Lambda Q_W(L)}\Psi_2\right),
\ee
where $g$ is the  B-model pairing,  $\Lambda$ is a real parameter and  $L$ is a localization function
\be
L =   \sum_{k=1}^N r^k \psi^k_\Phi.
\ee 
\newline
{\bf Note}:  The exponent in the pairing evaluates into 
\be\label{oscill_exponent}
Q_W(L) = \sum_{k=1}^N \psi^k_\Phi \psi^k_R  + 2\pi i  \sum_{k=1}^N  r^k \frac{\p W}{\p Y_k}.
\ee
Hence the $Q_W(L)$ in (\ref{oscill_exponent}) is an oscillating function in parity-even variables $r$ and $Y$ what makes  the radial  integral converging. 
\newline\newline
The pairing $g^\Lambda_W$ matches with the B-model pairing (\ref{B_htqm_pairing}) for $\Lambda =0$. The pairing $g^\Lambda_W$ is $Q_W$-invariant, indeed
\be
 \begin{split}
g^\Lambda_W(Q_W\Psi_1, \Psi_2) &=g\left(Q_W\Psi_1, e^{\Lambda Q_W(L)}\Psi_2\right)  =- (-1)^{|\Psi_1|}g\left(\Psi_1, Q_We^{\Lambda Q_W(L)}\Psi_2\right) \\
&  = -  (-1)^{|\Psi_1|} g^\Lambda_W(\Psi_1, Q_W \Psi_2).
 \end{split}
\ee 
We used the $Q_W$-invariance of the B-model pairing and an identity  
\be\label{id_q_w_commut}
Q_W e^{Q_W(L)} =Q_W e^{Q_WL +L Q_W} =  e^{Q_W(L)} Q_W.
\ee
The $Q_W$-invariance of the pairing implies  that it is well-defined on $H^\ast (Q_W, V_B)$.
\newline\newline
{\bf Lemma}: On $Q_W$-closed states the pairing $g^\Lambda_W$ is independent of $\Lambda$. 
\newline\newline
{\bf Proof}: The derivative of the pairing evaluates into
\be
 \begin{split}
\frac{d}{d\Lambda} & g^\Lambda_W(\Psi_1, \Psi_2)  = \frac{d}{d\Lambda} g\left(\Psi_1, e^{\Lambda Q_W(L)} \Psi_2\right) =  g\left(\Psi_1,  (Q_W L + L Q_W) e^{\Lambda Q_W(L)} \Psi_2\right) \\
&= g\left(\Psi_1,  L  e^{\Lambda Q_W(L)} Q_W \Psi_2\right) - (-1)^{|\Psi_1|} g\left(Q_W \Psi_1,    e^{\Lambda Q_W(L)} \Psi_2\right)  =0
\end{split}
\ee
We used the $Q_W$ invariance of the B-model pairing and $Q_W$-closeness of both $\Psi_1$ and $\Psi_2$ and an identity (\ref{id_q_w_commut}).
$\hfill\blacksquare$
\newline\newline
The $\Lambda$-independence means that the pairing $g^\Lambda_W$ matches with the  B-model pairing $g$ on $H^\ast (Q_W, V_B)$. The B-model pairing $g$ is non-degenerate on $V_B$ and $Q_W$-invariant, hence $g$ is non-degenerate on   $Q_W$-cohomology, so is the  $g^\Lambda_W$.
\newline\newline
{\bf Proposition}: The $Q_W$-closed state $\Psi$ and its holomorphic germ $\Phi$  are in the same $Q_W$-cohomology class i.e.
\be\label{b_model_state_hol_germ}
\Psi = \Phi \in H^\ast (Q_W, V_B).
\ee
{\bf Proof}:  Since $Q_W(L)$ in (\ref{oscill_exponent}) is an oscillating function in parity-even variables $r$ and $Y$, then  the integral will  localize near critical points of  $Q_W(L)$.  
Moreover, the  localization exponent is scaled by  $\Lambda$, so we can choose large $\Lambda$  to eliminate the subleading corrections to the  saddle point  formula.    
 The critical points of $Q_W(L)$  are determined from
 \be
 \begin{split}
\frac{\p}{\p r^k} Q_W(L) & =  2\pi i  \frac{\p W}{\p Y_k}=0,\;\;\; \frac{\p}{\p Y_m} Q_W(L) = 2\pi   i  \sum_{k=1}^N  r^k \frac{\p^2 W}{\p Y_k  \p Y_m}=0, \\
\frac{\p}{\p \psi^k_\Phi} Q_W(L) & =  \psi^k_R=0,\;\;\; \frac{\p}{\p \psi^k_R} Q_W(L) = -  \psi^k_\Phi=0.
\end{split}
 \ee
Under the  assumption  that the critical points  of $W$  are isolated,   the rank of  $\frac{\p^2W}{\p Y_l \p Y_m}$ is maximal, hence the integral in the limit $\Lambda \to \infty$  localizes to the critical  points of $W$ and origin in all other variables
\be
r^k =\psi^k_R = \psi^k_\Phi=0,\;\; Y = Y_0,\;\; dW(Y_0) = 0.
\ee
The ratio of determinants for the integration around the critical point is
\be
\det \left( \frac{\p^2 Q_W(L)}{\p \psi^k_\Phi \p \psi^l_R} \right) \cdot  \frac{(2\pi)^N}{\det  \left(\frac{\p^2 Q_W(L)}{\p Y_k \p r^l} \right) }  = \frac{(2\pi)^N \Lambda^{N}}{  (2\pi  \Lambda)^{N}  \det  \left(\frac{\p^2W}{\p Y_k \p Y_l} \right)}.
\ee
The leading order saddle point formula for the pairing  integral  localizes the  $g^\Lambda_W$-pairing to 
\be\label{W_pairing}
g^\Lambda_W( \Psi_1, \Psi_2) =   \sum_{ dW= 0}  \frac{1}{  \det  \left(\frac{\p^2W}{\p Y_k \p Y_l} \right)} \;\; \Psi_1 \cdot \Psi_2 \Big|_{r=\psi=0}. 
\ee
Note  that the $\Lambda$-dependence drops out from the leading saddle point formula.  
We observe that the pairing $g^\Lambda_W$ for B-model states   localizes to the corresponding holomorphic germs. In particular, we have an equality
\be
0 = g^\Lambda_W( \Psi, \Psi') - g^\Lambda_W( \Phi, \Psi') =   g^\Lambda_W( \Psi-\Phi, \Psi').
\ee
The equality holds for all   $Q_W$-closed  states $\Psi'$ and   pairing $g_W^\Lambda$ is non-degenerate on $H^\ast(Q_W, V_B)$, hence  $\Psi$  and $\Phi$ represent the same class in $Q_W$-cohomology, i.e.
\be
\Psi = \Phi  \in H^\ast (Q_W, V_B),
\ee
what completes the proof of the proposition. $\hfill\blacksquare$

\subsection{Higher pairings in B-model}\label{sec_b_model_high_pairing}
{\bf Definition}: For a B-model with superpotential  $W$ we can  introduce a $\mathbb{C}[[z]]$-valued  pairing  on B-model states
\be\label{def_higher_par_b_model}
\mathcal{K}_W(\Psi_1, \Psi_2) =  \int d\mu\; \Psi_1 \cdot  e^{\Lambda \{Q_W+zG_-, L\}} \Psi_2=g\left(\Psi_1, e^{\Lambda \{Q_W+zG_-, L\}}  \Psi_2\right),
\ee
with the  integration measure  (\ref{QM_W-pairing}), real parameter $\Lambda$  and localization function 
\be
L =   \sum_{k=1}^N r^k \psi^k_\Phi.
\ee 
\newline\newline
The pairing $\mathcal{K}_W$ matches with the B-model pairing (\ref{B_htqm_pairing}) for $\Lambda =0$. The pairing $\mathcal{K}_W$  is a pairing between $H^\ast (Q_W-zG_-)$  and $H^\ast (Q_W+zG_-)$. Indeed,  the conjugation formula 
\be
 \begin{split}
\mathcal{K}_W((Q_W-zG_-)\Psi_1, \Psi_2) &= g\left((Q_W-zG_-)\Psi_1, e^{\Lambda \{Q_W+zG_-, L\}}  \Psi_2\right)\\
& =- (-1)^{|\Psi_1|} g\left(\Psi_1, (Q_W+zG_-)e^{\Lambda \{Q_W+zG_-, L\}}  \Psi_2\right) \\
& =- (-1)^{|\Psi_1|} g\left(\Psi_1, e^{\Lambda \{Q_W+zG_-, L\}} (Q_W+zG_-) \Psi_2\right) \\
&=- (-1)^{|\Psi_1|}  \mathcal{K}_W(\Psi_1, (Q_W+zG_-)\Psi_2).
 \end{split}
\ee 
{\bf Lemma}:  On $H^\ast (Q_W-zG_-)\otimes H^\ast (Q_W+zG_-)$ the pairing (\ref{def_higher_par_b_model}) is independent on $\Lambda$. 
\newline\newline
{\bf Proof}: The derivative of the pairing evaluates into 
\be\nn
\begin{split}
\frac{d}{d\Lambda}&\mathcal{K}_W(\Psi_1, \Psi_2) = \frac{d}{d\Lambda} g( \Psi_1, e^{\Lambda \{Q_W+zG_-, L\}} \Psi_2) = g( \Psi_1, \{Q_W+zG_-, L\} e^{\Lambda \{Q_W+zG_-, L\}} \Psi_2)\\
&=g( \Psi_1, (Q_W+zG_-) Le^{\Lambda \{Q_W+zG_-, L\}} \Psi_2) +g( \Psi_1, L(Q_W+zG_-) e^{\Lambda \{Q_W+zG_-, L\}} \Psi_2)  \\
&=- (-1)^{|\Psi_1|}g( (Q_W-zG_-) \Psi_1, Le^{\Lambda \{Q_W+zG_-, L\}} \Psi_2) +g( \Psi_1, L e^{\Lambda \{Q_W+zG_-, L\}} (Q_W+zG_-) \Psi_2)  \\
& = 0.
\end{split}
\ee
In the relation we used $Q_W$-invariance of the B-model pairing (\ref{Q_preserve}),  $G_-$-invariance of the B-model  pairing (\ref{def_G_inv_pairing}) an equality
\be
(Q_W+zG_-) e^{\Lambda \{Q_W+zG_-, L\}} =e^{\Lambda \{Q_W+zG_-, L\}} (Q_W+zG_-). 
\ee
to complete the proof. $\hfill\blacksquare$
\newline\newline
The $\Lambda$-independence means that the pairing $\mathcal{K}_W$ matches with the  B-model pairing $g$ on $H^\ast (Q_W, \ker G_-)$. The B-model pairing $g$ is non-degenerate on $V_B$ and $Q_W$-, $G_-$-invariant, hence $g$ is non-degenerate on   $H^\ast (Q_W, \ker G_-)$, so is the  $\mathcal{K}_W$.
\newline\newline
{\bf Remark}: We can introduce an  expansion for the pairing in  $z$ to define higher pairings 
\be
\mathcal{K}_W(\Psi_1, \Psi_2) = \sum_{k=0}^\infty z^k\; \mathcal{K}_W^{(k)}(\Psi_1, \Psi_2),
\ee
such that the   $\mathcal{K}_W^{(0)}$ is identical to the $g^\Lambda_W$ from the section \ref{sec_pairing}.  
\newline\newline
The  argument of exponent in (\ref{def_higher_par_b_model})  is a sum of two terms:  a function 
\be
\{Q_W, L\} = Q_W(L) =  \sum_{k=1}^N  \psi^k_\Phi \psi^k_R  + 2\pi  i  \sum_{k=1}^N r^k \frac{\p W}{\p Y_k}
\ee
and first-order differential operator 
\be
\{G_-, L\}  = - i  \sum_{k=1}^N r^k \frac{\p }{\p Y_k}.
\ee
We can use  the Zassenhaus formula
\be
e^{A+B} = e^A \;e^B \;e^{-\frac12 [A,B]} \; e^{\frac16 (2[B,[A,B]]+ [A,[A,B]])} \dots
\ee
to express the localization exponent  as product of the $Q_W(L)$-localizaton exponent and  differential operator with coefficients in $z$, i.e.
\be\label{loc_z_loc_no_z}
\begin{split}
e^{ \Lambda \{Q_W+zG_-, L\} }= e^{\Lambda Q_W( L)}  \cdot \mathcal{D} (z, \Lambda, r, Y, \p_Y).
\end{split}
\ee
The representation (\ref{loc_z_loc_no_z}) allows us to conclude that the  higher pairings also localize to $\psi = r = 0$, i.e.  the value of the pairing $\mathcal{K}_W$ is the same for $Q_W$-, $G_-$-closed  states $\Psi, \Psi'$ and their holomorphic germs $\Phi, \Phi'$, i.e. 
\be\label{QM_z_pairing_localiz}
\mathcal{K}_W(\Psi, \Psi') = \mathcal{K}_W (\Phi, \Phi').
\ee
The  localization exponent to first  order in $z$ 
\be
\begin{split}
e^{ \Lambda \{Q_W+zG_-, L\} }= e^{\Lambda Q_W( L)}  \left(1+\pi  z \Lambda^2\; \sum_{k,l=1}^N r^k r^l  \frac{\p^2 W}{\p Y_k \p Y_l} - i z \Lambda \sum_{k=1}^N r^k \frac{\p }{\p Y_k}+ \cO(z^2) \right)
\end{split}
\ee
allows us to evaluate the $\mathcal{K}_W^{(1)}$-pairing 
\be\label{K_1_B-model}
\begin{split}
\mathcal{K}_W^{(1)}(\Psi_1, \Psi_2)& =    \int d\mu \;\; \Psi_1 \cdot  \;e^{\Lambda Q_W( L)}  \left( \pi \Lambda^2\; r^k r^l  \frac{\p^2 W}{\p Y_k \p Y_l} - i  \Lambda r^k \frac{\p }{\p Y_k}  \right)\Psi_2 \\
&= -\pi \sum_{dW = 0}  \;\;   (\p_{Y_k} \p_{Y_{l}} W)^{-1}   \frac{ \left(  \Psi_1 \p_{Y_k}  \p_{Y_l}\Psi_2  - \Psi_2 \p_{Y_k} \p_{Y_l}  \Psi_1 \right)}{\det \p_{Y_k} \p_{Y_{l}} W}   \Big|_{r=\psi=0}.
\end{split} 
\ee
{\bf Remark}: In case of single $Y$-variable  the pairing simplifies into 
\be
\begin{split}
\mathcal{K}_W^{(1)}(\Psi_1, \Psi_2)& =  -2\pi \sum_{W' = 0}  \;\;  \frac12  \frac{ \Psi_1 \Psi_2''  - \Psi_1'' \Psi_2}{(W'')^2}   \Big|_{r=\psi=0}.
\end{split} 
\ee
{\bf Remark}: The  pairings  $\mathcal{K}^{(0)}_W$ and $\mathcal{K}^{(1)}_W$ on holomorphic functions match with the corresponding  higher pairing components   (\ref{LGS_0_pairing})  and (\ref{LGS_1_pairing})  for  the LGS theory.
\newline\newline
 In proposition (\ref{b_model_state_hol_germ})  we showed that  $Q_W$-closed  B-model state  $\Psi$  and its holomorphic germ  represent the  same class in $H^\ast (Q_W, V_B)$.  The pairing $\mathcal{K}_W$ allows us  to make a  stronger statement:
\newline\newline
{\bf Proposition}: The $Q_W$-, $G_-$-closed state $\Psi$ and its holomorphic germ $\Phi$  are in the same  class of $H^\ast (Q_W+zG_-, V_B)$   and 
 there exists  a tropical form ($z$-independent)  $\chi$ such that 
\be
\Psi = \Phi + Q_W \chi,\;\;\; G_-\chi=0.
\ee
{\bf Proof}: The pairing of both representatives  with arbitrary $\Psi'\in H^\ast (Q_W-zG_-)$ are identical i.e.
\be
\mathcal{K}_W(\Psi', \Psi) = \mathcal{K}_W (\Psi', \Phi)\;\;\Longrightarrow \mathcal{K}_W(\Psi', \Psi-\Phi) = 0. 
\ee
The non-degeneracy of the pairing leads to 
\be
\Psi - \Phi   =0 \in H^\ast (Q_W+zG_-),
\ee
what implies the the  existence of the tropical form $\chi$ such that 
\be
\Psi-\Phi  = (Q_W+ zG_-)\chi.
\ee
By construction $\Psi$ and $\Phi$ are independent of $z$ and  we can choose $\chi$, which is $z$-independent, hence $G_-\chi = 0$, what concludes the proof. $\hfill\blacksquare$
\newline\newline
{\bf Example}: The holomorphic germ  for the  $\mathbb{P}^1$-mirror  state (\ref{P_1_mirror_state})  is 
\be
\Phi_\g = \Psi^{\mathbb{P}^1}_\g \Big|_{r=\psi=0} = q_+ e^{iY} \Theta (-r_0) +  q_- e^{-iY}  \Theta(r_0).
\ee
For $r_0<0$ it coincides with the $\Psi_+$-divisor, while for $r_0>0$  it coincides with $\Psi_-$-divisor.
For both cases we showed that the corresponding tropical forms $\chi_\pm$ are $G_-$-closed.

\subsection{B-model deformation by a  holomorphic function}\label{sec_b_model_deform_hol_fun}

In this section we will   describe a  one-parameter family of  deformations  for  the B-type    HTQM $(V_B, Q_W, G_\pm, \mu_2, g)$  by a holomorphic function $\Phi$. 
 The holomorphic function belongs to the B-model space of states, hence we can use  construction from section \ref{sec_htqm_deform} for the leading order deformation 
\be\label{b_model_def_function}
\begin{split}
Q^\e_W  &= Q_W-2\pi [G_-, \mu_2(\e\Phi,\cdot)]  = Q_W +\e\; 2\pi i\frac{\p \Phi}{ \p Y_k} \frac{\p }{\p \psi^k_\Phi}   \\
&= Q +  2\pi i \frac{\p W}{ \p Y_k} \frac{\p }{\p \psi^k_\Phi}   +2\pi i \e\; \frac{\p \Phi}{ \p Y_k} \frac{\p }{\p \psi^k_\Phi}  = Q_{W^\e}.
\end{split}
 \ee
 The last equality allows us to describe the deformed differential  $Q^\e_W$ in the form of    differential in B-model  with deformed  superpotential 
 \be
 W^\e  =  W +\e\; \Phi. 
 \ee
 The  deformation of a B-model  state is defined as (a finite)  expansion in $W$  and A-model propagators $K$
\be
\begin{split}
\Psi^\e & = \Psi  + 2\pi  K_{W} G_- \mu_2 (\e\Phi ,\Psi) +\cO(\e^2) \\
&  = \Psi  + 2\pi  K G_- \mu_2 (\e\Phi ,\Psi)   + 2\pi   K G_- \mu_2 (W, 2\pi KG_- \mu_2 (\e\Phi ,\Psi))... +\cO(\e^2).
  \end{split}
\ee
The deformation of a mirror state   $\Psi = \Psi_\g^W$ simplifies  into
\be\nn
\begin{split}
[ \Psi_\g^W]^\e &=\Psi_\g^W  +2\pi  K_{W} G_- \mu_2 (\e \Phi, \Psi_\g^W ) +\cO(\e^2) \\
&  = \Psi_\g +2\pi  KG_- \mu_2 (W, \Psi_\g)  +2\pi  K G_- \mu_2 (\e\Phi, \Psi_\g )  \\
&\;\;\;\;\;\;\;\;\;+2\pi  K G_- \mu_2 (W, 2\pi KG_- \mu_2 (W ,\Psi_\g)) + 2\pi K G_- \mu_2 (\e \Phi, 2\pi KG_-\mu_2(W,\Psi_\g))  \\
 &\;\;\;\;\;\;\;\;\;+2\pi  K G_- \mu_2 (W, 2\pi KG_- \mu_2 (\e \Phi, \Psi_\g)) +...+\cO(\e^2)\\
& = \Psi_\g  +  2\pi K G_- \mu_2 (W+\e \Phi, \Psi^W_\g) \\
&\;\;\;\;\;\;\;\;\;+2\pi KG_- (W+\e \Phi, 2\pi K G_- \mu_2 (W+\e \Phi, \Psi^W_\g))  +...+\cO(\e^2) \\
& = \Psi_\g^{W_\e}
 \end{split}
\ee
Hence we conclude that the deformation $[ \Psi_\g^W]^\e$  of the mirror state  $\Psi_\g^W$ by a holomorphic function  in B-model  with superpotential $W$   is a   mirror state $ \Psi_\g^{W_\e}$ for the same  $\g$   but in  B-model with deformed superpotential $W^\e$.

\subsection{Higher pairing for mirror states}
The holomorphic germs for  mirror states in B-model can be used to construct a good section in the corresponding  LGS theory. 
\newline\newline
{\bf Definition}: The  {\it tropical  good section} in mirror LGS theory for toric space $X$ with superpotential $W$ is  a linear span of holomorphic germs for mirror states 
\be\label{def_trop_good_sec}
\hbox{Im}\; S^{trop}_W = \mathbb{C}\< \Phi^W_\g\;|\; \g\in H_{dR}^\ast (X) \>.
\ee 
\newline
{\bf Remark}: A similar construction of the good section from holomorphic germs of harmonic states was described in section 7.2 and 7.3 of \cite{Losev:1998dv}.
\newline\newline
The  definition  (\ref{def_good_section}) of a good section in  LGS    requires vanishing of higher pairings.
\newline\newline 
{\bf Proposition (higher pairing for tropical good section)}:  Higher LGS pairings (\ref{Pairing_hz_cohom}) for tropical good section  vanish. Moreover 
\be
{\bf K}_W (\Phi^W_{\g_1}, \Phi^W_{\g_2}) =  \int_X \; \g_1\wedge \g_2,\;\;\;\;  \;\; \forall \g_1, \g_2 \in H_{dR}^\ast (X).
\ee
{\bf Proof}:  By construction in (\ref{B_type_QM_obs_deform}) the mirror states $\Psi_\g^W$  are  $G_-$- and $Q_W$-closed, hence $\Psi_\g^W \in H^\ast (Q_W\pm zG_-)$ and  the  higher B-model pairing (\ref{def_higher_par_b_model})  for such states 
is independent of $\Lambda$ so we can relate it at different values of $\Lambda$. In particular, we use 
\begin{itemize}
 \item  $\Lambda\to\infty$  limit gives us  the LG pairing for holomorphic functions i.e.
\be\label{B_model_to_LG}
\mathcal{K}_W (\Psi^W_{\g_1}, \Psi^W_{\g_2}) = \mathcal{K}_W (\Phi^W_{\g_1}, \Phi^W_{\g_2}) ={\bf K}_W (\Phi^W_{\g_1}, \Phi^W_{\g_2}); 
\ee
\item  $\Lambda=0$ gives us the B-model pairing $g$ i.e.
\be\label{B_model_higher_to_normal}
\mathcal{K}_W (\Psi^W_{\g_1}, \Psi^W_{\g_2}) = g (\Psi^W_{\g_1}, \Psi^W_{\g_2}). 
\ee
\end{itemize}
The mirror states $\Psi^W_{\g_k}$ can be written in the following schematic form 
\be\label{mirror_decom_for_pairing}
\Psi^W_{\g} = \Psi_{\g} + G_- \chi^W_{\g}
\ee 
for the A-model state
\be
\chi^W_{\g} =- 2\pi K \mu_2 (W, \Psi_\g)-2\pi K \mu_2 (W, 2\pi KG_- \mu_2(W,  \Psi_\g))+\dots
\ee
The B-model pairing $g$ on representation (\ref{mirror_decom_for_pairing}) simplifies into
\be\label{A_to_B_mirror_pairing}
\begin{split}
g(\Psi^W_{\g_1}, \Psi^W_{\g_2}) &=g(\Psi_{\g_1}, \Psi_{\g_2}) +  g (\Psi_{\g_1} ,  G_-\chi^W_{\g_2}) + g(   G_-\chi_{\g_1}^W, \Psi_{\g_2} )  +g(G_-\chi_{\g_1}^W, G_-\chi_{\g_2}^W)\\
&=g(\Psi_{\g_1}, \Psi_{\g_2}) +  g (G_-\Psi_{\g_1} ,  \chi^W_{\g_2}) - g(   \chi_{\g_1}^W, G_- \Psi_{\g_2} )  - g(\chi_{\g_1}^W, G^2_-\chi_{\g_2}^W) \\
&=g(\Psi_{\g_1}, \Psi_{\g_2}).
\end{split}
\ee 
We used the $G_\pm$-invariance  of the pairing (\ref{def_G_inv_pairing})  to simplify the expression 

The A-model pairing on evaluation states $\Psi_\g$ is the intersection of the cohomology classes on $X$ i.e.
\be
g(\Psi_{\g_1}, \Psi_{\g_2}) = \int_X \; \g_1\wedge \g_2. 
\ee
The proof of the proposition is the following: We use   (\ref{B_model_to_LG}) to turn LGS higher pairing into the B-model higher pairing on mirror states. We use  (\ref{B_model_higher_to_normal}) to turn the B-model higher pairing into ordinary pairing 
which according to (\ref{A_to_B_mirror_pairing})  simplifies to the intersection of cohomology classes.  The higher parings are  $z$-expansion of ${\bf K}_W $ hence 
\be
 {\bf K}^{(k)}_W (\Phi_{\g_1}^W, \Phi^W_{\g_2}) =0,\;\; \forall\; k>0.
\ee
what completes the proof of the proposition.
$\hfill\blacksquare$
\newline\newline
{\bf Example}: The toric description of $\mathbb{P}^2$  consists of three compactifying divisors  with primitive normal vectors:  $(1,0)$, $(0,1)$ and $(-1,-1)$ so the mirror superpotential 
\be
W_{\mathbb{P}^2} = e^{iY_1} + e^{i Y_2} +  q e^{-i Y_1-i Y_2}
\ee
The de Rham cohomology of $\mathbb{P}^2$ are one-dimensional in degrees 0, 2, and 4.  The image of tropical good section is 
\be
\hbox{Im}\; S^{trop}_{\mathbb{P}^2} = \mathbb{C}\<1, q e^{-iY_1-iY_2}, e^{iY_1+iY_2}\>.
\ee

\section{Correlation functions for mirror states}\label{sec_corr_funct}

The correlation function invariance theorem from section  \ref{corr_inv} tells us that 
\be
\< \Psi_{1},...,\Psi_{n}, Q_X \chi\>_{Q_X} =0
\ee
for $Q_X$- $G_-$-closed states $\Psi_a$  and a $G_-$-closed  state $\chi$.  In previous section we showed that there exists a tropical form $\chi$ such that it can turn a mirror state into a holomorphic germ. Moreover, in section \ref{sec_b_model_deform_hol_fun} we saw  that the deformation of   B-model by a holomorphic function leaves it in the same class but with different superpotential.  Hence, our strategy for evaluating  correlation functions of mirror states  in B-model will be the following:
\begin{enumerate}
\item Replace  one mirror state by the corresponding  holomorphic germ.
\item Deform the superpotential and mirror states  by the holomorphic germ.
\item Apply the recursion formula for correlation functions to reduce the number of arguments by one.
\item Repeat steps 1-3 till the number of arguments reaches 3.
\end{enumerate} 

The simplification of the theory deformation while using the holomorphic germ rather than the mirror state itself suggest an alternative  strategy for the correlation function evaluation:  We replace all mirror states by the corresponding holomorphic germs and evaluate the correlation function.  The application of this strategy to the 4pt functions immediately gives zero answer. The reason for that is our definition of the B-model amplitudes via an expansion in  A-model   amplitudes. In A-model the propagator acts trivially on the holomorphic functions what  is the  source of zero answers.  

To resolve  the puzzle  we recall that the A-model amplitudes for tropical GW invariants have a particular number of divisor states, determined by the degree. The mirror states can provide additional divisor states, but if we already have too many (from holomorphic germs) the amplitude vanishes due to the degree selection for the tropical GW invariants. Since, all non-trivial GW invariants have degree bigger than zero, the corresponding A-model amplitudes have at least one divisor state, hence we can always replace a single mirror state by the holomorphic germ.   Later we will see that a single replacement  is just enough to prove   our main theorem. 

\subsection{4-point function invariance and holomorphic representatives}
Using the symmetries of the amplitudes we can rewrite the 4pt function in the form 
\be
\begin{split}
2 &\<\Psi_1, \Psi_2, \Psi_3, \Psi_4\>_{Q_W} =  g (\Psi_4 , \sum_{\sigma \in S_3} \mu_2( \Psi_{\sigma(1)},  K_WG_-\mu_2(\Psi_{\sigma(2)},\Psi_{\sigma(3)} ))) \\
 &= g (\Psi_4 , \Psi_{123})=g (\Phi_4 +Q_W \chi_+, \Psi_{123}) =  g (\Phi_4 , \Psi_{123}) =2 \<\Psi_1, \Psi_2, \Psi_3, \Phi_4\>_{Q_W} 
 \end{split}
\ee
The equality on the second line   requires
\be
g (Q_W \chi_+, \Psi_{123}) =0.
\ee
In (\ref{cp_1_dress_+})  we showed that  $\chi$ for $\mathbb{P}^1$-model does not belong to  tropical forms on $\mathbb{P}^1$, hence  the difference
\be
g (Q_W \chi_+, \Psi_{123}) -g ( \chi_+,  Q_W\Psi_{123})
\ee
might not be  zero. The difference is controlled by the boundary term of the  $r$-integration 
\be
\begin{split}
g &(Q_W \chi_+, \Psi_{123}) -g ( \chi_+,  Q_W\Psi_{123}) = \int d\mu\;  Q (\chi_+ \Psi_{123})   = \int d\mu\; d_R (\chi_+ \Psi_{123})  \\
&= \lim_{r\to -\infty} \int_{S^1} dY\;  \Theta (r_4-r)\ \Psi_{123}\Big|_{\psi=0}  = \lim_{r\to -\infty} \int_{S^1} dY\;  \Psi_{123}\Big|_{\psi=0}. 
\end{split}
\ee
We can   repeat the analysis for the case  different holomorphic representative   $\Phi_4'  = q e^{-iY}$  and $\chi_-$
\be
\begin{split}
g &(Q_W \chi_-, \Psi_{123}) -g ( \chi_-,  Q_W\Psi_{123}) = \lim_{r\to +\infty} \int_{S^1} dY\;  \Psi_{123}\Big|_{\psi=0}. 
\end{split}
\ee
Both  boundary terms vanish if  $Y$-independent part of $\Psi_{123}\Big|_{\psi=0}$  has finite support. We  can evaluate the $\Psi_{123}$ for $\mathbb{P}^1$-mirror states 
\be\nn
\begin{split}
 \int_{S^1} &dY\;\Psi_{123}\Big|_{\psi=0} = \frac12  \sum_{\sigma \in S_3 } \int_{S^1} dY\;\mu_2 (\Psi_{\sigma(3)}, 2\pi K_W G_- \mu_2 (\Psi_{\sigma(1)}, \Psi_{\sigma(2)}) )\Big|_{\psi=0}\\
& =  \pi q  \sum_{\sigma \in S_3 }  \left[ \Theta (\min(r_{\sigma(1)}, r_{\sigma(2)})-r) ) \Theta (r-r_{\sigma(3)})  + \Theta (r-\max(r_{\sigma(1)},r_{\sigma(2)}) )  \Theta(r_{\sigma(3)}-r)\right]
\end{split} 
\ee
and observe that the products of $\Theta$-functions indeed have the finite support.

 We can perform the same analysis for $\<\Psi_1, \Psi_2, \Phi_3, \Psi_4\>_{Q_W}$, where we  replaced the mirror state  $\Psi_3$ by the holomorphic  function $\Phi_3 = e^{iY}$.  The boundary term is controlled by the  function
\be
\begin{split}
 \int_{S^1} & dY\;\tilde{\Psi}_{123}\Big|_{\psi=0}  =   \int_{S^1} dY\;\mu_2 (\Phi_3, 2\pi K_W G_- \mu_2 (\Psi_1, \Psi_2) )\Big|_{\psi=0} \\
 &+ \int_{S^1} dY\; \mu_2 (\Psi_2, 2\pi K_W G_- \mu_2 (\Psi_1, \Phi_3) )\Big|_{\psi=0}  + \int_{S^1} dY\; \mu_2 (\Psi_1, 2\pi K_W G_- \mu_2 (\Psi_2, \Phi_3) )\Big|_{\psi=0}  \\
& =2\pi q \Theta (\min(r_1,r_2)-r) +2\pi q  \Theta(r_2-r)  \Theta (r-r_1) +2\pi q  \Theta(r_1-r) \Theta (r-r_2).
\end{split}
\ee
which    does not have the finite support and  the limit is finite and non-zero 
\be
\lim_{r\to -\infty} \int_{S^1} dY\;  \tilde{\Psi}_{123}\Big|_{\psi=0}  = 2\pi q \neq 0.
\ee
Hence  we conclude that  the simultaneous  replacement of two $\mathbb{P}^1$-mirror states $\Psi_3$ and $\Psi_4$ by the same holomorphic germs $q e^{-iY}$ does not preserve the invariance of the correlation function, i.e.
\be
\<\Psi_1, \Psi_2, \Phi_3, \Psi_4\>_{Q_W} - \<\Psi_1, \Psi_2, \Phi_3, \Phi_4\>_{Q_W} \neq 0.
\ee
{\bf Remark}: If we replace $\Psi_4$ by the the different  holomorphic  function $\Phi_4 =q e^{-iY}$  then the boundary term is controlled by the different limit 
\be
\lim_{r\to +\infty} \int_{S^1} dY\;  \tilde{\Psi}_{123}\Big|_{\psi=0}  =0.
\ee
Hence we conclude that  for the choice of holomorphic functions $\Phi_3 =  e^{iY}$ and $\Phi_4 =  q e^{-iY} $ preserves the 4-point function, i.e.
\be\label{corr_fun_p_1_hol_rep_repl}
\<\Psi_1, \Psi_2, \Psi_3, \Psi_4\>_{Q_W}  =  \<\Psi_1, \Psi_2, \Phi_3, \Phi_4\>_{Q_W} =  \<\Psi_1, \Psi_2, e^{iY},q  e^{-iY}\>_{Q_W}
\ee
Note that the equality (\ref{corr_fun_p_1_hol_rep_repl}) agrees with our heuristic analysis for the possible  $\mathbb{P}^1$-mirror states replacement, based on the A-model divisor counting. The  GW invariants for $\mathbb{P}^1$ are non-trivial only  for degree-1 maps,  hence  the A-model correlation functions have exactly  two divisor states $\Psi_+ =  e^{iY}$ and $\Psi_- =  e^{-iY}$.

\subsection{n-point function invariance and holomorphic representative}\label{sec_inv_conjecture}

{\bf Conjecture}:  For the mirror states $\Psi_1,..,\Psi_n$ for general toric  $X$  the state $\Psi_{1..n}$, evaluated on the sum (\ref{tree_state}) over   rooted trees with leaves $\Psi_1,...,\Psi_{n}$  and $\chi$, such that  
\be
Q_W \chi = \Psi_{n+1} - \Psi_{n+1}\Big|_{\psi=r=0},\;\;\; G_-\chi =0
\ee
satisfy
\be\label{conj}
g(\chi, d_R\Psi_{1..n}) - g(d_R\chi,\Psi_{1..n}) = 0,
\ee
i.e. the boundary term vanishes for all boundary divisors of $X$.
\newline\newline
 {\bf Proposition}:  The conjecture holds for $X = \mathbb{P}^1$.
 \newline\newline 
  {\bf Proof}:  For the $(n+1)$-point correlation function   the  boundary term  is 
 \be
g(\chi_\mp, d_R\Psi_{1..n}) - g(d_R\chi_\mp,\Psi_{1..n}) = \lim_{r\to \pm \infty}  \int_{S^1} dY\;\Psi_{1..n}\Big|_{\psi=0}
  \ee
In particular, if $Y$-independent part of  $\Psi_{1..n}\Big|_{\psi=0}$ has finite support for any  $\mathbb{P}^1$-mirror states  $\Psi_1,...,\Psi_{n}$ then the boundary term vanishes.
\newline\newline 
For any pair of $\mathbb{P}^1$ B-model states  $\Psi_1$ and $\Psi_2$ we can simplify
\be
K_{\mathbb{P}^1} G_- \mu_2 (\Psi_1, \Psi_2)= K G_- \mu_2 (\Psi_1, \Psi_2).
\ee
This equality follows from the degree counting. The B-model states are at most $(1,1)$-forms in $\psi$, so is the product. The $KG_-$ lowers degree of the state by $(1,1)$, so there are no higher order terms in expansion of the B-model 
propagator  $K_{\mathbb{P}^1} G_- $ in A-model propagators $KG_-$.

We can simplify the $KG_-$  for product of two $\mathbb{P}^1$-mirror states    into
\be\nn
\begin{split}
\Psi_{ab}'  &=2\pi  K G_- \mu_2 (\Psi^{\mathbb{P}^1}_a, \Psi^{\mathbb{P}^1}_b)   = e^{iY} \Theta (r-\max(r_a,r_b) ) + q e^{-iY} \Theta (\min(r_a, r_b)-r) 
\end{split}
\ee
and observe that it is degree-0  tropical form, hence  the product of two such expressions is also a  zero form. The $KG_-$-action on any zero form is zero, i.e. 
\be
\Psi'_{abcd} = K  G_- \mu_2 (\Psi'_{ab}, \Psi'_{cd}) = 0.
\ee 
This equality allows us to simplify the sum over rooted trees to 
\be
\Psi_{1..n} = \frac14 \sum_{k=1}^{n-1}  \mu_2 (\Psi'_{(1..k},  \Psi'_{k+1..n)}) ,\;\; \Psi'_1 = \Psi_1^{\mathbb{P}^1}
\ee
using the $\Psi_{1..n}'$-expressions defined recursively  
\begin{itemize}
\item $n=2$
\be\nn
\begin{split}
\Psi_{12}'  & =2\pi KG_-\mu_2( \Psi^{\mathbb{P}^1}_{1}, \Psi^{\mathbb{P}^1}_2) =  e^{iY} \Theta (r-\max(r_1,r_2) ) + q e^{-iY} \Theta (\min(r_1, r_2)-r); 
\end{split}
\ee
\item   $n>2$ 
\be
\Psi_{1..n}' =2\pi KG_-\mu_2 (\Psi^{\mathbb{P}^1}_{(n},  \Psi'_{1..n-1)}).
\ee
\end{itemize}
We can prove by induction that 
\be
\Psi'_{1..n}  =  e^{iY}  \Theta (r- \max(r_1,.., r_n) )+ q e^{-iY} \Theta (\min(r_1,.., r_n) -r)
\ee
and evaluate  
\be
\begin{split}
\Psi_{1..n}\Big|_{\psi=0}  &=  \frac{1}{4\cdot n!} \sum_{\sigma \in S_n} \sum_{k=1}^{n-1}  (e^{iY}  \Theta (r- \max(r_{\sigma(1)},.., r_{\sigma(k)}) )+ q e^{-iY} \Theta (\min(r_{\sigma(1)},.., r_{\sigma(k)})-r )) \\
 &\qquad \times  (e^{iY}  \Theta (r- \max(r_{\sigma(k+1)},.., r_{\sigma(n)}) )+ q e^{-iY} \Theta (\min(r_{\sigma(k+1)},.., r_{\sigma(n)})-r )).
 \end{split}
\ee
Each $Y$-independent  term  in the sum has finite support so is the whole sum, what completes the proof of a proposition.$\hfill\blacksquare$

\subsection{3-point functions}\label{sec_3pt_b_model}

{\bf Proposition (3-point localization formula)}: For  $Q_W$-closed  B-model  states $\Psi_1, \Psi_2, \Psi_3$ the  3-point function   equals to the LGS 3-point  function for corresponding holomorphic germs and superpotential $W$, i.e.
\be
  \begin{split}
 \< \Psi_1,\Psi_2,\Psi_3\>_{Q_W} &  =  \< \Phi_{1},\Phi_{2},\Phi_{3}\>_W.
\end{split}
\ee
{\bf Proof}:   The 3pt function   in B-model is 
\be
\< \Psi_{1},\Psi_{2},\Psi_{3}\>_{Q_W} = g(\Psi_1, \mu_2(\Psi_2, \Psi_3)).
\ee
The product of two $Q_W$-closed states is $Q_W$-closed, i.e.
\be
Q_W \mu_2(\Psi_2, \Psi_3) = \mu_2(Q_W \Psi_2, \Psi_3)+\mu_2(\Psi_2,  Q_W\Psi_3) =0,
\ee
hence we can replace B-model pairing $g$ by the pairing $g^\Lambda_W$
\be
\< \Psi_{1},\Psi_{2},\Psi_{3}\>_{Q_W} = g^\Lambda_W(\Psi_1, \mu_2(\Psi_2, \Psi_3))
\ee
and use the localization  
\be
\begin{split}
\< \Psi_{1},\Psi_{2},\Psi_{3}\>_{Q_W} & = \sum_{dW=0} \frac{\Psi_{1}\Psi_{2}\Psi_{3}}{\det \p_i \p_jW}\Big|_{r=\psi_\Phi = \psi_R = 0}  =  \sum_{dW=0} \frac{\Phi_{1}\Phi_{2}\Phi_{3}}{\det \p_i \p_jW} = \< \Phi_{1},\Phi_{2},\Phi_{3}\>_{W}.
\end{split}
\ee
Hence we completed the proof of the proposition. $\hfill\blacksquare$
\newline\newline
{\bf Example}: For $X = \mathbb{P}^1$  the mirror states for  cohomology representatives   $1, \g_P \in H^\ast(Q_{\mathbb{P}^1})$  have the holomorphic germs $1, e^{iY}$:
\begin{itemize}
\item {\bf  three 0-forms}: The B-model 3-point function  vanishes, due to the insufficient degree of the form
\be
\<\Psi^{\mathbb{P}^1}_1, \Psi^{\mathbb{P}^1}_1, \Psi^{\mathbb{P}^1}_1 \>_{Q_{\mathbb{P}^1}} =\int d\mu\; 1  =  0 = \< 1,1,1\>_{W_{\mathbb{P}^1}};
\ee
\item{\bf single 1-form}: The B-model  3-point function is the integral of the $\g_P$ over $\mathbb{P}^1$
\be
\<\Psi^{\mathbb{P}^1}_1, \Psi^{\mathbb{P}^1}_1, \Psi^{\mathbb{P}^1}_{P} \>_{Q_{\mathbb{P}^1}}  =\int d\mu\;  \Psi^{\mathbb{P}^1}_{P} =  \int_{\mathbb{P}^1} \g_P=  1 = \< 1,1, e^{iY}\>_{W_{\mathbb{P}^1} };
\ee
\item{\bf two 1-forms}: The B-model 3-point function vanishes, due to degree selection
\be
\<\Psi^{\mathbb{P}^1}_1, \Psi^{\mathbb{P}^1}_{P}, \Psi^{\mathbb{P}^1}_{P} \>_{Q_{\mathbb{P}^1}} = 0 = \< 1, e^{iY} , e^{iY}\>_{W_{\mathbb{P}^1}};
\ee
\item {\bf three 1-forms}: For three 1-forms, Poincare dual to the three distinct points $r_1, r_2, r_3$  the B-model 3-point function is 
\be
  \begin{split}
\<\Psi^{\mathbb{P}^1}_{P}, \Psi^{\mathbb{P}^1}_{P}, \Psi^{\mathbb{P}^1}_{P} \>_{Q_{\mathbb{P}^1}} &= \int d\mu\; \Psi^{\mathbb{P}^1}_{P}\Psi^{\mathbb{P}^1}_{P} \Psi^{\mathbb{P}^1}_{P}    =q \sum_{\sigma\in S_3} \Theta (r_{\sigma(1)},r_{\sigma(2)},r_{\sigma(3)})\\
&=q=  \< e^{iY} , e^{iY} , e^{iY} \>_{W_{\mathbb{P}^1}}.
\end{split}
\ee 
\end{itemize}

\subsection{4-point functions}\label{sec_4pt_mirr_LG}

{\bf Proposition (4-point localization formula)}: If conjecture \ref{sec_inv_conjecture} holds then  the 4-point correlation function of mirror states  $\Psi^W_a$   equals to the  4-point correlation function of the corresponding holomorphic germs  in LGS  theory 
with superpotential $W$ and tropical good section (\ref{def_trop_good_sec}),  i.e.
\be
  \begin{split}
 \< \Psi^W_1,\Psi^W_2,\Psi^W_3, \Psi^W_4\>_{Q_W} &  =  \< \Phi_{1},\Phi_{2},\Phi_{3}, \Phi_4\>^{S^{trop}}_W.
\end{split}
\ee
\newline
{\bf Proof}: The invariance theorem from section \ref{corr_inv}, extended by a conjecture  \ref{sec_inv_conjecture} allows us   to replace the mirror state $\Psi_4^W$ by the holomorphic germ $\Phi_4$ in the   4-point function of mirror states 
\be
\begin{split}
\< \Psi^W_{1},\Psi^W_2,\Psi^W_{3}, \Psi^W_4\>_{Q_W}  &=\< \Psi^W_{1},\Psi^W_2,\Psi^W_{3}, \Phi_4\>_{Q_W}.  
\end{split}
\ee
The holomorphic germ $\Phi_4$ is $Q_W$- and $G_-$-closed, hence we can use the recursion relation theorem from section  \ref{corr_recurs}
\be
\< \Psi^W_{1},\Psi^W_2,\Psi^W_{3}, \Phi_4\>_{Q_W} =  \frac{d}{d\e}\Big|_{\e=0}  \< \Psi_{1}^\e,\Psi_2^\e,\Psi_{3}^\e\>_{Q_{W^\e}  }   
\ee
for   deformed superpotential 
\be
W^\e = W +\e \Phi_4.
\ee
The  deformed states 
\be
\Psi_k^\e   = \Psi^W_k  +\e\; 2\pi K_{W} G_- \mu_2 (\Psi^W_k , \Phi_{4})
\ee
are $Q_{W^\e}$-closed hence the  3-point function  can be evaluated in terms of LGS theory 
\be
\< \Psi_{1}^\e,\Psi_2^\e,\Psi_{3}^\e\>_{Q_{W^\e}  }     =  \sum_{dW^\e =0} \frac{\Phi^\e_1\cdot\Phi^\e_2 \cdot\Phi^\e_2  }{\det \p_j \p_k W^\e}\Big|_{Y=Y_0} = \< \Phi_1^\e, \Phi_2^\e, \Phi_3^\e\>_{W^\e}
\ee
for holomorphic germs of deformed states 
\be\label{3pt_contact_terms}
\Phi^\e_k = \Psi^\e_k\Big|_{\psi=r=0}  = \Phi_k  +2\pi \e \;K_{W} G_- \mu_2 (\Psi^W_k , \Phi_{4})\Big|_{\psi=r=0}  = \Phi_k +  \e \; C^{trop}_W (\Phi_k, \Phi_4).
\ee
The equality (\ref{3pt_contact_terms}) defines  tropical  contact term $C^{trop}_W (\Phi_k, \Phi_4)$ and later we will show that these contact terms match with LGS  contact term   (\ref{def_conn_good_section})  for tropical good section (\ref{def_trop_good_sec}).

We can substitute the $\Phi^\e_k$ in terms of $\Phi_k$  and contact term  into the 3-point function derivative
\be
\begin{split}
\< \Psi^W_{1},\Psi^W_2,\Psi^W_{3}, \Psi^W_4\>_{Q_W}  & =  \frac{d}{d\e}\Big|_{\e=0} \< \Phi_1^\e, \Phi_2^\e, \Phi_3^\e\>_{W^\e} \\
& = \frac{d}{d\e}\Big|_{\e=0} \< \Phi_1, \Phi_2, \Phi_3\>_{W^\e} + \< C^{trop}_W (\Phi_1, \Phi_4), \Phi_2, \Phi_3\>_{W}\\
&\qquad+ \< \Phi_1, C^{trop}_W( \Phi_2, \Phi_4), \Phi_3\>_{W}+ \< \Phi_1, \Phi_2, C^{trop}_W( \Phi_3, \Phi_4)\>_{W} \\
& = \< \Phi_{1},\Phi_2,\Phi_{3}, \Phi_4\>^{S^{trop}}_{W} 
\end{split}
\ee
The last equality is the  the LGS recursion formula (\ref{n_pt_rec})  for the 4pt functions with contact terms defined  by (\ref{3pt_contact_terms}).
$\hfill\blacksquare$

\subsection{Contact terms}

{\bf Definition}: For a superpotential $W$,  holomorphic function $\Phi_2 $  and   mirror state  $\Psi^W_1$ with holomorphic germ  $ \Phi_1$
 we define  tropical contact term 
 \be
 C^{trop}_W \left( \Phi_1, \Phi_2\right) = \frac{d}{d\e}\Big|_{\e=0} \; \Psi_1^\e\Big|_{\psi=r=0} =  2\pi K_{W} G_- \mu_2 (\Psi^W_1 ,\Phi_2) \Big|_{\psi=r=0}. 
 \ee
 \newline
 {\bf Proposition (contact terms for tropical good section)}:\label{prop_trop_section} The tropical contact term equals (as classes in $H^\ast ({\bf Q}_W+z {\bf G}_-)$) to the LGS contact term for the tropical good section, 
i.e. 
   \be\label{trop_cont_terms_equality}
   \begin{split}
 C^{trop}_W (\Phi_1, \Phi_2)&  =  C^{S^{trop}}_W (\Phi_1, \Phi_2).  
 \end{split}
 \ee
{\bf Proof}:  The product  $\mu_2 (\Psi^W_1, \Phi_2) $ is a $Q_W$-closed state due to 
\be
Q_W \mu_2 (\Psi^W_1, \Phi_2) = \mu_2 (Q_W \Psi^W_1, \Phi_2)+\mu_2 (\Psi^W_1, Q_W\Phi_2) =0.
\ee
Hence it represents some class in $H^\ast (Q_W)$. Moreover,  the same class can be expressed via 
\be
\mu_2 (\Psi^W_1, \Phi_2) = \mu_2 (\Phi_1, \Phi_2)  = \Phi_1\Phi_2 \in H^\ast (Q_W).
\ee
Using the  map $\pi_W: R_{\mathbb{C}^{\ast N}} \to J_W$ we can write the class for the product of two holomorphic functions $\Phi_1\Phi_2$ in the form $\pi_W(\Phi_1\Phi_2)$.
Applying  an isomorphism   $J^{-1}:J_W= H^\ast (Q_W) \to H^\ast_{dR}(X): \;\Psi \mapsto J^{-1}(\Psi)$ from section \ref{sec_mirror_rel_corr} to the class  $\pi_W(\Phi_1\Phi_2) $    we can construct class $J^{-1}\pi_W(\Phi_1\Phi_2)$ in $H^\ast_{dR}(X)$. 
Let us choose  representatives $\g $ for each class of cohomology $H_{dR}^\ast(X)$, so the class $J^{-1}\pi_W(\Phi_1\Phi_2)$ is represented by a  tropical form $\g J^{-1}\pi_W(\Phi_1\Phi_2)$.
This tropical form  defines a  mirror state  $\Psi^W_{\g J^{-1} \pi_W(\Phi_1\Phi_2)}$, which    is  $G_-$-closed  and represents the same class  to $ \mu_2 (\Psi^W_1, \Phi_2)$ in $H^\ast (Q_W)$.  Hence there exists a tropical form $\chi$ such that  
 \be
 \mu_2 (\Psi^W_1, \Phi_2)  - \Psi^W_{\g J^{-1}\pi_W(\Phi_1 \Phi_2)}= Q_W\chi.
 \ee
We can use $\chi$ to evaluate 
 \be\label{k_g_trop}
 \begin{split}
 K_{W} G_- \mu_2 (\Psi^W_1, \Phi_2)& =  K_{W}G_- \Psi^W_{\g J^{-1}\pi_W(\Phi_1 \Phi_2)} +  K_{W} G_-   Q_W \chi \\
&=   - G_- \chi    + (Q_W+zG_- ) K_W G_-  \chi .
\end{split}
 \ee
 The tropical contact term is the holomorphic germ of (\ref{k_g_trop})
   \be
   \begin{split}
 C^{trop}_{W} (\Phi_1, \Phi_2) &=2\pi \left[ K_{W} G_- \mu_2 (\Psi^W_1 , \Phi_{2})  \right]\Big|_{\psi=r=0}  \\
 &  =  -2\pi G_- \chi\Big|_{\psi=r=0}  + ({\bf Q}_W+z{\bf G}_-) (...).
 \end{split}
 \ee
We can represent the $Q_W$ as a sum of two graded-commuting differentials 
\be
Q_W = 2\pi i {\bf Q}_W + d_R
\ee
to  write a solution to 
\be
Q_W \chi   =(2\pi i {\bf Q}_W + d_R) \chi = \Psi
\ee
using the homotopy ${\bf \Sigma}_W$ for ${\bf Q}_W$
\be
\chi  =\frac{1}{2\pi i} {\bf  \Sigma}_W \Psi - \frac{1}{(2\pi i)^2} {\bf \Sigma}_W d_R {\bf \Sigma}_W \Psi+ \dots
 \ee
The radial de Rham $d_R$ adds powers in $\psi_R$,  hence  the holomorphic germs for  higher terms in the sum vanish, i.e.
 \be
 G_-{\bf  \Sigma}_W d_R {\bf \Sigma}_W\left( \mu_2 (\Psi^W_1, \Phi_2)  - \Psi^W_{\g J^{-1}\pi_W(\Phi_1\Phi_2)} \right) \Big|_{\psi_R=0} = 0.
 \ee
The tropical  contact term  simplifies  to
 \be
    \begin{split}
  C^{trop}_{W} (\Phi_1, \Phi_2) & = -2\pi G_- \chi\Big|_{\psi=r=0}  = i  G_- {\bf \Sigma}_W   \left( \mu_2 (\Psi^W_1, \Phi_2)  - \Psi^W_{\g J^{-1} \pi_W (\Phi_1\Phi_2)} \right)  \Big|_{\psi=r=0}
 \end{split}
 \ee
 Let us recall the relation $ i G_- = {\bf G}_-$ between the B-model $G_-$ and LGS ${\bf G}_-$. 
Both  $G_-$ and  $ {\bf Q}_W$  act trivially on  the radial variables $r, \psi_R$, hence  we can simplify
 \be
     \begin{split}
 iG_- &   {\bf \Sigma}_W \left( \mu_2 (\Psi^W_1, \Phi_2)  - \Psi^W_{\g J^{-1} \pi_W(\Phi_1\Phi_2)} \right)  \Big|_{\psi=r=0}  \\
 & = {\bf G}_-  {\bf \Sigma}_W \left[ \left( \mu_2 (\Psi^W_1, \Phi_2)  - \Psi^W_{\g J^{-1} \pi_W(\Phi_1\Phi_2)} \right)  \Big|_{\psi_R=r=0}  \right] \Big|_{\psi_\Phi=0}.
     \end{split}
 \ee
The  mirror states are Hodge type  tropical forms,  i.e.  contain the same number of $\psi_\Phi$ and $\psi_R$ for each degree, hence we can further simplify
 \be
   \begin{split}
\left( \mu_2 (\Psi^W_1, \Phi_2)  - \Psi^W_{\g J^{-1} \pi_W(\Phi_1\Phi_2)} \right)  \Big|_{\psi_R=r=0}  &= \left( \mu_2 (\Psi^W_1, \Phi_2)  - \Psi^W_{\g J^{-1} \pi_W(\Phi_1\Phi_2)} \right) \Big|_{\psi_R=r=\psi_\Phi =0} \\
& = \mu_2 (\Phi_1, \Phi_2)  - \Phi^W_{\g J^{-1} \pi_W(\Phi_1\Phi_2)}.
  \end{split}
 \ee
For any function $\Phi$ we construct another function, defined via
\be
S_W \pi_W(\Phi) = \Phi^W_{\g J^{-1} \pi_W(\Phi)}
\ee
to express the tropical contact term in the form 
 \be
    \begin{split}
   C^{trop}_{W} (\Phi_1, \Phi_2) & ={\bf G}_-  {\bf \Sigma}_W \left( \mu_2 (\Phi_1, \Phi_2)  - \Phi^W_{\g J^{-1} \pi_W(\Phi_1\Phi_2)} \right) \\
   &={\bf G}_-  {\bf \Sigma}_W \left( \mu_2 (\Phi_1, \Phi_2)  - S_W \pi_W(\Phi_1\Phi_2) \right)
     \end{split}
 \ee
Earlier in a section \ref{good_sect} we saw that, in particular,   $S_W$  defines a section $H^\ast ({\bf Q}_W) \to H^\ast ({\bf Q}_W+z {\bf G}_-)$ hence we can drop exact terms in (\ref{k_g_trop}).   $\hfill\blacksquare$
\newline\newline
 {\bf Example}: Let  $X = \mathbb{P}^1$, $\Psi^W_1$ is a  mirror state for the point observable at $r=r_1$
\be
\Psi^W_1 =\Psi_P^{\mathbb{P}^1}=    \delta (r-r_1)\psi_\Phi\psi_R +  e^{iY} \Theta (r-r_1) +  q e^{-iY}  \Theta(r_1-r)
\ee
 and  holomorphic function
\be
\Phi_2  = q e^{-iY}.
\ee 
The  tropical contact term is the holomorphic germ of 
\be
\begin{split}
K_{W_{\mathbb{P}^1}} G_- \mu_2 (\Psi^W_1, \Phi_2)  &  = K G_- \mu_2 (\Psi^W_1, \Phi_2)  +  K G_- \mu_2(W_{\mathbb{P}^1}, KG_- \mu_2 (\Psi^W_1, \Phi_2)   )+...  \\
&=K G_- \mu_2 (\Psi_1, \Phi_2)= KG_- (q e^{-iY}  \delta (r-r_1)\psi_\Phi\psi_R  )\\
&= \int^\infty_0 dt\; e^{-tH} G_+ G_- (q e^{-iY} \delta (r-r_1)\psi_\Phi\psi_R) \\
& =q e^{-iY}   \int^\infty_0 dt\;  \delta (r-r_1+t)    = q e^{-iY} \Theta (r_1-r),  
\end{split}
\ee
while the germ evaluation gives us 
\be
 C^{trop}_{W_{\mathbb{P}^1}} (\Phi_1, \Phi_2)  = K_{W_{\mathbb{P}^1}} G_- \mu_2 (\Psi^W_1, \Phi_2)   \Big|_{\psi=r=0} = q e^{-iY} \Theta (r_1).  
\ee
The holomorphic representative of the mirror state 
\be
\Phi_1  = \Psi^W_1\Big|_{\psi=r=0} =  e^{iY} \Theta (-r_1) +  q e^{-iY}  \Theta(r_1)
\ee
gives us  contact terms of the form 
\be
\begin{split}
C^{trop}_{W_{\mathbb{P}^1}} ( q e^{-iY} ,  q e^{-iY} ) &=  q e^{-iY},  \\
C^{trop}_{W_{\mathbb{P}^1}} (  e^{iY} ,  q e^{-iY} ) &=  0.
\end{split}
\ee
The tropical contact terms  matches with the $\mathbb{P}^1$-mirror  LGS contact terms  for tropical good section for $\mathbb{P}^1$ 
\be
\hbox{Im}\; S^{trop} = \mathbb{C}\<1, e^{iY}\>.
\ee

\subsection{5-point function}
{\bf Proposition (5-point localization formula)}: If conjecture \ref{sec_inv_conjecture} holds then  the 5-point correlation function of mirror states  $\Psi^W_a$   equals to the  5-point correlation function of the corresponding holomorphic germs  in LGS  theory 
with superpotential $W$ and tropical good section (\ref{def_trop_good_sec}),  i.e.
\be
  \begin{split}
 \< \Psi^W_1,\Psi^W_2,\Psi^W_3, \Psi^W_4, \Psi^W_5\>_{Q_W} &  =  \< \Phi_{1},\Phi_{2},\Phi_{3}, \Phi_4, \Phi_5\>^{S^{trop}}_W.
\end{split}
\ee
{\bf Proof}: We can rewrite the  5-point function in B-model using the recursion formula for the B-model deformation by a holomorphic germ of the mirror state $\Psi_5^W$
\be\nn
\begin{split}
 \< \Psi^W_1,\Psi^W_2,\Psi^W_3, \Psi^W_4, \Psi^W_5\>_{Q_W}  &= \< \Psi^W_1,\Psi^W_2,\Psi^W_3, \Psi^W_4, \Phi_5\>_{Q_W} =  \frac{d}{d\e}\Big|_{\e=0}  \< \Psi_{1}^\e,\Psi_2^\e,\Psi_{3}^\e, \Psi_4^\e\>_{Q_{W^\e}  }.   
\end{split}
\ee
The deformed superpotential is
\be
W^\e = W +\e \Phi_5
\ee
and  deformed states  are
\be
\Psi_k^\e   = \Psi^W_k  +\e K_{W} G_- \mu_2 (\Psi^W_k , \Phi_{5}) = \Psi^{W^\e}_k,
\ee
 while  the  holomorphic  germs are  
\be
\Phi_k^\e = \Psi_k^\e \Big|_{\psi=r=0}  =\Phi_k  +2\pi \e \;K_{W} G_- \mu_2 (\Psi^W_k , \Phi_{5})\Big|_{\psi=r=0}  = \Phi_k +  \e \; C^{trop}_{W} (\Phi_k, \Phi_5).
\ee
The deformed states are  mirror states for $W^\e$  hence we  can repeat the  4-point function   analysis  from previous section
\be
 \< \Psi_{1}^{W_\e},\Psi_2^{W_\e},\Psi_{3}^{W_\e}, \Psi_4^{W_\e}\>_{Q_{W^\e}  }    = \< \Psi_{1}^\e,\Psi_2^\e,\Psi_{3}^\e, \Phi_4^\e\>_{Q_{W^\e}  }    = \frac{d}{d\lambda}\Big|_{\lambda=0}  \< \Psi_{1}^{\e,\lambda},\Psi_2^{\e,\lambda},\Psi_{3}^{\e,\lambda}\>_{Q_{W^{\e,\lambda}}  }   
\ee
with   deformed superpotential 
\be
W^{\e,\lambda} = W^\e +\lambda \Phi^\e_4   = W + \e \Phi_5 + \lambda \Phi_4  + \lambda \e\; C^{trop}_W (\Phi_4 , \Phi_{5})
\ee
and  deformed states 
\be
\Psi_k^{\e,\lambda}   = \Psi^\e_k  +\lambda \;K_{W^\e} G_- \mu_2 (\Psi^\e_k , \Phi^\e_{4}).
\ee
\newline\newline
{\bf Remark}: The appearence of the $\lambda \e\; C^{trop}_W (\Phi_4 , \Phi_{5})$ is a signature of the non-linear relation between the linear times $\e, \lambda$ on the image of good section and coordinates $T$ in the superpotential deformation, introduced by K. Saito.
\newline\newline
The states $\Psi_k^{\e,\lambda}$ are mirror states hence we can express the  3-point function in terms of  the LGS theory 
\be
\< \Psi_{1}^{\e,\lambda},\Psi_2^{\e,\lambda},\Psi_{3}^{\e,\lambda}\>_{Q_{W^{\e,\lambda}}  }      =  \sum_{dW^{\e,\lambda} =0} \frac{\Phi^{\e,\lambda}_1\cdot\Phi^{\e,\lambda}_2 \cdot\Phi^{\e,\lambda}_2  }{\det \p_j \p_k W^{\e,\lambda}}
 =\< \Phi_{1}^{\e,\lambda},\Phi_2^{\e,\lambda},\Phi_{3}^{\e,\lambda}\>_{W^{\e,\lambda}  } 
\ee
for holomorphic germs
\be
\begin{split}
\Phi^{\e,\lambda}_k &= \Psi^{\e,\lambda}_k\Big|_{\psi=r=0}  = \Phi^\e_k  +\lambda \;K_{W^\e} G_- \mu_2 (\Psi^\e_k , \Phi^\e_{4})\Big|_{\psi=r=0}   \\
&= \Phi^\e_k +  \lambda \; C^{trop}_{W^\e} (\Phi^\e_k, \Phi^\e_4).
\end{split}
\ee
The $C^{trop}_{W^\e} (\Phi^\e_k, \Phi^\e_4)$ is a contact term for the good section for deformed superpotential $W_\e$.  Hence, we   can apply the LSG recursion formula  (\ref{n_pt_rec}) for the 4-point function
\be
\begin{split}
\< \Psi^\e_{1}&,\Psi^\e_2,\Psi^\e_{3}, \Psi^\e_4\>_{Q_{W^\e}}   =  \frac{d}{d\lambda}\Big|_{\lambda=0} \< \Psi_{1}^{\e,\lambda},\Psi_2^{\e,\lambda},\Psi_{3}^{\e,\lambda}\>_{Q_{W^{\e,\lambda}}  }   \\
& = \frac{d}{d\lambda}\Big|_{\lambda=0} \< \Phi^\e_1, \Phi^\e_2, \Phi^\e_3\>_{W^{\e, \lambda}} + \< C^{trop}_{W^\e} (\Phi^\e_1, \Phi^\e_4), \Phi^\e_2, \Phi^\e_3\>_{W^\e}\\
&\qquad+ \< \Phi^\e_1, C^{trop}_{W^\e}( \Phi^\e_2, \Phi^\e_4), \Phi^\e_3\>_{W^\e}+ \< \Phi^\e_1, \Phi^\e_2, C^{trop}_{W^\e}( \Phi^\e_3, \Phi^\e_4)\>_{W^\e} \\
& = \< \Phi^\e_{1},\Phi^\e_2,\Phi^\e_{3}, \Phi^\e_4\>^{S^{trop}}_{W^\e} 
\end{split}
\ee
Similarly we can apply the LSG recursion  (\ref{n_pt_rec}) for the 5-point function
\be\nn
\begin{split}
 \< \Psi^W_1&,\Psi^W_2,\Psi^W_3, \Psi^W_4, \Psi^W_5\>_{Q_W}   =  \frac{d}{d\e}\Big|_{\e=0} \< \Psi_1^\e, \Psi_2^\e, \Psi_3^\e, \Psi^\e_4\>_{Q_{W^\e}}  = \frac{d}{d\e}\Big|_{\e=0} \< \Phi^\e_{1},\Phi^\e_2,\Phi^\e_{3}, \Phi^\e_4\>^{S^{trop}}_{W^\e} \\
& = \frac{d}{d\e}\Big|_{\e=0} \< \Phi_{1},\Phi_2,\Phi_{3}, \Phi_4\>^{S^{trop}}_{W^\e} + \< C^{trop}_W (\Phi_1, \Phi_5), \Phi_2, \Phi_3, \Phi_4\>^{S^{trop}}_{W}+ \< \Phi_1, C^{trop}_W( \Phi_2, \Phi_5), \Phi_3, \Phi_4\>^{S^{trop}}_{W} \\
&\qquad+ \< \Phi_1, \Phi_2, C^{trop}_W( \Phi_3, \Phi_5), \Phi_4\>^{S^{trop}}_{W} + \< \Phi_1, \Phi_2, \Phi_3, C^{trop}_W( \Phi_4, \Phi_5)\>^{S^{trop}}_{W}  \\
& = \< \Phi_{1},\Phi_2,\Phi_{3}, \Phi_4, \Phi_5\>^{S^{trop}}_{W} 
\end{split}
\ee
what completes the proof of the proposition.
$\hfill\blacksquare$

\subsection{Parallel transport of a good section}
In LGS theory, given a good section $S_W$ for superpotential $W$ we can extend it to a good section $S_{W_t}$ for superpotential $W_t$ via the parallel transport.  
For a given superpotential $W$ and $\g\in H^\ast_{dR}(X) $ we construct a mirror state 
\be
\Psi^W_\g = \Psi_\g +  KG_- \mu_2 (W, \Psi_\g)+  KG_-\mu_2 (W, KG_- \mu_2 (W, \Psi_\g))+\dots
\ee
which  define an image of good section
\be
Im \; S^W = \mathbb{C}\< \Phi^W_\g\;| \g \in H_{dR}^\ast (X)\>.
\ee
Let $W \to   W + \delta W$,  then the  change  of the mirror state  to the leading order is given by 
\be
\begin{split}
 \Psi^{W+\delta W}_\g - \Psi_\g^W  = K_{W} G_- \mu_2 (\delta W, \Psi^{W}_\g) +\cO(\delta W)^2.
\end{split}
\ee
The corresponding change of holomorphic germ
\be
 \Phi^{W+\delta W}_\g - \Phi_\g^W =K_{W} G_- \mu_2 (\delta W, \Psi^{W}_\g)\Big|_{r=\psi=0} = C^{trop}_{W} (\delta W, \Phi^{W}_\g) = C^{S^{trop}}_{W} (\delta W, \Phi^{W}_\g)
\ee
Hence we demonstrated that the  tropical good section is parallel with respect to connection determined by it.

\subsection{Localization of correlation functions} \label{mirr_symmetry}
{\bf Theorem (localization of correlation functions)}: The   B-model correlation function  for the mirror states constructed for  observables $\g_k \in H_{dR}^\ast (X)$ on toric space $X$ equal to the   LGS correlation function 
\be
\< \Psi^X_{\g_1},...,\Psi^X_{\g_n}\>_{Q_X}  = \< \Phi^X_{\g_1},...,\Phi^X_{\g_n}\>_{W_X} 
\ee
The   mirror  LGS theory has   the following data
\begin{enumerate}
\item The LGS theory is on $\mathbb{C}^{\ast N}$, where $N$ is the complex dimension of $X$.
\item The holomorphic top form on $\mathbb{C}^{\ast N}$ written in terms of cylindrical coordinates $(r, Y)$ is
\be
\Omega  = dY_1\wedge ...\wedge dY_N.
\ee
\item The  LGS superpotential is a mirror superpotential  for $X$, written in terms of the compactifying divisors for $X$, i.e. 
\be
W_X  =  \sum _{\vec{b} \in B_X} q_{\vec{b}}\; e^{i \< \vec{b}, Y\>}.
\ee
\item The  image of a good section  
\be
Im \; S_W^{trop} = \mathbb{C}\<\Phi^W_\g|\; \g \in H^\ast_{dR}(X)\>.
\ee
\item The  LSG observables  are holomorphic germs for mirror states 
\be
\Phi^X_{\g} = \Psi^X_\g\Big|_{\psi=r=0}.
\ee 
\end{enumerate}
{\bf Proof}: We are going to embed the statement  of the theorem into more general equality 
\be
 \< \Psi^{W}_{\g_1},...,\Psi^{W}_{\g_n}\>_{Q_{W}}  = \< \Phi^{W}_{\g_1},...,\Phi^{W}_{\g_n}\>_{W}.  
\ee
The theorem is the case when  $W = W_X$. Such rewriting allows us to use an  induction in the number of states  $n$. The proof for $n=3$  follows from proposition on 3-point localization from section \ref{sec_3pt_b_model}.  We  use the invariance  theorem \ref{corr_inv}  under assumption of  conjecture \ref{sec_inv_conjecture}  to replace the  mirror state by its holomorphic germ in  the $(n+1)$-point correlation function 
\be
\< \Psi^W_{\g_1},...,\Psi^W_{\g_n}, \Psi^W_{\g_{n+1}}\>_{Q_W} =\< \Psi^W_{\g_1},...,\Psi^W_{\g_n}, \Phi^W_{\g_{n+1}}\>_{Q_W}.  
\ee
 We  apply   the recursion formula  theorem \ref{corr_recurs} for  deformation of  HTQM with differential $Q_W$ by a holomorphic germ state  $\Phi^W_{\g_{n+1}}$  and  express the $n+1$-point function as a derivative of n-point function in deformed theory, i.e.
\be
\begin{split}
\< \Psi^W_{\g_1},...,\Psi^W_{\g_n}, \Phi^W_{\g_{n+1}}\>_{Q_W}  &= \frac{d}{d\e}\Big|_{\e=0}  \< \Psi^{W^\e}_{\g_1},...,\Psi^{W^\e}_{\g_{n}}\>_{Q^\e_W  }.    
\end{split}
\ee
Using our expression (\ref{b_model_def_function}) for the B-model deformation by a holomorphic function  we  replace $Q_W^\e$ by $Q_{W^\e}$   for   superpotential, deformed by the holomorphich germ of $\Psi_{n+1}^W$
\be
W^\e = W +\e \Phi^W_{\g_{n+1}}.
\ee
Using an  assumption of induction that the equality holds  for  $n$-point correlation functions
\be
 \< \Psi^W_{\g_1},...,\Psi^W_{\g_n}\>_{Q_W} =  \< \Phi^W_{\g_1},...,\Phi^W_{\g_n}\>_{W} 
\ee
we can express the $(n+1)$-point correlation function in B-model in terms of $n$-point functions in LSG theory
\be
\begin{split}
\< \Psi^W_{\g_1},...,\Psi^W_{\g_n}, \Psi^W_{\g_{n+1}}\>_{Q_W}  &=\frac{d}{d\e}\Big|_{\e=0}  \< \Phi^{W^\e}_{\g_1},...,\Phi^{W^\e}_{\g_{n}}\>_{W^\e }    
\end{split}
\ee
for  holomorphic functions given by the holomorphic germs of deformed states
\be
 \Phi^{W^\e}_{\g_k} =  \Psi^{W^\e}_{\g_k} \Big|_{\psi=r=0} = \Phi_{\g_k}^W +2\pi K_WG_- \mu_2(\Phi^W_{\g_{n+1}}, \Psi^W_{\g_k})\Big|_{\psi=r=0}.
\ee
According to the  tropical good section proposition  \ref{prop_trop_section} we can identify the second term in the expression above with tropical contact term
\be
2\pi K_WG_-\mu_2 (\Phi^W_{\g_{n+1}}, \Psi^W_{\g_k})\Big|_{\psi=r=0} = C^{S^{trop}}_W (\Phi^W_{\g_k}, \Phi^W_{\g_{n+1}}). 
\ee
The recursion formula  (\ref{n_pt_rec}) for the LGS correlation functions  allows us to rewrite  the derivative of $n$-point function as $(n+1)$-point correlation function in LSG theory with superpotential $W$, i.e.
\be
\< \Psi^W_{\g_1},...,\Psi^W_{\g_n}, \Psi^W_{\g_{n+1}}\>_{Q_W}  =\frac{d}{d\e}\Big|_{\e=0}  \< \Phi^{W^\e}_{\g_1},...,\Phi^{W^\e}_{\g_{n}}\>_{Q_{W^\e}  }   =   \< \Phi^W_{\g_1},...\Phi^W_{\g_{n}},\Phi^W_{\g_{n+1}}\>_{W} 
\ee
and  complete the proof of the theorem.
$\hfill\blacksquare$

\section*{Acknowledgments}

We are grateful to  Yasha Neiman   for many discussions on the topics presented in this paper.   The work of  A.L. is supported  by the Basic Research Program of the National Research University Higher School of Economics and by Wu Wen-Tsun Key Lab of Mathematics. The work of V.L. is   supported by the Quantum Gravity Unit of the Okinawa Institute of Science and Technology Graduate University (OIST).

\bibliography{LGS_mirr_ref}{}
\bibliographystyle{utphys}

\end{document}